\documentclass[preprint,draft,showkeys,
showpacs,tightenlines,preprintnumbers,prc,amsmath,amssymb]{revtex4}

\def\AnswerYes{y}
\def\ShowLabelsVersion{n}         
\def\ShowChangesVersion{y}        
\def\ShowAnnotationsVersion{n}    

\usepackage{dcolumn}
\usepackage{bm}

\usepackage[dvips]{color}                 
\usepackage[final]{graphicx}              
\graphicspath{{figures/}{./}}

\usepackage{multirow}                     
\usepackage{amssymb}
\usepackage{amsmath}
\usepackage{xspace}               

\ifx\ShowLabelsVersion\AnswerYes
   \usepackage
{showkeys}
\fi

\ifx\ShowAnnotationsVersion\AnswerYes
   \newcommand{\comment}[1]{{\scriptsize\sffamily\bfseries{#1}}}
   \newcommand{\margin}[1]{\marginpar{\scriptsize\sffamily\bfseries{#1}}}
\else
   \newcommand{\comment}[1]{}
   \newcommand{\margin}[1]{}
   
\fi

\ifx\ShowChangesVersion\AnswerYes
   \usepackage{ulem}
    
   \newcommand{\delete}[1]{\sout{#1}}            
   \renewcommand{\emph}[1]{\textit{#1}}           
\else

   \newcommand{\sout}[1]{}
   \newcommand{\xout}[1]{}

   \newcommand{\delete}[1]{}
\fi

\newcommand{\hq}{\hspace{0.5em}}

\newcommand{\dd}{\mathrm{d}}
\newcommand{\ii}{\mathrm{i}}
\newcommand{\bra}{\langle}
\newcommand{\ket}{\rangle}
\newcommand{\de}{\partial}
\newcommand{\mpi}{m _\pi}
\newcommand{\ChiEFT}{$\chi$EFT\xspace}
\newcommand{\HIGS}{HI$\gamma$S\xspace}
\newcommand{\MeV}{\ensuremath{\mathrm{MeV}}}

\begin{document}
\preprint{arXiv:1006.4849 [nucl-th]}
\preprint{INT-PUB-10-021}
\preprint{\textbf{revised version with erratum 27 March 2012}}

\title{Nucleon Spin-Polarisabilities from Polarisation Observables\\
  in Low-Energy Deuteron Compton Scattering}

\author{Harald W.~Grie{\ss}hammer} \email{hgrie@gwu.edu} \affiliation{Center
  for Nuclear Studies, Department of Physics, George Washington University,
  Washington, DC 20052, USA.}

\author{Deepshikha Shukla}
\email{dshukla@physics.unc.edu} \affiliation{Center for Nuclear Studies,
  Department of Physics, George Washington University, Washington, DC 20052,
  USA.}\affiliation{Department of Physics and Astronomy, The University of
  North Carolina, Chapel Hill, NC,USA.}

\date{\today}

\begin{abstract}
  We investigate the dependence of polarisation observables in elastic
  deuteron Compton scattering below the pion production threshold on the
  spin-independent and spin-dependent iso-scalar dipole polarisabilities of
  the nucleon.  The calculation uses Chiral Effective Field Theory with
  dynamical $\Delta(1232)$ degrees of freedom in the Small Scale Expansion at
  next-to-leading order. Resummation of the $NN$ intermediate rescattering
  states and including the $\Delta$ induces sizeable effects. The analysis
  considers cross-sections and the analysing power of linearly polarised
  photons on an unpolarised target, and cross-section differences and
  asymmetries of linearly and circularly polarised beams on a vector-polarised
  deuteron. An intuitive argument helps one to identify kinematics in which
  one or several polarisabilities do not contribute. Some double-polarised
  observables are only sensitive to linear combinations of two of the
  spin-polarisabilities, simplifying a multipole-analysis of the data.
  Spin-polarisabilities can be extracted at photon energies
  $\gtrsim100\;\MeV$, after measurements at lower energies of
  $\lesssim70\;\MeV$ provide high-accuracy determinations of the
  spin-independent ones.  An interactive \emph{Mathematica 7.0} notebook of
  our findings is available from hgrie@gwu.edu.

  \textbf{Note:} The original publication contained three errors. The Erratum
  published in Eur.~Phys.~J.~\textbf{A} (2012) is reprinted in its entirety on
  the following page. The changes outlined there have been made to the text in
  this arXiv version, and the variations of $\gamma_{E1E1}$ in figs.~6, 11,
  13, 16, 18, 20, 22 have been replaced by those which indicated the variation
  as in the captions.
\end{abstract}



\maketitle

\newpage

\begin{center}
  \textbf{\large Erratum to Europ.~Phys.~J.~\textbf{A46}, 249
    (2010):\\
    Nucleon Spin-Polarisabilities from Polarisation\\ Observables in
    Low-Energy Deuteron Compton Scattering}

\vspace{2ex}

\textbf{Harald W.~Grie{\ss}hammer$^{a}$}\footnote{Email:
    hgrie@gwu.edu; corresponding author}

\textit{and}

\textbf{Deepshikha Shukla$^{b}$}

\vspace{2ex}

\textit{$^a$ Institute for Nuclear Studies, Department of Physics,\\ George
  Washington University, Washington, DC 20052, USA.}

\textit{$^b$ Department of Physics \& Astronomy, \\James Madison University,
  Harrisonburg, VA 22807, USA}

\date{\today}
\end{center}


We correct three errors in the original publication. 

First, eq.~(21) on p.~255 describing the implementation of varying the
individual polarisabilities contains an incorrect kinematic pre-factor
$\sqrt{s_{\gamma N}}/M$. This should be deleted, as should the references to
it in the text immediately following. The numerical implementation does not
contain this factor and thus remains unchanged. 

Second, the original version implies that the sensitivity of observables is
analysed by varying the \emph{iso-scalar} values of the scalar and
spin-polarisabilities by an iso-scalar amount of $\pm2$ canonical units.  This
is not the case. Instead, setting $\delta\alpha_{E1}$, $\delta\beta_{M1}$,
$\delta\gamma_{E1E1}$, $\delta\gamma_{M1E2}$, $\delta\gamma_{M1M1}$,
$\delta\gamma_{E1M2}$ to $\pm2$ units varies the polarisabilities of only
\emph{one} of the nucleons by $\pm2$ units, while that of the other nucleon is
kept at the iso-scalar value.
Two paragraphs after eq.~(21) (starting at bottom of left column on p.~255),
the first sentence is thus replaced by:

\emph{In the next step, these contributions are independently varied by $\pm
  2$ canonical units for one nucleon to analyse the effect of each on the
  various observables. The polarisabilities of the other nucleon are kept
  fixed at the iso-scalar value. This corresponds to a change of the
  iso-scalar polarisabilities by half as much, i.e.~by $\pm1$ unit. Since
  deuteron Compton scattering is sensitive only to iso-scalar quantities,
  varying either the proton or neutron polarisabilities leads to the same
  result. In practise, the scalar polarisabilities of the proton are better
  constrained, and deuteron Compton scattering experiments are more likely
  focused on extracting neutron polarisabilities. In that case, these studies
  can be interpreted as providing the sensitivities on varying the neutron
  polarisabilities by $\pm2$ units, with fixed proton polarisabilities. The
  spin-independent polarisabilities\dots}

Third, the variation of the spin-polarisability $\gamma_{E1E1}$ was
implemented with an incorrect sign. The correct results are obtained by
re-interpreting the plots only of $\delta\gamma_{E1E1}$ in figs.~6, 11, 13,
16, 18, 20, 22: Dotted lines represent a change by $+2$ units, dashed ones by
$-2$ units.

Except for the last modification, the figures and conclusions remain
unchanged. The corresponding \emph{mathematica} notebook, available from
\texttt{hgrie@gwu.edu}, has been adjusted. It now reflects variations of the
\emph{iso-scalar} polarisabilities by $\pm2$ units (i.e.~of one individual
nucleon polarisability by $\pm4$ units, with those of the other kept fixed),
thus complementing the article's perspective.

\textbf{Note:} The changes outlined in this erratum have been made to the text
in this arXiv version, and the variations of $\gamma_{E1E1}$ in figs.~6, 11,
13, 16, 18, 20, 22 have been replaced by those which indicated the variation
as in the captions.  \newpage

\section{Introduction}
\label{sec:introduction}

Polarisabilities quantify in detail the two-photon response of the charge and
current distributions of the effective low-energy degrees of freedom inside
the nucleon to external electro-magnetic fields, see
e.g.~\cite{hgrieproc,barry,Report,Sc05,Phillips:2009af} for reviews. In particular at
\HIGS~\cite{Weller:2009zza,Miskimen,Miskimentalk,Weller,Gao,Ahmed},
MAMI~\cite{AhrensBeckINT08} and MAXlab~\cite{Feldman:2008zz,Feldman2}, a host of
Compton scattering experiments on the proton, deuteron and ${}^3$He are
planned or under way to test the predictions of theories
and models. The goal of this article is to support and trigger
experimental planning and analysis by a model-independent assessment of
the sensitivity of elastic deuteron Compton scattering observables with
polarised beams and/or targets at energies $\lesssim130\;\MeV$ on nucleon
polarisabilities. An interactive \emph{Mathematica 7.0} notebook of our
findings is available from Grie\3hammer (hgrie@gwu.edu).

At such energies, the dominant response terms are
the six polarisabilities in which at least one of the photons couples to a
dipole. They are canonically parameterised starting from the most
general interaction between the nucleon $N$ with spin $\vec{\sigma}/2$ and an
electro-magnetic field of fixed, non-zero energy $\omega$ in the
centre-of-mass (cm) frame, see e.g.~\cite{barry,hgrieproc}:
\begin{eqnarray}
  \mathcal{L}_\text{pol}&=&2\pi\;N^\dagger \;\big[{\alpha_{E1}(\omega)}\;\vec{E}^2\;+
  \;{\beta_{M1}(\omega)}\;\vec{B}^2\; +\;{\gamma_{E1E1}(\omega)}
  \;\vec{\sigma}\cdot(\vec{E}\times\dot{\vec{E}})\;
  +\;{\gamma_{M1M1}(\omega)}
  \;\vec{\sigma}\cdot(\vec{B}\times\dot{\vec{B}}) \nonumber \\
  &&\;\;\;\;\;\;\;\;\;\;\;\;
  -\;2{\gamma_{M1E2}(\omega)}\;\sigma_i\;B_j\;E_{ij}\;+
  \;2{\gamma_{E1M2}(\omega)}\;\sigma_i\;E_j\;B_{ij} \;+\;\dots \big]\;N
  \label{eq:pols-lag}
\end{eqnarray} 
The electric or magnetic photon undergoes a transition ${Xl\to Yl^\prime}$
(${X,Y=E,M}$) of definite multipolarity ${l,l^\prime=l\pm\{0;1\}}$;
${T_{ij}:=\frac{1}{2} (\de_iT_j + \de_jT_i)}$. The electric and magnetic
polarisabilities $\alpha_{E1}(\omega)$ and $\beta_{M1}(\omega)$ are
independent of the nucleon spin. In contradistinction, the four
spin-polarisabilities couple to the nucleon spin.  In the ``pure''
spin-polarisabilities $\gamma_{E1E1}(\omega)$ and $\gamma_{M1M1}(\omega)$,
only the electric or magnetic dipoles of the photon probe the nucleon.  On the
other hand, $\gamma_{M1E2}(\omega)$ and $\gamma_{E1M2}(\omega)$ are ``mixed''
polarisabilities of dipole and quadrupole transitions. At very low photon
energies $\omega\ll30\;\MeV$, the contributions of $\alpha_{E1}(\omega)$ and
$\beta_{M1}(\omega)$ to the scattering amplitude vary as $\omega^2$, while the
spin-polarisabilities behave as $\omega^3$ and are thus suppressed. The static
values themselves, ${\alpha}_{E1}\equiv\alpha_{E1}(\omega=0)$ and
${\beta}_{M1}\equiv\beta_{M1}(\omega=0)$, are non-zero and often simply called
``the polarisabilities''.

In elastic deuteron Compton scattering $\gamma d\to\gamma d$
(see~\cite{Phillips:2009af} for a review) only the average nucleon
polarisabilities
${\alpha}_{E1}=\frac{1}{2}({\alpha}^{(p)}_{E1}+{\alpha}^{(n)}_{E1})$ etc.~are
directly accessible, since the deuteron is an iso-scalar.  A recent analysis
of all elastic deuteron Compton scattering data in Chiral Effective Field
Theory \ChiEFT with dynamical $\Delta(1232)$ degrees of freedom
quotes~\cite{Hi05,Hi05b}
\begin{equation}
  \label{eq:robertpols}
  {\alpha}_{E1}=11.3\pm0.7_\mathrm{stat}\pm0.6_\mathrm{Baldin}
  \pm1_\mathrm{theory}
  \;,\; 
  {\beta}_{M1} =3.2\mp0.7_\mathrm{stat}\pm0.6_\mathrm{Baldin}
  \pm1_\mathrm{theory}\;\;.
\end{equation}
The error-bars represent uncertainties in the Baldin sum rule, statistics, and
higher-order effects in the systematic \ChiEFT expansion. Other extractions
and the value quoted by the Particle Data Group are compatible, but with
larger error-bars~\cite{Ko03,Lu03,Sc05,Hi05a,Be99,Be02,Be04}.  Throughout,
the values are quoted in the ``canonical units'' of $10^{-4}$~fm$^3$ for the
spin-independent polarisabilities and $10^{-4}$~fm$^4$ for the
spin-polarisabilities. Lattice QCD simulations at pion masses which allow for
chiral extrapolations are also pursued, albeit many hurdles have still to be
overcome; for recent progress,
see~\cite{Alexandru:2009id,Engelhardt:2010tm,Detmold:2010ts}.

In contrast to the static values, the \emph{energy-dependent} or
\emph{dynamical polarisabilities} are identified \emph{at fixed energy} only
by a multipole-analysis, i.e.~by their different angular dependence in Compton
scattering~\cite{Griesshammer:2001uw,Hi04,hgrieproc}. While quite different
frameworks could reproduce the zero-energy values, the underlying mechanisms
are only properly revealed by investigating their energy-dependence.  For
example, detailed implications of chiral symmetry on the pion cloud as well as
effects from the $\Delta(1232)$ as lowest-lying nuclear excitation have a
substantial impact in particular at energies around and above the pion mass,
as demonstrated
in~\cite{Hi04,Hildebrandt:2003md,Hi05,Hi05a,Hi05b,Pascalutsa:2002pi,Lensky:2009uv}.
Compton scattering on nucleons and light nuclei provides thus a wealth of
information about the internal structure of the nucleon itself.

Of particular interest are the four so far practically un-determined
spin-polarisabilities. They parameterise the response of the nucleon
\emph{spin} to the photon field, in analogy to rotations of the polarisation
plane of linearly polarised photons propagating through a medium in the
presence of a constant magnetic field (Faraday effect). In the nucleon, the
permanent magnetic moment of the nucleon spin provides the necessary magnetic
background field.

For both the proton and neutron, only linear combinations of the four
spin-polarisabilities have been constrained in forward and backward
scattering. We concentrate here on the neutron. With
$\gamma_i\equiv\gamma_i(\omega=0)$, quasi-free Compton scattering on the
deuteron~\cite{Ko03} reports
\begin{eqnarray}
  \gamma_{\pi}^{(n)} &=&
  -{\gamma}_{E1E1}^{(n)}+{\gamma}_{M1M1}^{(n)}-{\gamma}_{E1M2}^{(n)}+
  {\gamma}_{M1E2}^{(n)} \nonumber\\&=& 
  58.6\pm  4.0_\text{model}-[42.7\dots46.7]_{\pi^0\text{-pole}}=
  [15.9\dots11.9]\pm4.0_\text{model}\;\;, 
  \label{eq:gpn}
\end{eqnarray}
where the $t$-channel $\pi^0$-pole piece (with uncertainties from different
extractions) is subtracted~\cite{Sc05,Report}.  Due to the large error-bars,
this is compatible with theory predictions of $11.8$~\cite{Pa03} and
$8.86\pm0.25_\text{stat}\pm1_\text{theory}$~\cite{Hi04} from two variants of
\ChiEFT with dynamical $\Delta(1232)$, and $13.7$ from fixed-$t$ dispersion
relation~\cite{Hi04}. The forward spin-polarisability is inferred as
compatible with zero from a VPI-FA93 multipole analysis~\cite{Sa94} of the
Gell-Mann Goldberger Thirring sum rule~\cite{ggt}, from
\ChiEFT~\cite{Pa03,Hi04,VijayaKumar:1999cd}, or from fixed-$t$ dispersion
relations~\cite{Hi04}:
\begin{equation}
  \gamma_0^{(n)} = -{\gamma}_{E1E1}^{(n)}-{\gamma}_{M1M1}^{(n)}-
  {\gamma}_{E1M2}^{(n)}-{\gamma}_{M1E2}^{(n)}
  \simeq0 \label{eq:g0n}
\end{equation}
The spread may be conservatively estimated as $\pm4$ since both theoretical
and experimental uncertainties are generally believed to be large, with for
example \ChiEFT results only poorly converging, see e.g.~\cite{Hemmert:2000dp}
for a more thorough discussion. Neither linear combination allows one
therefore to interpret the detailed dynamics of charges and currents in the
nucleon. To test theories and models, reliable experiments are needed which
determine all polarisabilities with minimal theoretical bias, and theoretical
uncertainties should be assessed with care.

Polarised Compton scattering off the free proton and neutron has been studied
in the same framework used here~\cite{Hildebrandt:2003md}. Due to the lack of
free neutron targets, Compton scattering on the lightest nuclei is however the
likely avenue to hunt for the elusive neutron polarisabilities. This work
focuses therefore on identifying deuteron observables that help to extract the
six iso-scalar nucleon polarisabilities. Previous work on un-polarised
deuteron Compton scattering include calculations using traditional potential
models~\cite{We83,We83a,Wi95a,Lv95,Lv00,Levchuk:2000mg,Ka99} and
\ChiEFT~\cite{Be99,Be02,Be04,Hi05a,Hi05,Hi05b}, see
e.g.~\cite{Phillips:2009af} for an overview. In addition, Compton scattering
off ${}^3$He provides an promising avenue to extract neutron
polarisabilities~\cite{mythesis,Ch07,Sh08}.

Polarisation variables with vector-polarised deuterons and circularly
polarised photons were to our knowledge first considered by Chen, Ji and Li in
``pion-less'' EFT~\cite{Chen:2004wwa}. We extend a previous study in \ChiEFT
for energies between $70$ and $135\;\MeV$~\cite{Ch05, mythesis} to include
both the $\Delta(1232)$ as dynamical degree of freedom in the Small Scale
Expansion variant~\cite{He97, He98} and the correct Thomson limit, as
described in detail in Refs.~\cite{Hi05,Hi05b}. This increases the region of
applicability to energies between zero and the pion-production threshold. As
in Refs.~\cite{Ch05, mythesis}, we investigate linearly polarised photons on
an un-polarised deuteron target and circularly polarised photons on a
vector-polarised target, but add a study of linearly polarised photons on a
vector-polarised target.  High-flux, high-accuracy,
high-degree-of-polarisation experiments are particularly suited for the newest
generation of facilities like \HIGS~\cite{Weller:2009zza}. Our results
indicate that several polarisation observables can be instrumental to extract
not only static and energy-dependent spin-independent nucleon
polarisabilities, but also the spin-dependent ones.

The contents are organised as follows: In Sec.~\ref{sec:calc}, we summarise
the essential ingredients of the calculation. The explicit $\Delta$ and
$NN$-intermediate states provide substantial corrections in observables,
Sec.~\ref{sec:comparison}.  Section~\ref{sec:lpol} contains the analysis of
single-polarisation observables. Results are reported in Sec.~\ref{sec:ldpol}
for the double-polarisation observables with linearly-polarised photons, and
in Sec.~\ref{sec:cpol} for double-polarisation observables with
circularly-polarised photons. Besides the customary summary, we conclude in
Sec.~\ref{sec:summary} by proposing a road-map to experimentally determine the
iso-scalar, spin-independent and spin-dependent nucleon polarisabilities to
high accuracy from experiment using deuteron targets. Preliminary findings
were published in a proceeding~\cite{Griesshammer:2009pq}.

\section{Methodology}
\label{sec:calc}

\subsection{Outline of Amplitude Calculation}
\label{sec:amplitudes}

Since the deuteron Compton scattering amplitudes are computed as described in
detail in Refs.~\cite{Hi05b,Hi05}, we here only briefly recapitulate the main
ingredients. The amplitudes are
\begin{equation} 
  \mathcal{M}_{\lambda_f, \lambda_i}^{M_f, M_i} = \bra
  M_f,\lambda_f|T_{\gamma NN}| M_i, \lambda_i \ket\;\;,
  \label{eq:M}
\end{equation}
where $\lambda_{i/f}=\pm$ denotes the circular polarisation of the
initial/final photon and $M_{i/f}\in\{0;\pm1\}$ the magnetic quantum number of
the initial/final deuteron spin. In \ChiEFT with explicit $\Delta(1232)$
degrees of freedom using the Small Scale Expansion (SSE)~\cite{He97,He98}, the
interaction kernel $T_{\gamma NN}$ is expanded up to next-to-leading order,
$\mathcal{O}({\epsilon}^3)$. The expansion parameter $\epsilon$ denotes a
typical low-energy momentum like the photon energy, the pion mass or the mass
difference between the real part of the position of the $\Delta(1232)$-pole in
the $S$-matrix and the nucleon mass, ${\mathrm{Re}}[M_{\Delta}]
-M=271.1\;\MeV$, each measured in powers of the typical break-down scale of
SSE, which is estimated to be $\lesssim1$~GeV. The kernel is finally
convoluted with deuteron wavefunctions to obtain the amplitudes
$\mathcal{M}_{\lambda_f, \lambda_i}^{M_f, M_i}$. All contributions to the
kernel are listed in~\cite{Hi05b,Hi05}.  In the one-nucleon sector, these are:
\begin{figure}[!htb]
  \begin{center}
    \includegraphics*[width=0.95\linewidth]{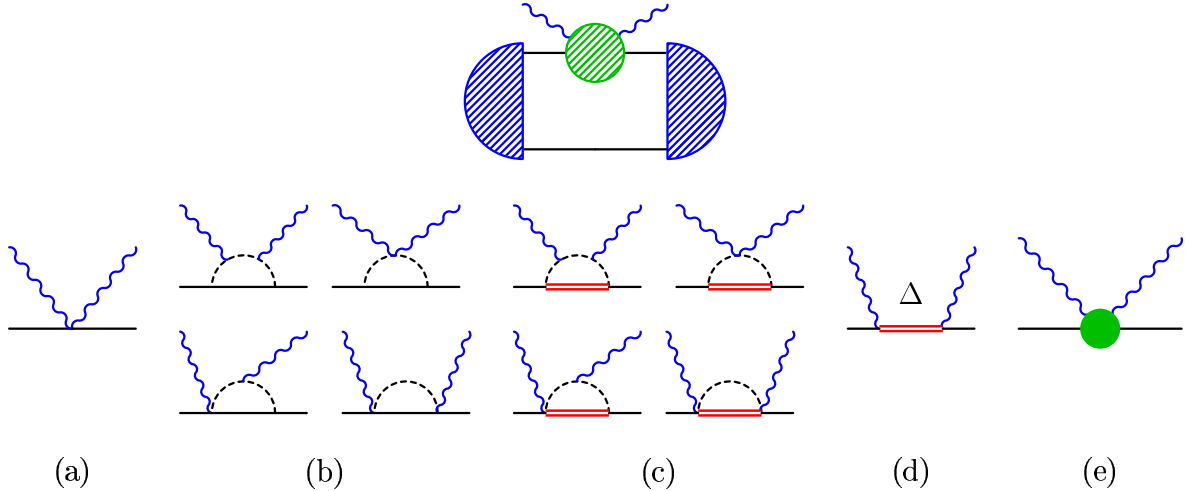}
    \caption{One-nucleon contributions to $\gamma d$ scattering in SSE \ChiEFT
      at order $\epsilon^3$ (permuted and crossed diagrams not shown). Top:
      embedding into the deuteron. Bottom: one-nucleon Thomson term (a); pion
      cloud around the nucleon (b) and $\Delta(1232)$ (double line; (c));
      excitation of an intermediate $\Delta$ (d); short-distance effects to
      $\alpha_{E1}$ and $\beta_{M1}$ (e).}
\label{fig:onenucleon}
\end{center}
\end{figure}
\begin{enumerate}
\item Thomson scattering on one nucleon, Fig.~\ref{fig:onenucleon} (a).
\item Coupling to the chiral dynamics of the pion cloud around one nucleon
  Fig.~\ref{fig:onenucleon} (b).
\item Excitation of the $\Delta(1232)$ intermediate state,
  Fig.~\ref{fig:onenucleon} (d), and coupling to the pion cloud around it
  Fig.~\ref{fig:onenucleon} (c). The relevance of the $\Delta(1232)$ will be
  discussed in Sec.~\ref{sec:comparison}.
\item Two energy-independent, iso-scalar short-distance coefficients,
  Fig.~\ref{fig:onenucleon} (e). They encode the contributions to the nucleon
  polarisabilities which come at this order neither from the deformation of
  the pion cloud around the nucleon or $\Delta$, nor from the excitation of
  the $\Delta$ itself. Strictly speaking, they are formally of higher order,
  $\mathcal{O}({\epsilon}^4)$. However, we follow
  Refs.~\cite{Hi04,Hi05a,Hi05b,Hi05} in modifying the SSE to account for the
  stark discrepancy between the experimental static spin-independent dipole
  polarisabilities and the $\mathcal{O}({\epsilon}^3)$ SSE values due to
  $\Delta$ effects. These ``off-sets'' are determined by data as described
  below, but the energy- and isospin-dependence of the spin-independent
  polarisabilities are at this order predicted in SSE \ChiEFT.
\item Since the deuteron is an iso-scalar, there is no contribution from the
  iso-vector $t$-channel $\pi^0$ coupling to two photons via the axial
  anomaly.
\end{enumerate}

Polarisabilities parameterise the ``structure'' part of the amplitudes and
contain -- as any multipole-expansion -- not more or less information than the
amplitude itself. However, decomposing into channels of different
multipolarity makes the information more easily accessible and interpretable.
For example, the strongly para-magnetic coupling of the photon and nucleon to
the dynamical $\Delta$ dominates the energy-dependence of the polarisabilities
$\beta_{M1}$ and $\gamma_{M1M1}$, while pion-cloud effects dominate
$\alpha_{E1}$ and $\gamma_{E1E1}$. At this order in SSE, the polarisabilities
are uniquely extracted from the sum of contributions 2 to 5, see~\cite{Hi04}.
While there is some ambiguity which makes it necessary to specify exactly how
the ``pole'' and ``structure'' parts are divided, this artificial separation
has of course no bearing on observables. We also found that contributions from
quadrupole and higher polarisabilities are negligible in the present analysis,
as was the case in unpolarised observables~\cite{Hi04,Hi05b,Hi05}.

The numerical values of all parameters were taken from
Ref.~\cite[Table~1]{Hi05a}. In particular, the strength of the $\gamma N
\Delta$ vertex is determined by the decay $\Delta \rightarrow \gamma N$. In
addition, the two short-distance contributions to the spin-independent
polarisabilities are fitted to the 28 data of elastic deuteron Compton
scattering below $100\;\MeV$, using the iso-scalar Baldin sum-rule
\begin{equation}
\alpha_{E1}+\beta_{M1}=14.5\pm0.6
\label{eq:baldin}
\end{equation}
as constraint~\cite{Olmos,Ko03}. This value is within error-bars of a SAID
analysis~\cite{Ar02} and an extraction by Levchuk and L'vov~\cite{Lv00}. The
results of the static spin-independent polarisabilities are given in
\eqref{eq:robertpols}, Refs.~\cite{Hi05,Hi05b}.  At this order in \ChiEFT, the
proton and neutron polarisabilities are identical, and the
spin-polarisabilities are parameter-free predictions. For reference, the
central values of the iso-scalar dipole polarisabilities are quoted from
Hildebrandt et al.~\cite{Hi04,Hi05,Hi05b} (with theoretical uncertainties of $\approx\pm1$ from
higher-order contributions):
\begin{eqnarray}
  {\alpha}_{E1} = 11.3\;\;&,& \;\;{\beta}_{M1} = 3.2 \nonumber \\
  \gamma_{E1E1} = -5.5\;\;, \;\;
  \gamma_{M1M1} = 3.1 \;\;&,& \;\;
  \gamma_{M1E2} = 1.0 \;\;, \;\;
  \gamma_{E1M2} = 1.0 
  \label{eq:cval}
\end{eqnarray}
There are two classes of diagrams in the two-nucleon sector, see
Fig.~\ref{fig:twonucleon}, in extension of the \ChiEFT $\mathcal{O}(Q^3)$
amplitudes without explicit $\Delta(1232)$ which were derived for
$\omega\sim\mpi$ in~\cite{Be99}:
\begin{figure}[!htb]
  \begin{center}
    \includegraphics*[width=0.95\linewidth]{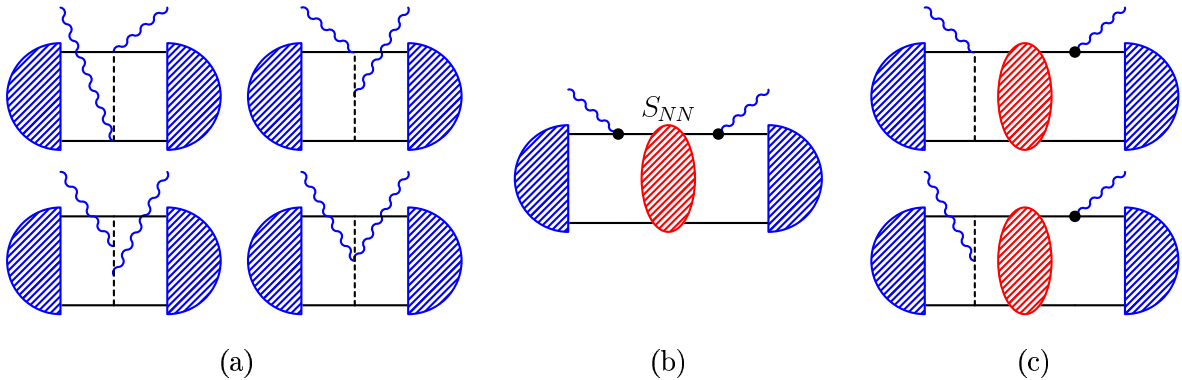}
    \caption{Two-nucleon contributions to $\gamma d$ scattering in SSE \ChiEFT
      at order $\epsilon^3$ (permuted and crossed diagrams not shown). Photons
      couple to the same pion (a); re-scattering contributions ((b,c);
      ellipse: two-nucleon $S$-matrix; dot: coupling via minimal substitution
      or magnetic moment).}
\label{fig:twonucleon}
\end{center}
\end{figure}
\begin{enumerate}
\item Both photons couple to the same pion-exchange current,
  Fig.~\ref{fig:twonucleon} (a)~\cite{Be99}.
\item Rescattering contributions, Fig.~\ref{fig:twonucleon} (b/c): The photons
  couple to the nucleon charge, magnetic moment and/or to different
  pion-exchange currents. Between the two photon couplings, the nucleons
  re-scatter arbitrarily often, via the full $NN$ S-matrix. These parts are
  computed in the ``Green's Function Hybrid Approach'' detailed in
  Refs.~\cite{Hi05b,Hi05,Ka99} and references therein.
\end{enumerate}
One can show that the latter contribution must be included in any consistent
power-counting of \ChiEFT for the two-nucleon
system~\cite{Griesshammer:2007aq,hgrieproc,hgrieforthcoming}. This can be understood
intuitively as follows: The initial coherent two-nucleon state is not
perturbed too much by absorbing or emitting a very-low-energy photon with
energy $\omega$. Due to the large $NN$ scattering lengths, the two nucleons
can interact multiple times over a typical time- and length-scale $1/\omega$
before another photon is emitted, producing a final state which contains an
outgoing photon and a deuteron. It has been demonstrated by Friar and
Arenh\"ovel that these rescattering contributions ensure the correct Thomson
limit of $\gamma d$ scattering. This exact low-energy theorem is a consequence
of current conservation, which turn is equivalent to gauge-invariance at this
order~\cite{Friar,We83,We83a}. In contradistinction, rescattering
contributions become small and hence perturbative at higher photon energies.
This is intuitively clear, as the struck nucleon has then only a very short
time $\sim1/\omega$ to scatter with its partner before the coherent final
state must be restored. Reference~\cite{Hi05,Hi05b} showed that
rescattering contributions are negligible in unpolarised Compton scattering
for $\omega\gtrsim60\;\MeV$, except for a markedly suppressed residual
dependence on the details of the deuteron wave function used. The importance
of rescattering for polarisation observables will be assessed in
Sec.~\ref{sec:comparison}.

With the correct Thomson limit and explicit $\Delta$-degrees of freedom, the
range of validity of SSE \ChiEFT spans therefore in principle from $\omega =
0$ to the the $\Delta$ resonance region, where finite $\Delta$-with effects
must be taken into account. In practise, the present formulation puts the
pion-production threshold $\gamma d\to\pi X$ not at the exact kinematically
correct position, but higher orders in $\epsilon$ include the deuteron recoil
effects in perturbation. Since the polarisabilities and observables around
threshold are quite sensitive to the exact threshold
position~\cite{Hi05a,Hi05}, we limit the discussion to
$\omega\lesssim120\;\MeV$, where such effects are small.

By convoluting phenomenological wave-functions with interactions from EFT, we
follow a ``hybrid approach'' advocated by Weinberg~\cite{Weinberg,Weinberg2}.  As
in~\cite{Hi05b,Hi05}, the Argonne V18 potential~\cite{AV18} is used to produce
the $NN$-rescattering matrix, together with the ``NNLO'' \ChiEFT deuteron
wave-function at cut-off $650\;\MeV$~\cite{Epelbaum,Epelbaum2}. Since the deuteron is an
iso-scalar, neither the deuteron wave-function nor the intermediate $NN$
potential contains a $\Delta N$ component, so that using wave-functions and
potentials without explicit $\Delta$ degrees of freedom is justified. We also
verified that using instead the AV18 or Nijmegen93~\cite{Nijm93} deuteron
wave-functions, or the LO chiral potential to generate the $NN$ S-matrix in
the intermediate state, induces only negligible differences $\lesssim3\%$ in
all observables even at $\omega\approx 125\;\MeV$, as found for unpolarised
cross-sections in~\cite{Hi05,Hi05b}. In a fully self-consistent EFT
calculation, the kernel, wavefunctions and potential should of course be
derived in the same framework.  However, issues of matching electro-magnetic
currents~\cite{Kolling:2009iq,Pastore:2009is}, wave-functions and potentials
only appear at two higher orders than considered here and may thus be taken as
estimate of the size of higher-order corrections. At lower energies, the
Thomson limit decreases this residual uncertainty even more, since Friar and
Arenh\"ovel demonstrated that for $\omega\to0$, the Thomson limit is restored
for any combination of potential and wave-function~\cite{Friar,We83,We83a}.
Numerically, the Thomson limit is fulfilled on the $0.6\%$-level.

We close by clearly stating the limitations of this analysis. The purpose is a
study emphasising the \emph{relative sensitivity} of Compton scattering
observables on \emph{varying} the polarisabilities. \emph{Credible
  predictions} of their absolute magnitudes are however only meaningful when
all systematic uncertainties are properly propagated into observables. Such
errors include: theoretical uncertainties from discarding contributions in SSE
\ChiEFT which are higher than order $\epsilon^3$; uncertainties from data and
the Baldin Sum rule in $\alpha_{E1}$ and $\beta_{M1}$ \eqref{eq:robertpols};
and to a lesser extend error-bars of the $\gamma N\Delta$ determination,
residual wave-function and potential dependence as well as numerical
uncertainties.  Such an effort is under way as part of a comprehensive
approach of Compton scattering on the proton, deuteron and ${}^3$He in \ChiEFT
from zero energy to beyond the pion-production threshold~\cite{allofus}.

\subsection{Observables}
\label{sec:obs}

With 2 photon helicities $\lambda_{i/f}=\pm1$ and 3 value for the deuteron
polarisation $M_{i/f}\in\{0;\pm1\}$ in both the in- and out-state,
$(2\times3)^2=36$ helicity amplitudes exist. As constructed by Chen, Ji and
Li, parity and time-reversal symmetry leave 12 independent structures
describing scattering of photons on a spin-$1$ target. A scalar or umpolarised
target probes $2$ of them, a vector-polarised one $4$, and a tensor-polarised
$6$. As each amplitude is complex and the overall phase is fixed, $23$
independent observables per energy and angle are needed to fully determine the
Compton scattering amplitude. Our goal is however not a comprehensive study of
all observables with polarised in and/or out states.  Rather, we only consider
elastic reactions in which the polarisations in the final state are not
detected and the initial deuteron is either unpolarised or vector-polarised.
Amongst these, several are not independent. We look for those with the
strongest signal of scalar and spin-polarisabilities.

We define our co-ordinate system by choosing the beam direction to be the
$z$-axis, the scattering plane to be the $xz$-plane, with the $y$-axis
perpendicular to it. The differential cross-section for unpolarised deuteron
Compton scattering is
\begin{equation}
  \frac{\dd\sigma}{\dd\Omega}= \Phi^2 \frac{1}{6} \sum \limits_{M_f, M_i ;
    \lambda_f, \lambda_i} |\mathcal{M}_{\lambda_f, \lambda_i}^{M_f, M_i}|^2\;\;. 
  \label{eq:dcs}
\end{equation}
The factor $\frac{1}{6}$ comes from averaging over the initial deuteron and
photon polarisations. In the centre-of-mass (cm) and laboratory (lab) frames,
the phase-space factor $\Phi$ is:
\begin{eqnarray}
  \Phi_{\rm cm} &=& \dfrac{M_d}{4 \pi \sqrt{s_{\gamma d}}}\;\;,\;\; \mbox{
    where }\, s_{\gamma d}= \omega + \sqrt{\omega^2 + M_d^2} \nonumber \\
  \Phi_{\rm lab} &=& \dfrac{\omega_f}{4\pi \omega_i}\;\;,\;\; \mbox{ where }\,\,
  \omega_f=\dfrac{M_d \omega_i}{M_d + \omega_i (1 - \cos \theta_{\rm lab})}\;\;.
  \label{eq:phase}
\end{eqnarray}
Here, $\theta_\text{lab}$ is the scattering angle in the lab frame, $M_d$ the
deuteron mass, $\omega_i$ ($\omega_f$) the initial (final) photon energy in
the lab frame, $\sqrt{s_{\gamma d}}$ the total energy in the $\gamma d$ cm
frame, and $\omega = \frac{\omega_i}{\sqrt{1+2\omega_i/M_d}}$ the photon
energy in the cm frame.

A host of observables can be defined in deuteron Compton scattering with
incoming circularly or linearly polarised photon beams and unpolarised or
polarised deuteron targets, but without detection of the final-state
polarisations. We consider here only subset involving an unpolarised or
vector-polarised deuteron.

\begin{figure}[!htb]
  \begin{center}
    \includegraphics*[width=0.6\linewidth]{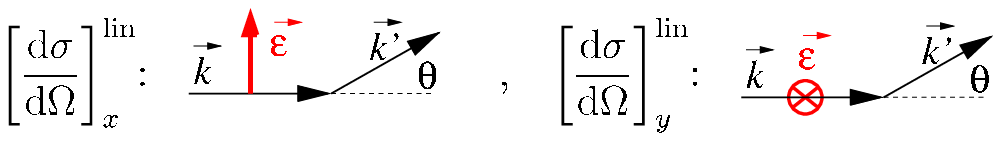}
    \caption {(Colour online) Observables for incoming
      linearly-polarised photon and unpolarised target.}
\label{fig:lin-nospin}
\end{center}
\end{figure}

Two independent observables exist for linearly-polarised photons and an
unpolarised target, Fig.~\ref{fig:lin-nospin}.
The cross-section for an incoming photon polarised in the scattering plane is
denoted by $\left[ \dfrac{\dd\sigma}{\dd\Omega}\right]^\text{lin}_x$, and by
$\left[ \dfrac{\dd\sigma}{\dd\Omega}\right]^\text{lin}_z$ for polarisation
perpendicular to the scattering plane:
\begin{eqnarray}
  \left[ \dfrac{\dd\sigma}{\dd\Omega}\right]^\text{lin}_x &=&
  \dfrac{\Phi^2}{3}\sum \limits_{M_f, M_i ; \lambda_f, \lambda_i = \hat{x}}
  |\mathcal{M}_{\lambda_f}^{M_f, M_i}|^2 
  \label{eq:dcsx} \\
  \left[ \dfrac{\dd\sigma}{\dd\Omega}\right]^\text{lin}_y &=& \dfrac{\Phi^2}{3}\sum
  \limits_{M_f, M_i ; \lambda_f, \lambda_i = \hat{y}} |{\mathcal
    M}_{\lambda_f}^{M_f, M_i}|^2. \label{eq:dcsy}
\end{eqnarray}
It was already shown in Ref.~\cite{Ch05} that the corresponding photon
polarisation asymmetry
\begin{equation}
  \Pi^\text{lin}=\left(
    \left[ \dfrac{\dd\sigma}{\dd\Omega}\right]^\text{lin}_x-
    \left[ \dfrac{\dd\sigma}{\dd\Omega}\right]^\text{lin}_y\right)
  /\dfrac{\dd\sigma}{\dd\Omega}
  \label{eq:polasym}
\end{equation}
is not a good observable to extract the polarisabilities.

There are four observables for which both beam and target are polarised.
\begin{figure}[!htb]
  \begin{center}
    \includegraphics*[width=0.65\linewidth]{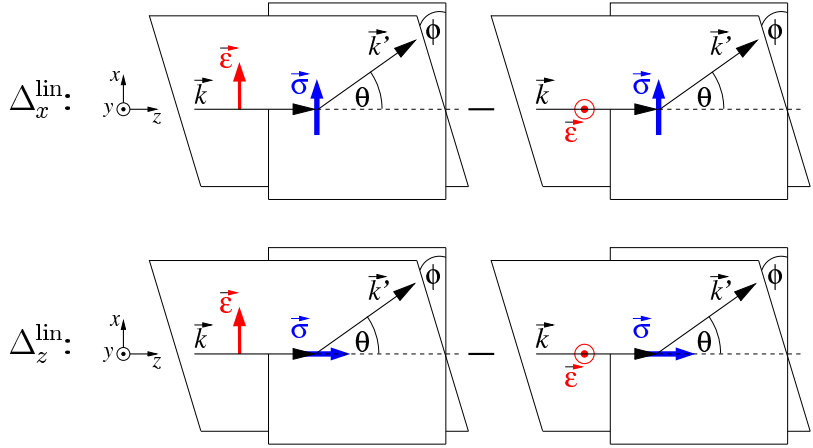}
    \caption {(Colour online) Observables for linearly-polarised photon on
      vector-polarised target.}
  \label{fig:lin-spin}
\end{center}
\end{figure}
The double-polarisation observables with linearly polarised photons $\left(
  \lambda_i \in \left\lbrace \hat{x}, \hat{y} \right\rbrace \right)$ are shown
in Fig.~\ref{fig:lin-spin}.  One is the 
cross-section difference on a target polarised in a plane perpendicular to the
beam direction and target polarisations parallel vs.~perpendicular to that of
the incoming photon:
\begin{equation}
  \Delta_{x}^\text{lin}=\left(\frac{\dd\sigma}{\dd\Omega}\right)_{\hat{x}
    \rightarrow} - \left(\frac{\dd\sigma}{\dd\Omega}\right)_{\hat{y}
    \rightarrow} \label{eq:deltaxlin}
\end{equation}
The subscripts $\hat{x}$ or $\hat{y}$ describe the initial photon
polarisation, and the arrows denote the initial target polarisation ($\uparrow
\equiv +\hat{z}$ and $\rightarrow \equiv +\hat{x}$). 
The other observable measures the same photon polarisation differences on a
target polarised along the beam direction:
\begin{equation}
  \Delta_{z}^\text{lin}
  =\left(\frac{\dd\sigma}{\dd\Omega}\right)_{\hat{x} \uparrow} -
  \left(\frac{\dd\sigma}{\dd\Omega}\right)_{\hat{y} \uparrow}
  \label{eq:deltazlin}
\end{equation} 
Both observables depend also on the angle $\phi$ between the scattering plane
and the plane formed by beam direction and the incoming photon polarisation.
The best signals are found when both planes coincide, $\phi=0$.

\begin{figure}[!htb]
  \begin{center}
    \includegraphics*[width=0.65\linewidth]{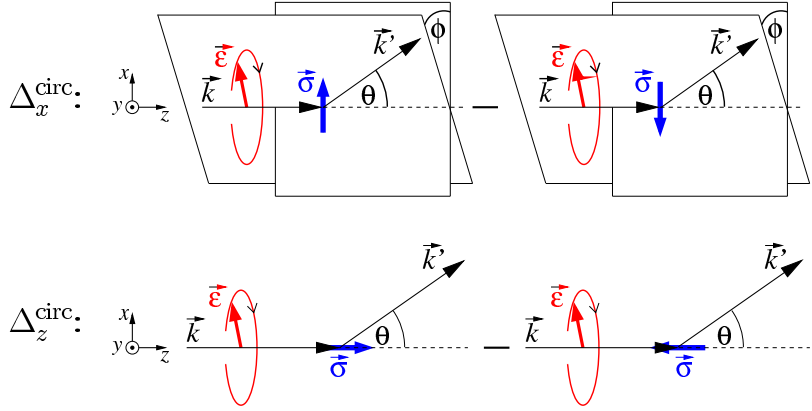}
    \caption {(Colour online) Observables for circularly-polarised photon on
      vector-polarised target.}
  \label{fig:circ-spin}
\end{center}
\end{figure}

The double-polarisation observables with circularly polarised photons $
\lambda_i =\mp\dfrac{\hat{x}+i\hat{y}}{\sqrt{2}}$, with $\lambda_i = \pm 1$
for photons with positive or negative helicity, are shown in
Fig.~\ref{fig:circ-spin}.
The deuteron target can again be vector-polarised in the scattering plane
along $\hat{x}$ or along the beam-direction $\hat{z}$.  Cross-section
differences are found by flipping the target polarisation. For any incoming
photon helicity:
\begin{eqnarray}
  \Delta_{x,(\lambda_i =
    \pm1)}^\text{circ}&=&\left(\frac{\dd\sigma}{\dd\Omega}\right)_{\uparrow
    \rightarrow} - \left(\frac{\dd\sigma}{\dd\Omega}\right)_{\uparrow
    \leftarrow} \label{eq:deltax}\\
  \Delta_{z,(\lambda_i = \pm1)}^\text{circ}
  &=&\left(\frac{\dd\sigma}{\dd\Omega}\right)_{\uparrow \uparrow} -
  \left(\frac{\dd\sigma}{\dd\Omega}\right)_{\uparrow \downarrow}
  \label{eq:deltaz} 
\end{eqnarray}
The first arrow of the subscript denotes the beam helicity; the second the
target polarisation. The perpendicular polarisation $\Delta_x^\text{circ}$ is
the difference of the differential cross-sections with the target polarised
along $+\hat{x}$ vs.~$-\hat{x}$, i.e.~perpendicular to the beam helicity. It
depends again on the angle $\phi$ between the polarisation and scattering
planes, with best results for $\phi=0$. Similarly, the parallel polarisation
$\Delta_z^\text{circ}$ is the difference with target polarised parallel
vs.~anti-parallel to the beam helicity and is independent of $\phi$. In
Ref.~\cite{Chen:2004wwa}, the same quantity is denoted by $2\Delta_1\sigma$.

Of the four quantities $\Delta^\text{lin/circ}_{x/z}$, only two are
  linearly independent, namely for example those which involve circularly
  polarised photons, or alternatively linearly polarised photons.  We look for
  those with the strongest polarisability signal.

The cross-section differences $\Delta$ set the scale for count-rates, and
hence for the beamtime necessary. Normalising to sums of cross-sections
removes many systematical experimental uncertainties. As pointed out
  in~\cite{Ch05}, the denominators are not total unpolarised cross-sections
  since one deals with a spin-$1$ target. The polarisation
asymmetries $\Sigma$ are~\footnote{Except for the super-script ``circ'',
    the notation is identical to that in Refs.~\cite{Ch05,Chen:2004wwa}}:
\begin{eqnarray}
  \Sigma_{x}^\text{lin}&=&
  \frac{\left(\frac{\dd\sigma}{\dd\Omega}\right)_{\hat{x} \rightarrow} -
    \left(\frac{\dd\sigma}{\dd\Omega}\right)_{\hat{y}
      \rightarrow}}{\left(\frac{\dd\sigma}{\dd\Omega}\right)_{\hat{x}
      \rightarrow} + \left(\frac{\dd\sigma}{\dd\Omega}\right)_{\hat{y}
      \rightarrow}} \label{eq:sigmaxlin} \\
  \Sigma_{z}^\text{lin}
  &=& \frac{\left(\frac{\dd\sigma}{\dd\Omega}\right)_{\hat{x} \uparrow} -
    \left(\frac{\dd\sigma}{\dd\Omega}\right)_{\hat{y} \uparrow}}{\left(
      \frac{\dd\sigma}{\dd\Omega}\right)_{\hat{x} \uparrow} +
    \left(\frac{\dd\sigma}{\dd\Omega}\right)_{\hat{y} \uparrow}}
  \label{eq:sigmazlin}\\
  \Sigma_{z,(\lambda_i = \pm1)}^\text{circ}
  &=&\frac{\left(\frac{\dd\sigma}{\dd\Omega}\right)_{\uparrow \uparrow} -
    \left(\frac{\dd\sigma}{\dd\Omega}\right)_{\uparrow
      \downarrow}}{\left(\frac{\dd\sigma}{\dd\Omega}\right)_{\uparrow
      \uparrow} + \left(\frac{\dd\sigma}{\dd\Omega}\right)_{\uparrow
      \downarrow}}
  \label{eq:sigmaz} \\
  \Sigma_{x,(\lambda_i =
    \pm1)}^\text{circ}&=&\frac{\left(\frac{\dd\sigma}{\dd\Omega}\right)_{\uparrow
      \rightarrow} - \left(\frac{\dd\sigma}{\dd\Omega}\right)_{\uparrow
      \leftarrow}}{\left(\frac{\dd\sigma}{\dd\Omega}\right)_{\uparrow
      \rightarrow} + \left(\frac{\dd\sigma}{\dd\Omega}\right)_{\uparrow
      \leftarrow}} \label{eq:sigmax}
\end{eqnarray}
Nevertheless, division by a small spin-averaged cross-section may enhance
theoretical uncertainties, or hide un-feasibly small count-rates. We also find
that asymmetries $\Sigma$ are usually by $\lesssim30$\% less sensitive to
variations of the polarisabilities than cross-section differences $\Delta$.
Sometimes, sensitivity to the nucleon structure is even lost entirely, while
in no case do we see an enhancement. It is the purview of our experimental
colleagues to determine whether these draw-backs outweigh the benefits. 

\subsection{Strategy}
\label{sec:stg}

In analysing the sensitivity of a given polarisation observable on the dipole
polarisabilities, we adopt the approach of Ref.~\cite{Ch05} to vary the static
central values as given in eq.~\eqref{eq:cval} by adding one
energy-independent parameter for each dipole polarisability, labelled $\delta
\alpha_{E1}$, $\delta \beta_{M1}$, $\delta \gamma_{E1E1}$, $\delta
\gamma_{M1M1}$, $\delta \gamma_{E1M2}$ and $\delta \gamma_{M1E2}$. In the
notation and definition of dynamical polarisabilities of Ref.~\cite{Hi04}, but
without a kinematical prefactor $(\omega+\sqrt{\omega^2+M^2})/M$, the complete
$\mathcal{O}(\epsilon^3)$ one-nucleon amplitudes of Sec.~\ref{sec:amplitudes}
in the cm frame of the $\gamma N$ system are thus supplemented by
\begin{eqnarray}
  {A}^{\mathrm{fit}}(\omega,\,z)&=& 4\pi\,\omega^2\,\bigg[
  [\delta\alpha_{E1}+z\,\delta\beta_{M1}]
  \,(\vec{\epsilon}^\prime\cdot\vec{\epsilon})
  -\delta\beta_{M1}\,
  (\vec{\epsilon}^\prime\cdot\hat{k})\,(\vec{\epsilon}\cdot\hat{k}^\prime)\nonumber\\
  &&-\ii\,[\delta\gamma_{E1E1}+z\,\delta\gamma_{M1M1}+\delta\gamma_{E1M2}
         +z\,\delta\gamma_{M1E2}]\,\omega\,\vec{\sigma}\cdot
         (\vec{\epsilon}^\prime\times\vec{\epsilon})\nonumber\\
  &&+\ii\,[\delta\gamma_{M1E2}-\delta\gamma_{M1M1}]\,\omega\,\vec{\sigma}\cdot
         \left(\hat{k}^\prime\times\hat{k}\right)
         (\vec{\epsilon}^\prime\cdot\vec{\epsilon})\nonumber\\
  &&+\ii\,\delta\gamma_{M1M1}\,\omega\,\vec{\sigma}\cdot
  \left[\left(\vec{\epsilon}^\prime\times\hat{k}\right)
    (\vec{\epsilon}\cdot\hat{k}^\prime)-
    \left(\vec{\epsilon}\times\hat{k}^\prime\right)
    (\vec{\epsilon}^\prime\cdot\hat{k})\right]\nonumber\\
  &&+\ii\,\delta\gamma_{E1M2}\,\omega\,\vec{\sigma}\cdot
  \left[\left(\vec{\epsilon}^\prime\times\hat{k}^\prime\right)
    (\vec{\epsilon}\cdot\hat{k}^\prime)-
    \left(\vec{\epsilon}\times\hat{k}\right)
    (\vec{\epsilon}^\prime\cdot\hat{k})\right]
\bigg]\;\;.
\label{fit}
\end{eqnarray}
The full amplitudes are then embedded into the deuteron
as extension of diagram (e) in Fig.~\ref{fig:onenucleon}.  Here,
$\vec{\epsilon}$ and $\hat{k}$ ($\vec{\epsilon}^\prime$ and $\hat{k}^\prime$)
are the polarisation and direction of the incoming (outgoing) photon in the
$\gamma N$ cm frame, and $z=
\hat{k}\cdot\hat{k}^\prime$.  Except for the kinematically convenient
pre-factor $\sqrt{s_{\gamma N}}/M$, this form is identical to that derived
from the interaction \eqref{eq:pols-lag}.

One can think of these contributions as parameterising the difference between
predicted and (so-far un-measured) experimental static values of the
polarisabilities, under the assumption that the energy-dependence of the
dipole polarisabilities is predicted correctly in SSE \ChiEFT. Conceptually,
this would imply that these additional interactions parameterise
energy-independent, short-distance contributions to the amplitudes,
i.e.~contributions which are at this order not explained by pion-cloud or
$\Delta$ effects. Alternatively, one can also see them as parameterising
deviations from the SSE \ChiEFT amplitudes at fixed nonzero energy, including
the theoretical uncertainties of higher-order effects. In that case, the
deviations themselves could be seen as energy-dependent. To reiterate, this
analysis is only sensitive to the average proton and neutron dipole
polarisabilities, since the deuteron is an iso-scalar.

In the next step, these contributions are independently varied by $\pm 2$
canonical units for one nucleon to analyse the effect of each on the various
observables. The polarisabilities of the other nucleon are kept fixed at the
iso-scalar value. This corresponds to a change of the iso-scalar
polarisabilities by half as much, i.e.~by $\pm1$ unit. Since deuteron Compton
scattering is sensitive only to iso-scalar quantities, varying either the
proton or neutron polarisabilities leads to the same result. In practise, the
scalar polarisabilities of the proton are better constrained, and deuteron
Compton scattering experiments are more likely focused on extracting neutron
polarisabilities. In that case, these studies can be interpreted as providing
the sensitivities on varying the neutron polarisabilities by $\pm2$ units,
with fixed proton polarisabilities. The spin-independent polarisabilities are
known to better accuracy, see eq.~\eqref{eq:robertpols}. But the goal is to
compare their sensitivity to those of the much less accurately known
spin-polarisabilities, for which different theoretical descriptions can
disagree by as much as $4$ units, see discussion below \eqref{eq:gpn} and
\eqref{eq:g0n}. We therefore see a variation of $\pm2$ units as a compromise
and note that variation by other amounts is easily obtained from the one given
since all observables are linear in the variations
$\delta(\alpha,\beta,\gamma_i)$. Quadratic contributions in the variations
contribute $\lesssim1\%$ to observables.

As hinted above, the polarisabilities are the multipoles of the structure part
of the amplitudes and thus do not contain more or less information than the
corresponding amplitudes. Therefore, determining the dipole polarisabilities
is now in principle reduced to a multipole-analysis of $6+1$ high-accuracy
scattering experiments using \eqref{fit}, as outlined
in~\cite{hgrieproc,Miskimentalk}.  

However, we do not present the result of the analysis in its entirety here.
Additional constraints must be considered in real life. The sum of electric
and magnetic polarisabilities is related by the Baldin sum rule
\eqref{eq:baldin}, and weaker constraints exist from the forward and backward
spin-polarisabilities (\ref{eq:gpn}/\ref{eq:g0n}). Experimental constraints
like detector and polarisation efficiencies and existing deuteron Compton
scattering data must be taken into account, too, to determine which
experiments will indeed show the greatest sensitivity on a given
polarisability and have the greatest impact in the network of data already
available.

\begin{figure}[!htb]
  \begin{center}
    \includegraphics*[width=\linewidth]{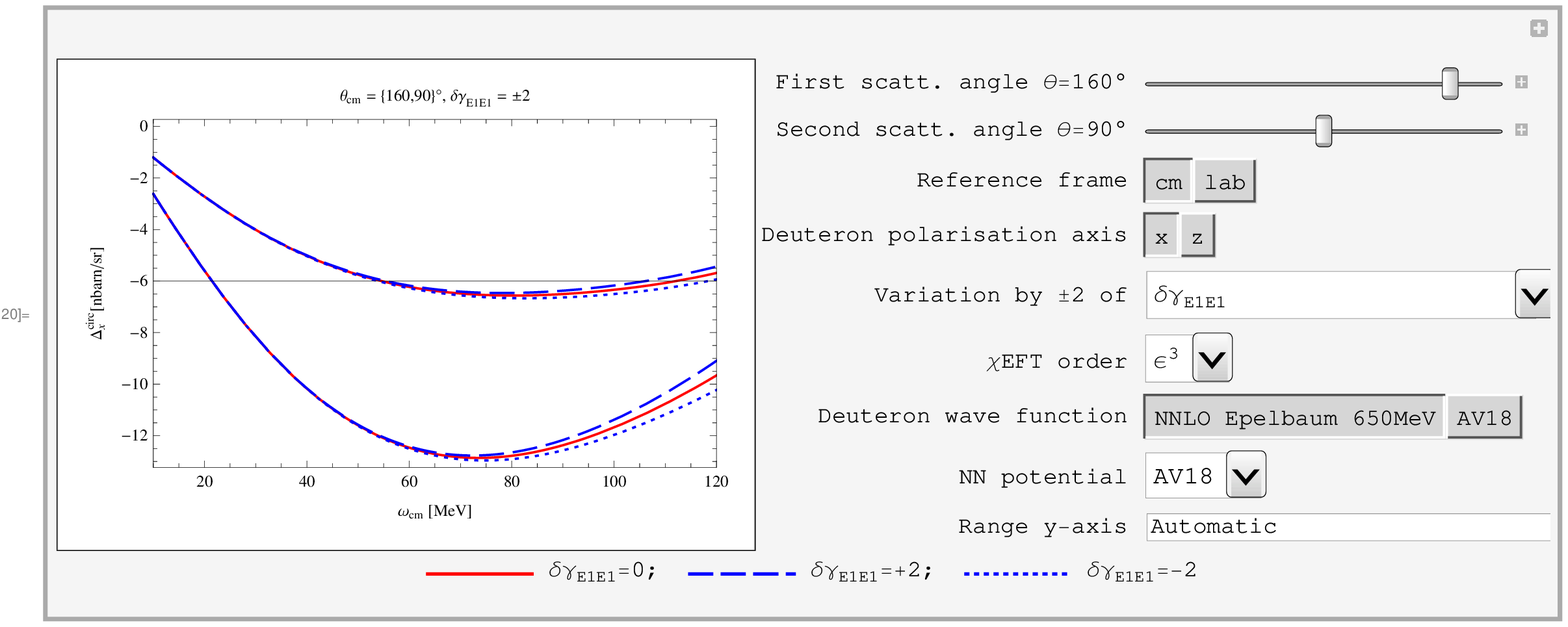}
    \caption {(Colour online) Screen-shot of part of the interactive
      \emph{Mathematica} notebook.}
  \label{fig:screenshot}
\end{center}
\end{figure}

Cross-section differences $\Delta$ and asymmetries $\Sigma$ for $6$
observables, depending on $6$ dipole polarisabilities and $3$ kinematic
variables (photon energy $\omega$ and scattering angles $\theta$ and $\phi$)
in the cm and lab frame, plus additional constraints, provide a large number
of parameters to explore.  Since the full information can not adequately be
conveyed in an article, we focus only on some prominent examples here and note
that in order to aide in planning and analysing experiments, the results for
all observables are available as an interactive \emph{Mathematica 7.0}
notebook from Grie{\ss}hammer (hgrie@gwu.edu). It contains both tables and
plots of energy- and angle-dependences of the cross-sections, cross-section
differences, analysing powers and asymmetries from $10$ to about $120$~MeV, in
both the cm and lab systems, including sensitivities to varying the
spin-independent and spin-dependent polarisabilities independently and subject
to the Baldin sum rule constraint \eqref{eq:baldin}.
Figure~\ref{fig:screenshot} shows a sample screen-shot.

In this article, we therefore focus on two representative energies in the
lab-frame as the one which is experimentally most relevant. At
$\omega_\text{lab}=45\;\MeV$, spin-polarisabilities should be suppressed since
they scale as $\omega^3$, see eq.~\eqref{fit}, while spin-independent
polarisabilities are already noticeable since they scale as $\omega^2$. On the
other hand, spin-polarisabilities should provide a substantial contribution at
$\omega_\text{lab}=125\;\MeV$, while staying below the pion-production
threshold avoids experimental and theoretical complications.

\section{Significance of $\Delta$-Isobar and $NN$-Rescattering}
\label{sec:comparison}

We now analyse the effect of the $\Delta$-isobar and intermediate
$NN$-rescattering contributions on polarisation observables. Without explicit
$\Delta$, the ingredients are that of Heavy-Baryon \ChiEFT at order $Q^3$,
where Q is again a typical small momentum scale, now measured in units of the
breakdown scale ${\mathrm{Re}}[M_{\Delta}] -M\approx300\;\MeV$ of the
$\Delta$-less theory.  Results with and without explicit $\Delta$ compare
HB\ChiEFT power-counting scheme at $\mathcal{O}(Q^3)$ with that of SSE at
$\mathcal{O}(\epsilon^3)$. The latter is a more
accurate description at higher energies due to its larger radius of
convergence.  Since effects from the dynamical $\Delta$ are small at low
energies, both results should agree inside the radius of convergence of
HB\ChiEFT. By comparing HB\ChiEFT and SSE, one can thus judge where HB\ChiEFT
does not converge any more.

On the other hand, leaving out $NN$-rescattering contributions is justified in
an order-by-order expansion of the amplitudes only for high photon energies,
$\omega\sim\mpi$~\cite{Be04,Griesshammer:2007aq,hgrieforthcoming}. As discussed in
Sec.~\ref{sec:amplitudes}, rescattering is mandatory in any consistent
power-counting at lower energies to restore the deuteron Thomson
limit~\cite{Griesshammer:2007aq,hgrieforthcoming}. Without it, the
calculation is inconsistent at low energies. At higher energies, the theory
remains perturbatively consistent since rescattering introduces then by design
a large number of two-body amplitudes which are formally of higher
order~\cite{Be04,Hi05,Hi05b,Griesshammer:2007aq}. The approaches
with and without rescattering should thus coincide for high enough energies.
Their comparison provides thus a tool to estimate a \emph{low-energy limit}
for expanding the rescattering amplitude. 

In unpolarised Compton scattering, Refs.~\cite{Hi05,Hi05a} showed that even
though ${\mathrm{Re}}[M_{\Delta}] -M\approx 300$~MeV, the $\Delta(1232)$
provides considerable strength at much lower energies ($\approx 100\;\MeV$),
due to its strong paramagnetism. This modifies the energy-dependent
polarisabilities $\beta_{M1}$ and $\gamma_{M1M1}$ dramatically and resolved
the ``SAL Puzzle''~\cite{Ka99,Lv00,Hornidge,Levchuk:2000mg,Be99,Be04} of
the apparent disagreement of both predictions and post-dictions with data
around $100\;\MeV$ at backward angles~\cite{Hi05a,Hi05,Hi05b}. In contrast,
including $NN$-rescattering in the intermediate state proved essential only
for $\omega \sim \frac{\mpi^2}{M}\lesssim50\;\MeV$, while rescattering
contributions provided only perturbative corrections at
$\omega_\text{lab}\gtrsim80\;\MeV$~\cite{Hi05,Hi05a,Hi05b}.

\begin{figure}[!htb]
  \begin{center}
    \includegraphics*[width=0.48\linewidth]{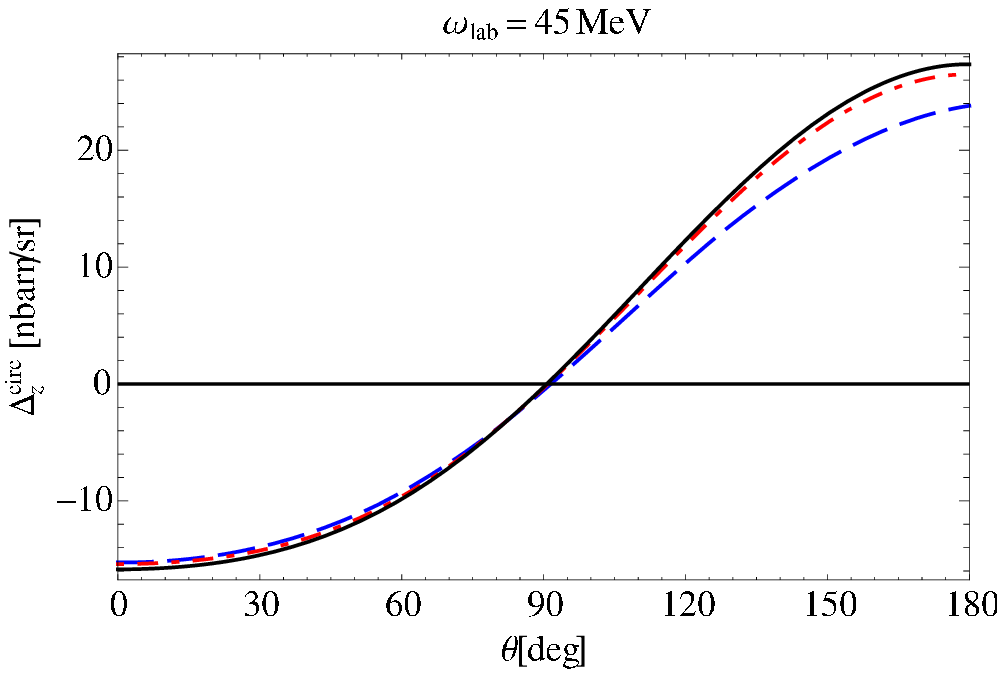}
    \hq\hq
    \includegraphics*[width=0.48\linewidth]{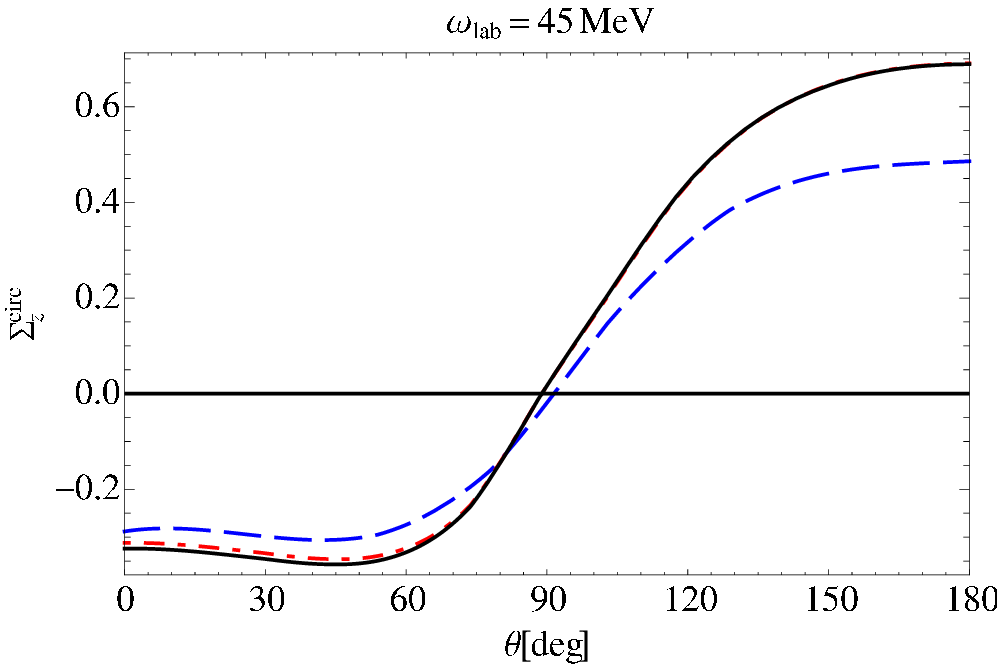}\\[1.5ex]
    \includegraphics*[width=0.48\linewidth]{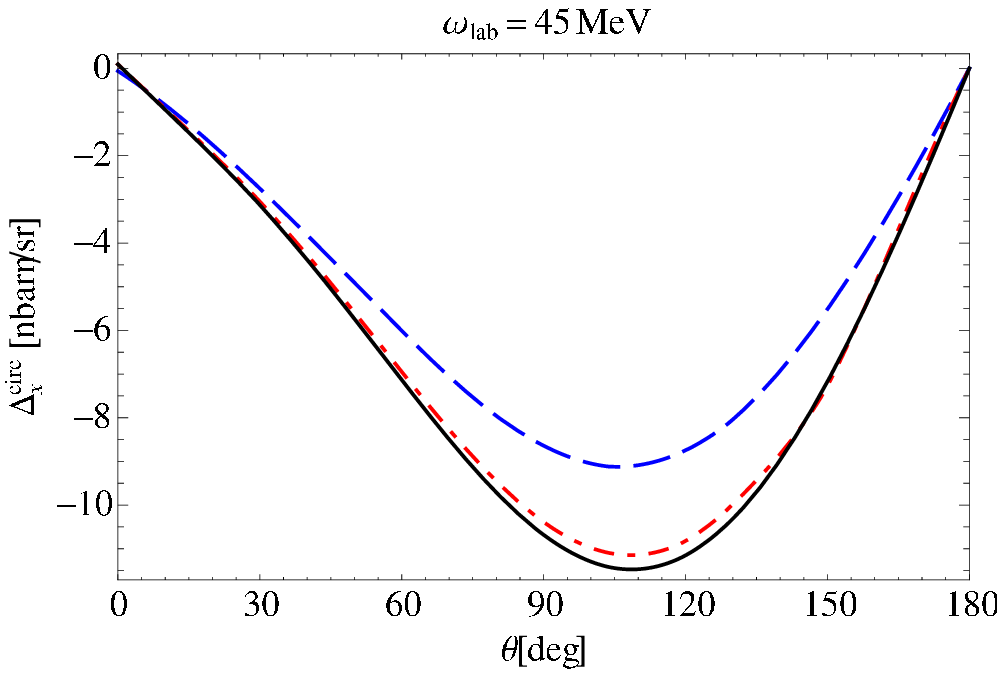}
    \hq\hq
    \includegraphics*[width=0.48\linewidth]{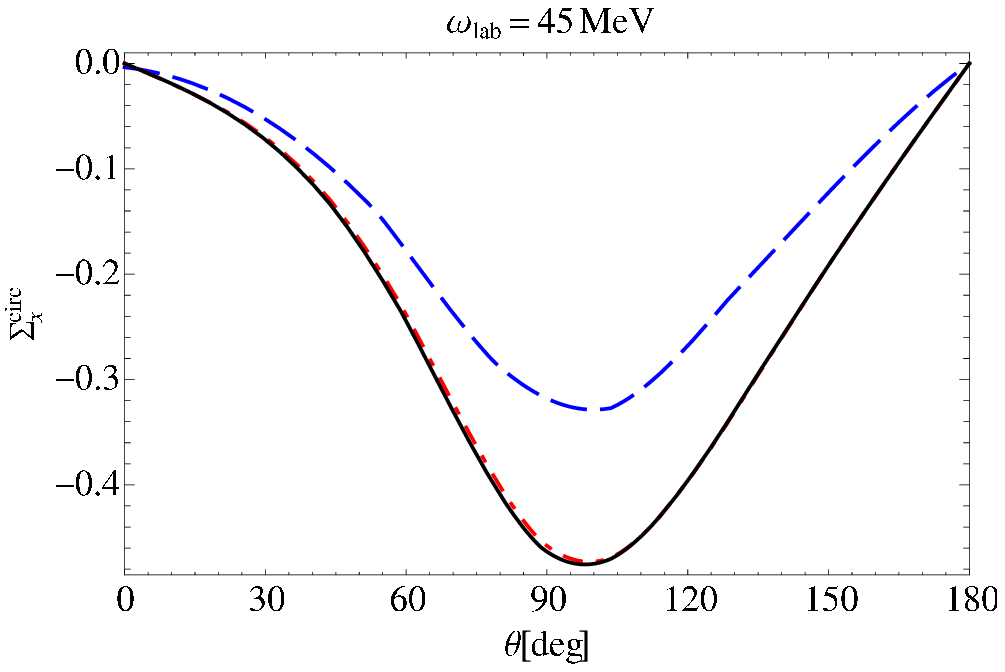}
    \caption {(Colour online) Dependence of typical polarisation observables
      at $\omega_\text{lab}=45$~MeV on rescattering and $\Delta$-effects.
      Left: $\Delta_z^\text{circ}$ (upper) and $\Delta_x^\text{circ}$ (lower);
      right: corresponding asymmetries $\Sigma_z$ and $\Sigma_x$. Dashed
      (blue): $\mathcal{O}(Q^3)$ HB\ChiEFT without rescattering (no $\Delta$);
      dot-dashed (red): complete $\mathcal{O}(Q^3)$ (with
      $NN$-rescattering, no $\Delta$); solid (black): full
      $\mathcal{O}(\epsilon^3)$ (with $\Delta$ and $NN$-rescattering).}
  \label{fig:comp45}
\end{center}
\end{figure}

\begin{figure}[!htb]
  \begin{center}
    \includegraphics*[width=0.48\linewidth]{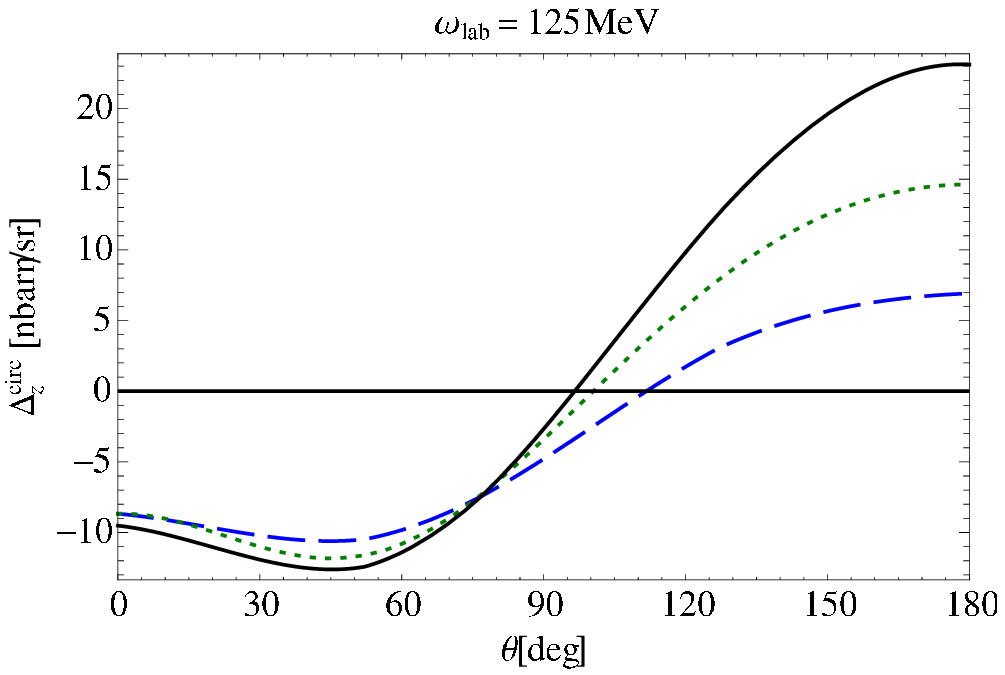}
    \hq\hq
    \includegraphics*[width=0.48\linewidth]{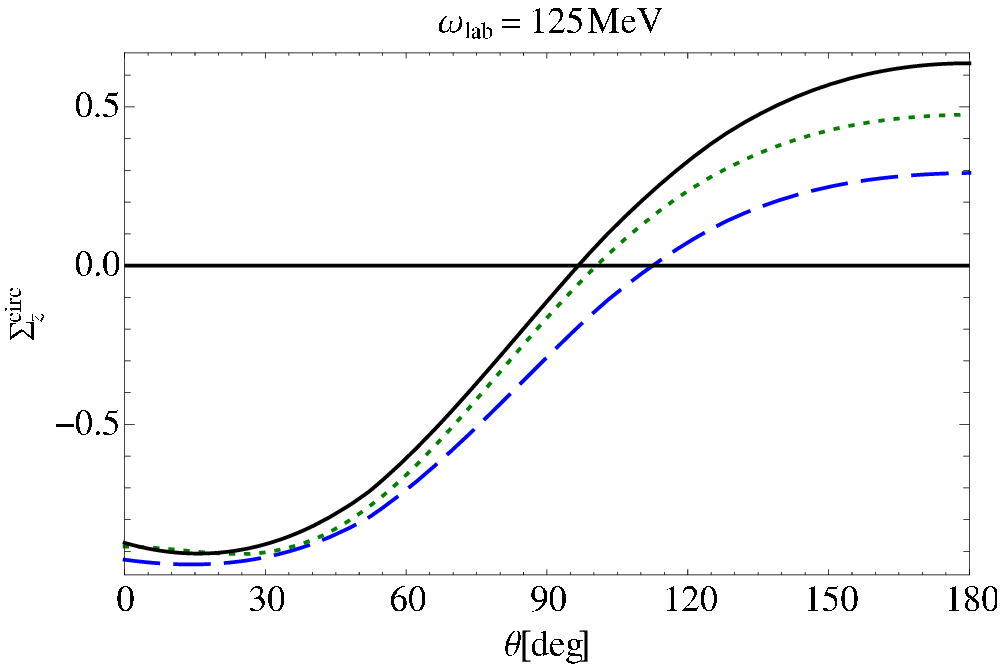}\\[1.5ex]
    \includegraphics*[width=0.48\linewidth]{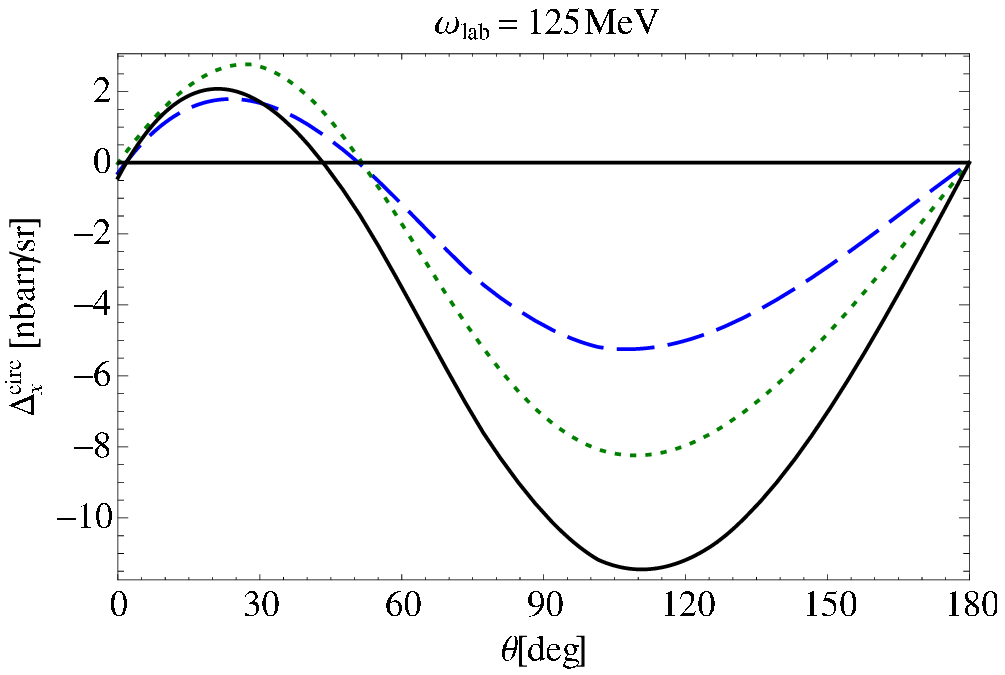}
    \hq\hq
    \includegraphics*[width=0.48\linewidth]{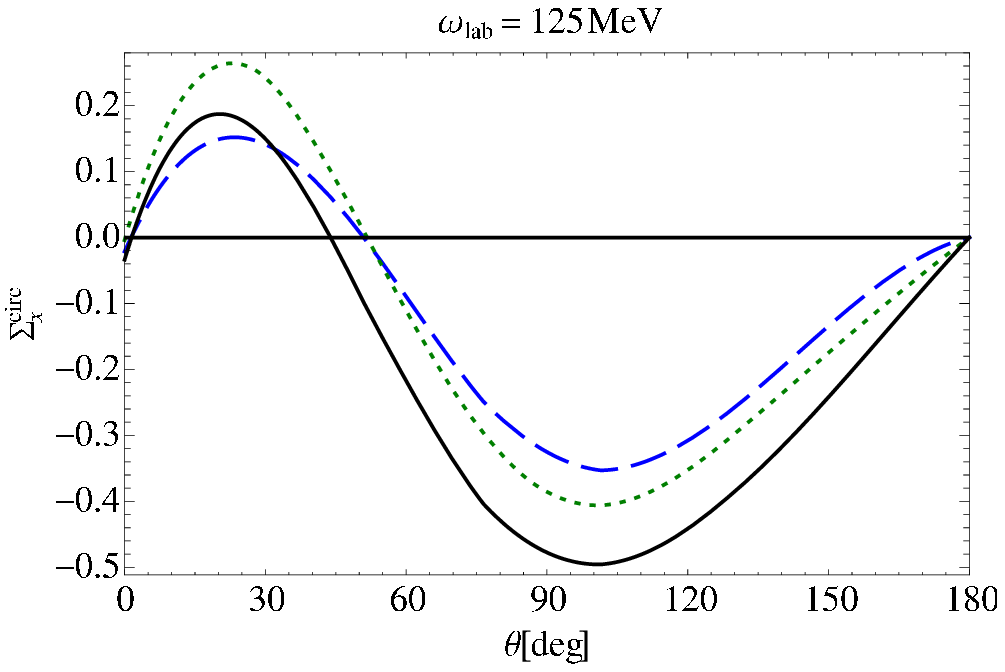}
    \caption{(Colour online) Dependence of typical polarisation observables at
      $\omega_\text{lab}=125$~MeV on rescattering and $\Delta$-effects. Left:
      $\Delta_z^\text{circ}$ (upper) and $\Delta_x^\text{circ}$ (lower);
      right: the corresponding asymmetries $\Sigma_z$ and $\Sigma_x$. Dashed
      (blue): $\mathcal{O}(Q^3)$ HB\ChiEFT without rescattering (no $\Delta$);
      dotted (green): $\mathcal{O}(\epsilon^3)$ (with $\Delta$) but no
      $NN$-rescattering; solid (black): full $\mathcal{O}(\epsilon^3)$ (with
      $\Delta$ and $NN$-rescattering).}
  \label{fig:comp125}
\end{center}
\end{figure}

Turning to polarisation observables, Figs.~\ref{fig:comp45} and
\ref{fig:comp125}
show cross-section differences $\Delta^\text{circ}_{x,z}$ and asymmetries
$\Sigma^\text{circ}_{x,z}$ with circularly-polarised photons in the lab frame
at $\omega_\text{lab}=45\;\MeV$ and $\omega_\text{lab}=125\;\MeV$,
respectively, with and without including the contributions from
$NN$-rescattering or from a dynamical $\Delta$. The other observables show a
similar behaviour.

In scattering at $45\;\MeV$, Fig.~\ref{fig:comp45}, effects from rescattering
are substantial. The dashed (blue) lines neither include rescattering nor a
dynamical $\Delta$, while the dot-dashed (red) lines include rescattering and
thus provide a consistent $\mathcal{O}(Q^3)$ HB\ChiEFT calculation (no
$\Delta$). By comparing the latter with the consistent
$\mathcal{O}(\epsilon^3)$ SSE result, solid (black) line, one concludes that a
dynamical inclusion of the $\Delta$ is not necessary at this energy. In
asymmetries (right panels), rescattering is slightly more pronounced than in
cross-section differences (left panels).  Within theoretical uncertainties,
$\Delta$-effects are the same in both.

As expected, $\Delta$-effects have become pronounced at $125\;\MeV$, see
Fig.~\ref{fig:comp125}.
A dynamical $\Delta$ without rescattering (dotted green) nearly doubles
$\Delta^\text{circ}_{x,z}$, compared to the HB\ChiEFT result without
rescattering (dashed blue). Surprisingly and in contradistinction to
unpolarised observables, rescattering effects provide an additional sizable
correction of up to $40\%$ in SSE (solid black). In asymmetries (right
panels), both effects are usually slightly smaller.  The large rescattering
correction suggests that contributions in which the photons couple to the same
pion-exchange current, Fig.~\ref{fig:twonucleon} (a), are not as prominent in
the polarisation observables as in unpolarised ones. For single-polarisation
observables, Ref.~\cite{mythesis, Ch05} pointed already out that this can be
attributed to the near-diagonal spin-structure of the two-body currents.
Clearly, the intermediate $NN$ rescattering matrix is highly off-diagonal in
angular-momentum space due to the electric and magnetic couplings and photon
multipolarity. Since rescattering should at these energies be perturbative as
discussed in Sect.~\ref{sec:amplitudes}, truncating the expansion at lowest
order, i.e. with only one intermediate pion exchange as performed in
Ref.~\cite{Be02,Be04} may suffice to obtain the same result. According to
  the power-counting in that r\'egime, such contributions are of higher order.
  This raises the question to which extend the \ChiEFT expansion converges.
  Only typical sizes of contributions are however estimated by power-counting.
  A comprehensive answer can only be given a complete next-order calculation
  in \ChiEFT.

Thus, both the dynamical $\Delta$-isobar and intermediate $NN$-rescattering
contributions are necessary in polarisation observables up to $125$~MeV. The
latter is a remnant of the Thomson limit even at high energies.
The former must be included since the $\gamma\Delta N$
coupling induces large paramagnetic effects whose energy-dependence is crucial
at high energies. At low energies, however, the $\Delta$ is indeed and as
expected not dynamical. Both effects increase the magnitude of the asymmetry
at back-angles for high energies, relative to a previous
investigation~\cite{Ch05}. In the following, we present only the complete SSE
results, i.e.~with dynamical $\Delta$ and $NN$-rescattering.

\section{Unpolarised Target and Linearly-polarised Photons}
\label{sec:lpol}

Experiments with linearly polarised photons and unpolarised target, as planned
or conducted at \HIGS~\cite{Weller:2009zza,Weller}, are not only more readily
realised than doubly-polarised experiments. They also provide an opportunity
to consider kinematics in which at least one of the dipole polarisabilities
does not contribute at all, facilitating extractions of the other ones.

\begin{figure}[!htb]
\begin{center}
\includegraphics*[width=0.15\linewidth]{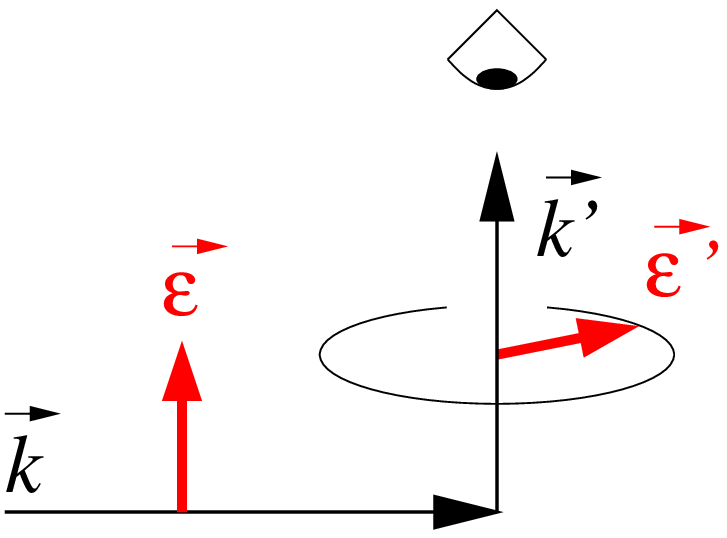}\hq\hq\hq\hq\hq\hq
\includegraphics*[width=0.15\linewidth]{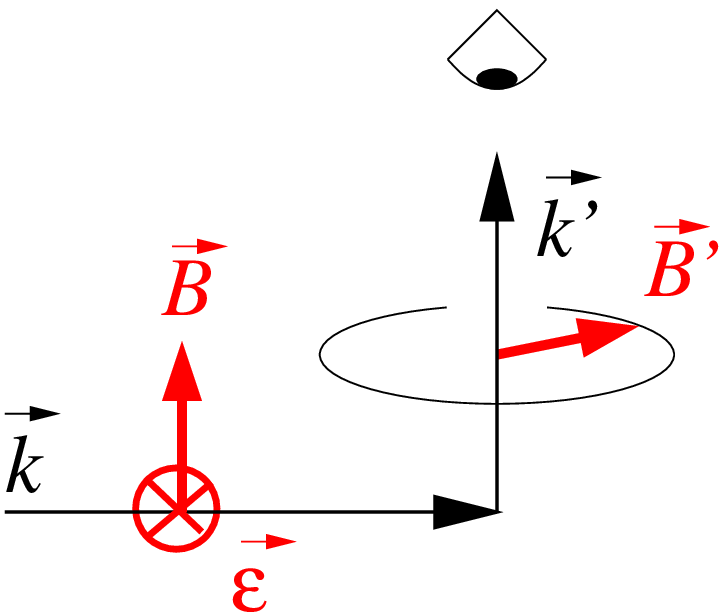}
\caption{\label{fig:switchingoffpols} Left: Configuration under which an
  induced dipole cannot radiate an $E1$ photon into a detector; right: same
  for radiation of an $M1$ photon. The ``eye'' represents an observer.}
\end{center}
\end{figure}

While none of the observables described in Sec.~\ref{sec:obs} is sensitive
only to one dipole polarisability, inspecting \eqref{eq:pols-lag} reveals
configurations in which some polarisabilities do \emph{not} contribute to
Compton scattering, as to our knowledge first pointed out by Maximon for the
proton~\cite{Maximon:1989zz}. Note that \eqref{eq:pols-lag} involves double
and triple scalar products between the electric and magnetic photon fields and
the nucleon spin. First, re-write the electric spin-independent dipole term in
\eqref{eq:pols-lag} as $2\pi\alpha_{E1}(\omega)\;N^\dagger
\,\vec{E}_\text{in}\cdot\vec{E}_\text{out}\,N$, where $ \vec{E}_\text{in}$
($\vec{E}_\text{out}$) is the electric field of the incoming (scattered)
photon.  There is no contribution when incoming and outgoing polarisations are
orthogonal to each other. In the cm frame, this is the situation when the
incoming photon polarisation is in the scattering plane and
$\theta_\text{cm}=90^\circ$, see Fig.~\ref{fig:switchingoffpols}.
Comparing to Fig.~\ref{fig:lin-nospin},
$\left[\frac{\dd\sigma}{\dd\Omega}\right]^\text{lin}_{x}(\theta_\text{cm}=90^\circ)$
is thus insensitive to $\alpha_{E1}$ when the photon is scattered off a
nucleon. More intuitively, the induced time-dependent electric dipole moment
$\vec{d}_\text{ind}=\alpha_{E1}\;\vec{E}_\text{in}$ in the nucleon leads to
radiation from a Hertz'ian dipole, which emits no radiation along its
axis.

Analogously,
$\left[\frac{\dd\sigma}{\dd\Omega}\right]^\text{lin}_{y}(\theta_\text{cm}=90^\circ)$
on a nucleon is insensitive to $\beta_{M1}$ because the induced magnetic
dipole does not radiate under that angle, see Fig.~\ref{fig:switchingoffpols}.
Similar arguments apply to the sensitivity of double-polarisation observables
on spin-polarisabilities, see Sec.~\ref{sec:ldpol}. 

\begin{figure}[!htb]
  \begin{center}
    \includegraphics*[width=0.48\linewidth]{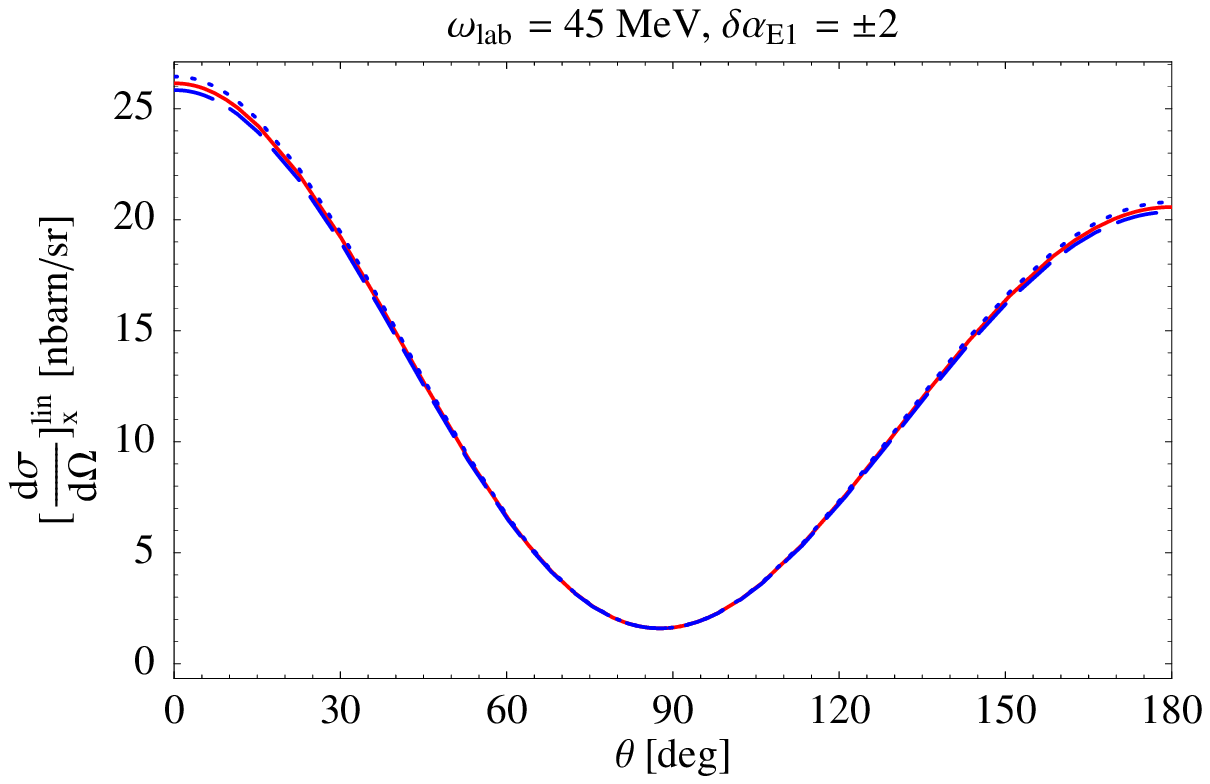}
    \hq\hq
    \includegraphics*[width=0.48\linewidth]{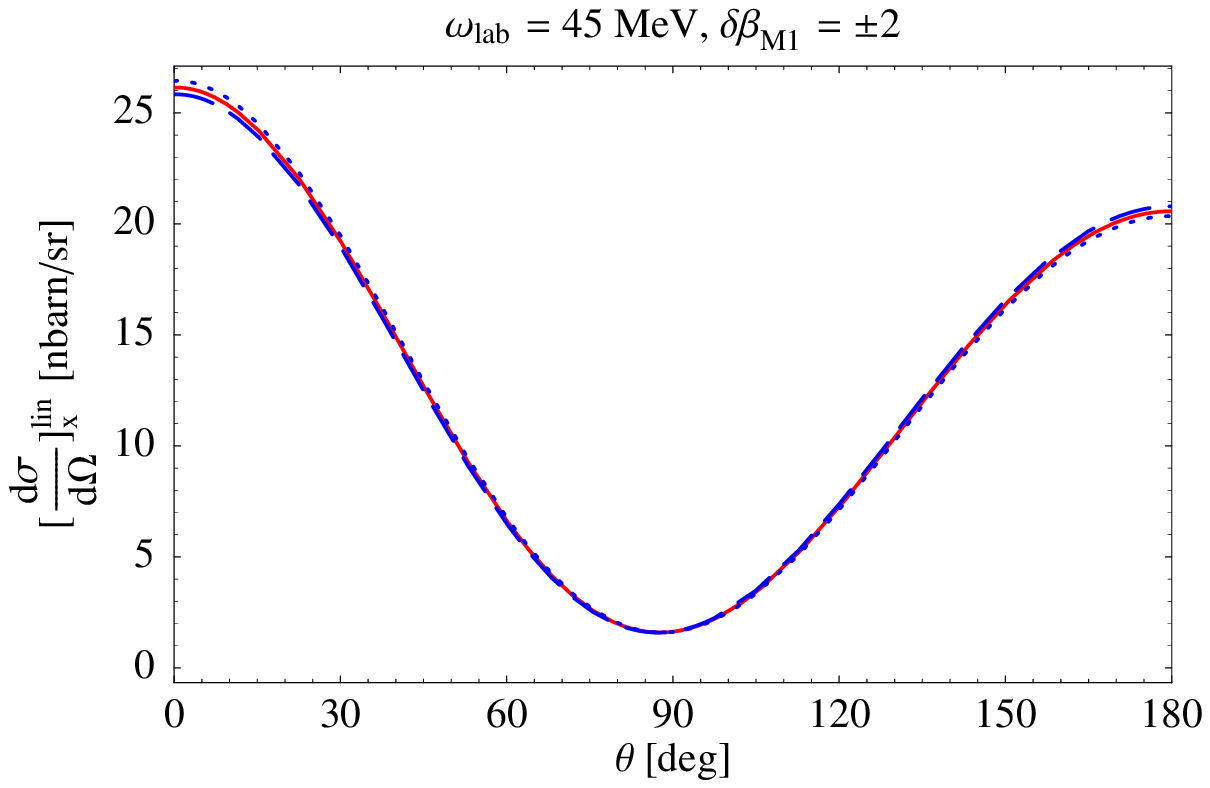}\\[1.5ex]
    \includegraphics*[width=0.48\linewidth]{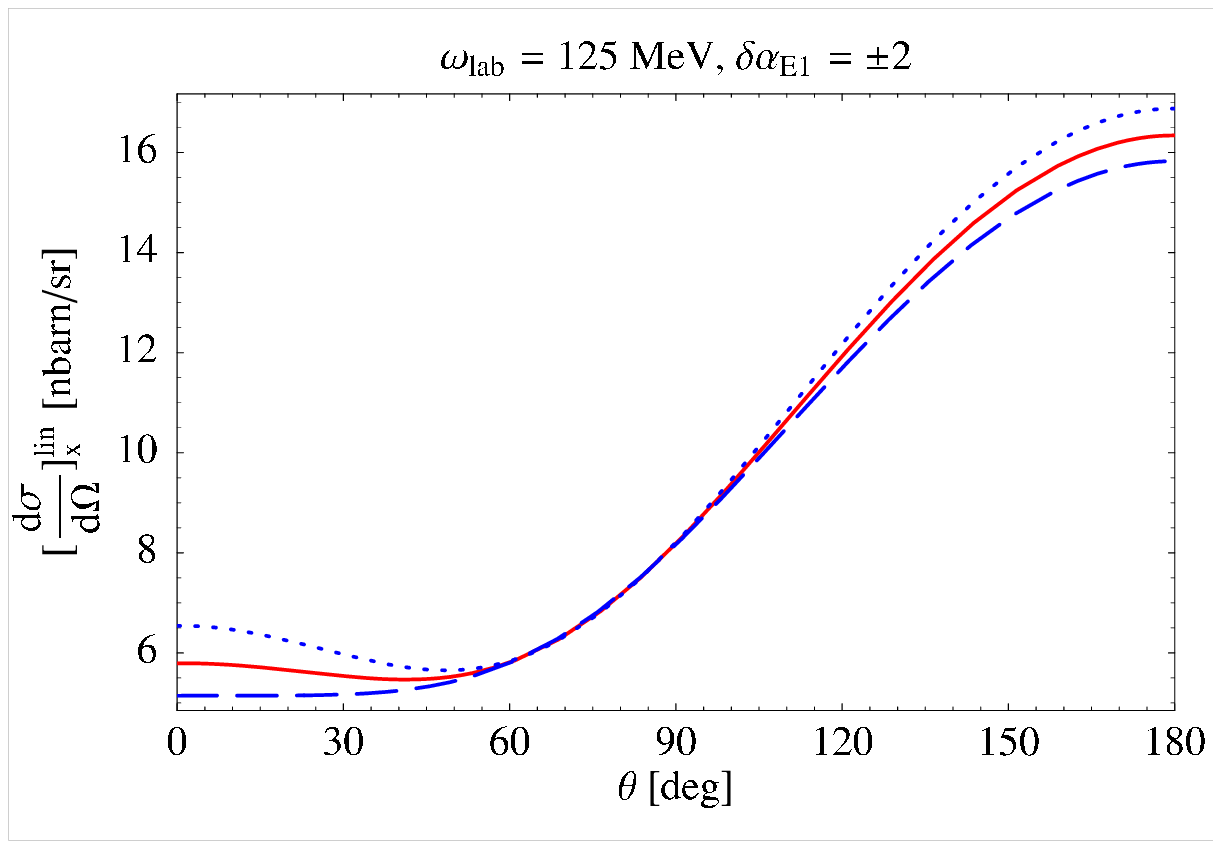}
    \hq\hq
    \includegraphics*[width=0.48\linewidth]{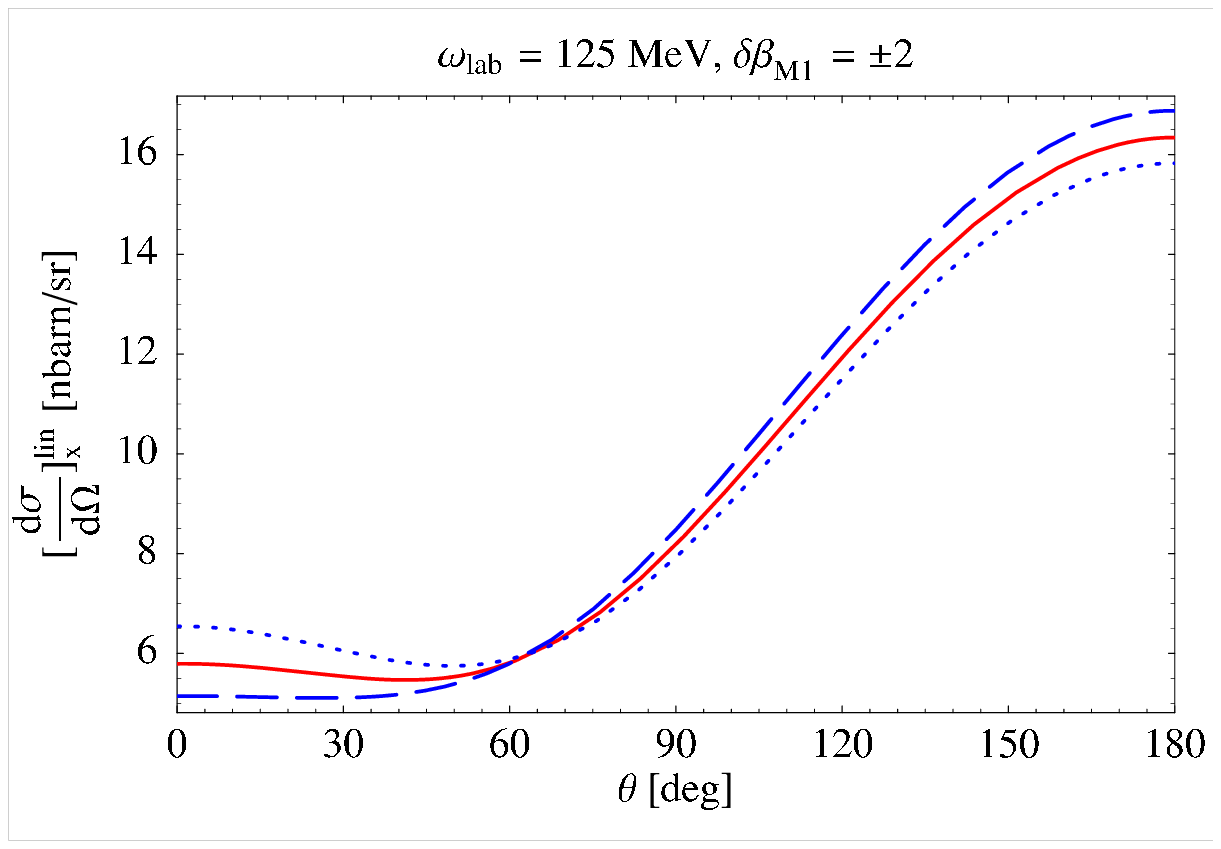}
    \caption {(Colour online) Dependence of $\left[
        \frac{\dd\sigma}{\dd\Omega}\right]^\text{lin}_{x}$ on the spin-independent
      dipole polarisabilities at $\omega_\text{lab}=45$~MeV (top) and
      $\omega_\text{lab}=125$~MeV (bottom) around the complete
      $\mathcal{O}(\epsilon^3)$ calculation (solid red line). Left:
      $\alpha_{E1}$ varied by $\delta \alpha_{E1}=+2$ (dashed blue line);
      $\delta \alpha_{E1}=-2$ (dotted blue line). Right: Same for
      $\beta_{M1}$. Notice the scale offset in the lower panels.}
\label{fig:dcsx_ab}
\end{center}
\end{figure}

\begin{figure}[!htb]
  \begin{center}
    \includegraphics*[width=0.48\linewidth]{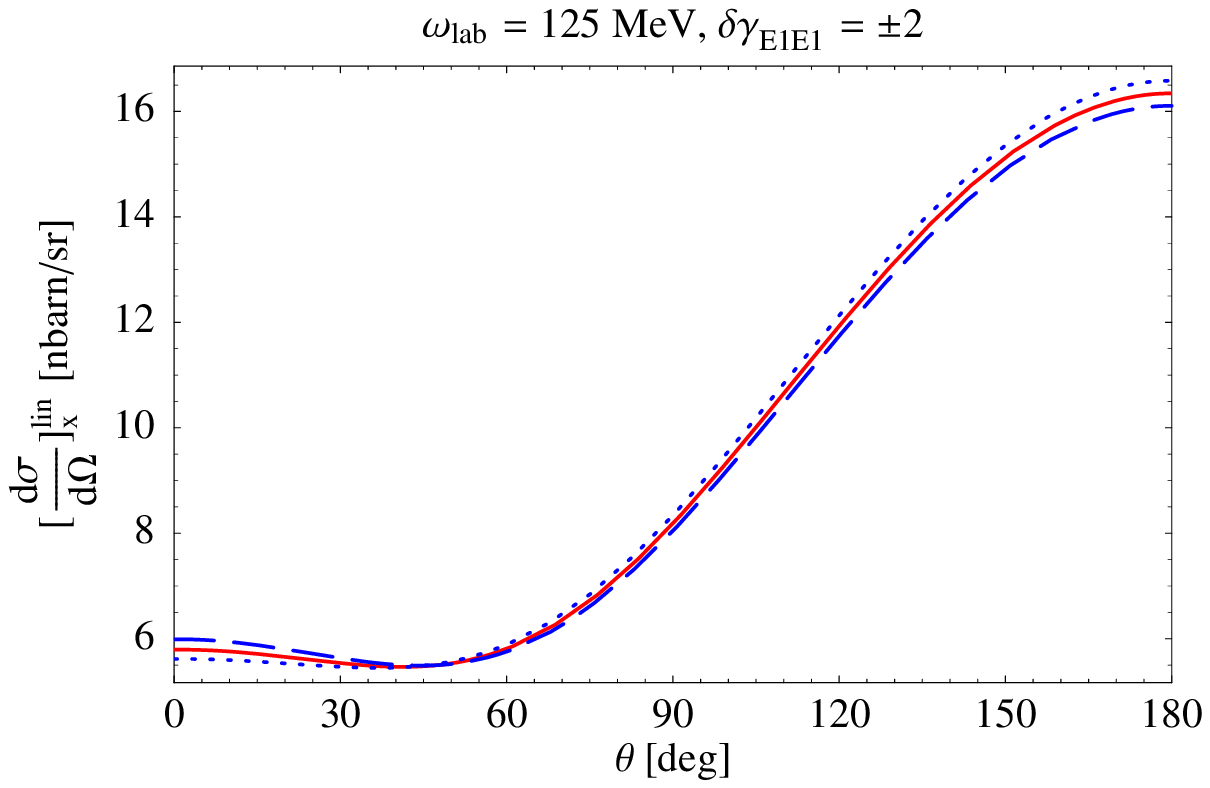}
    \hq\hq
    \includegraphics*[width=0.48\linewidth]{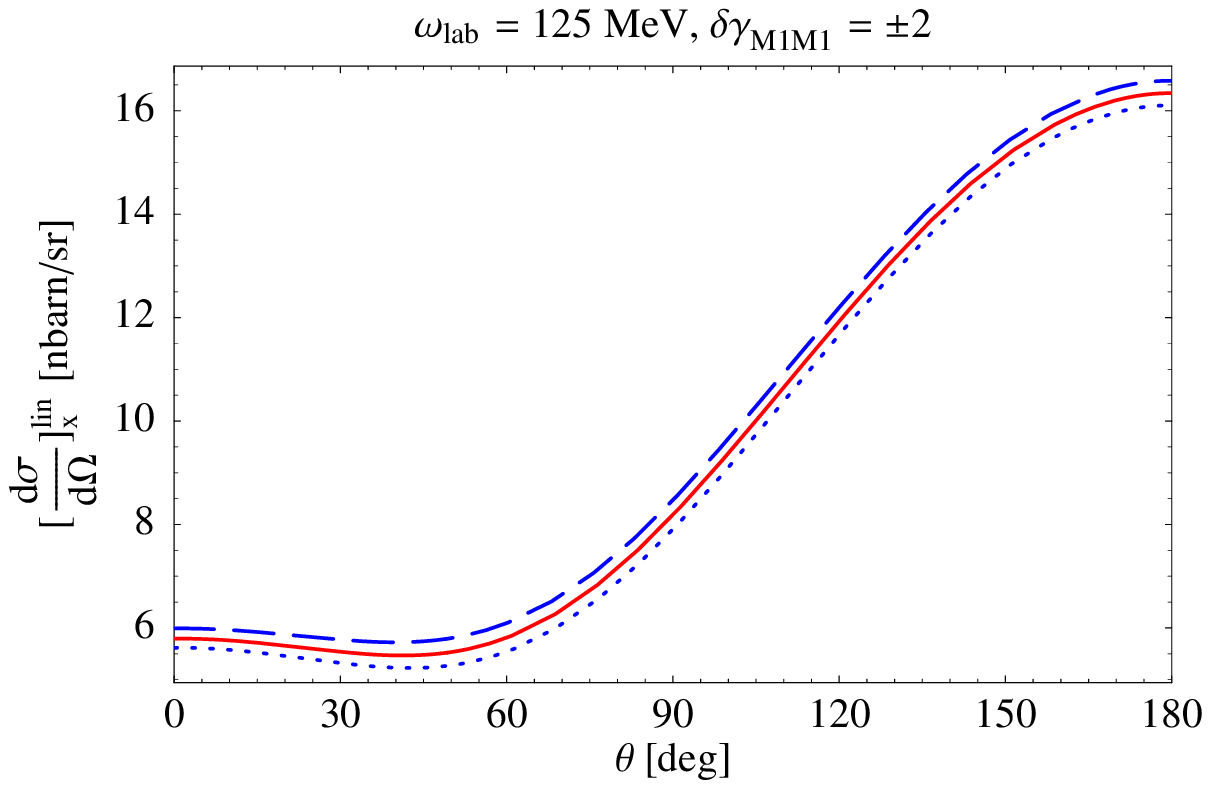}\\[1.5ex]
    \includegraphics*[width=0.48\linewidth]{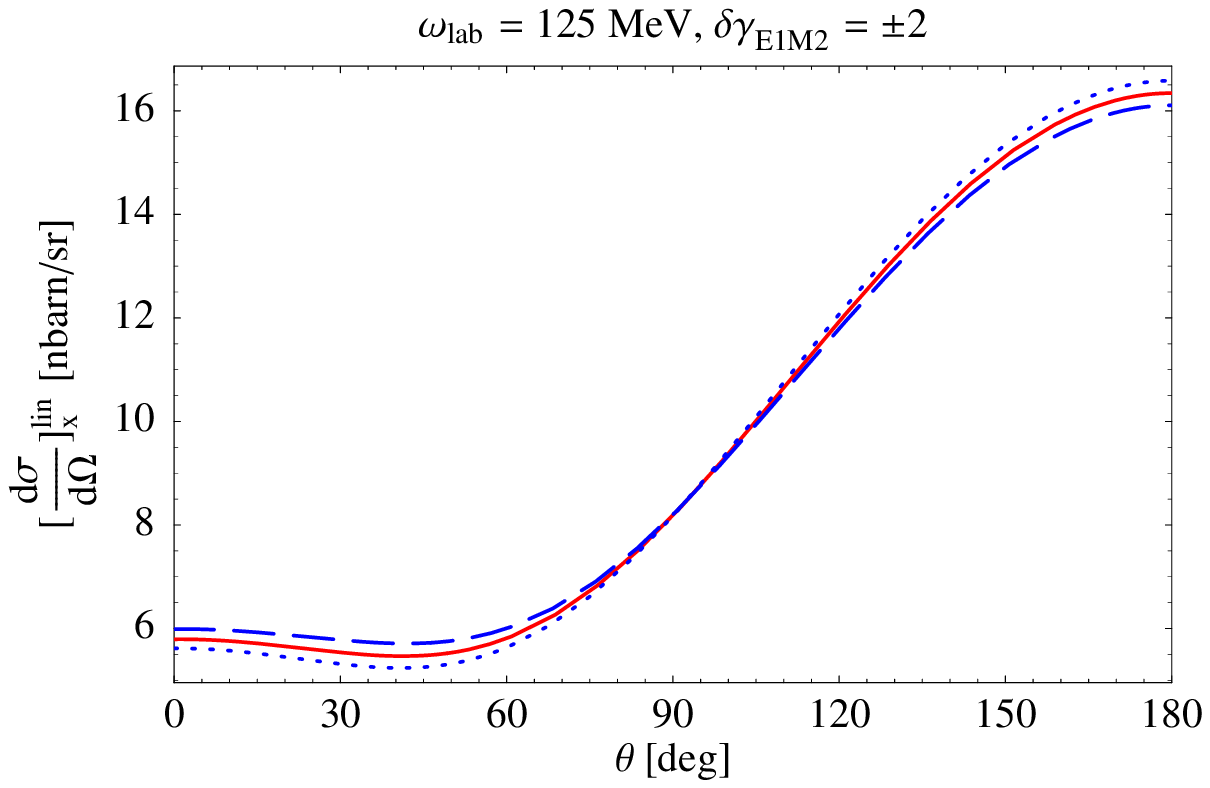}
    \hq\hq
    \includegraphics*[width=0.48\linewidth]{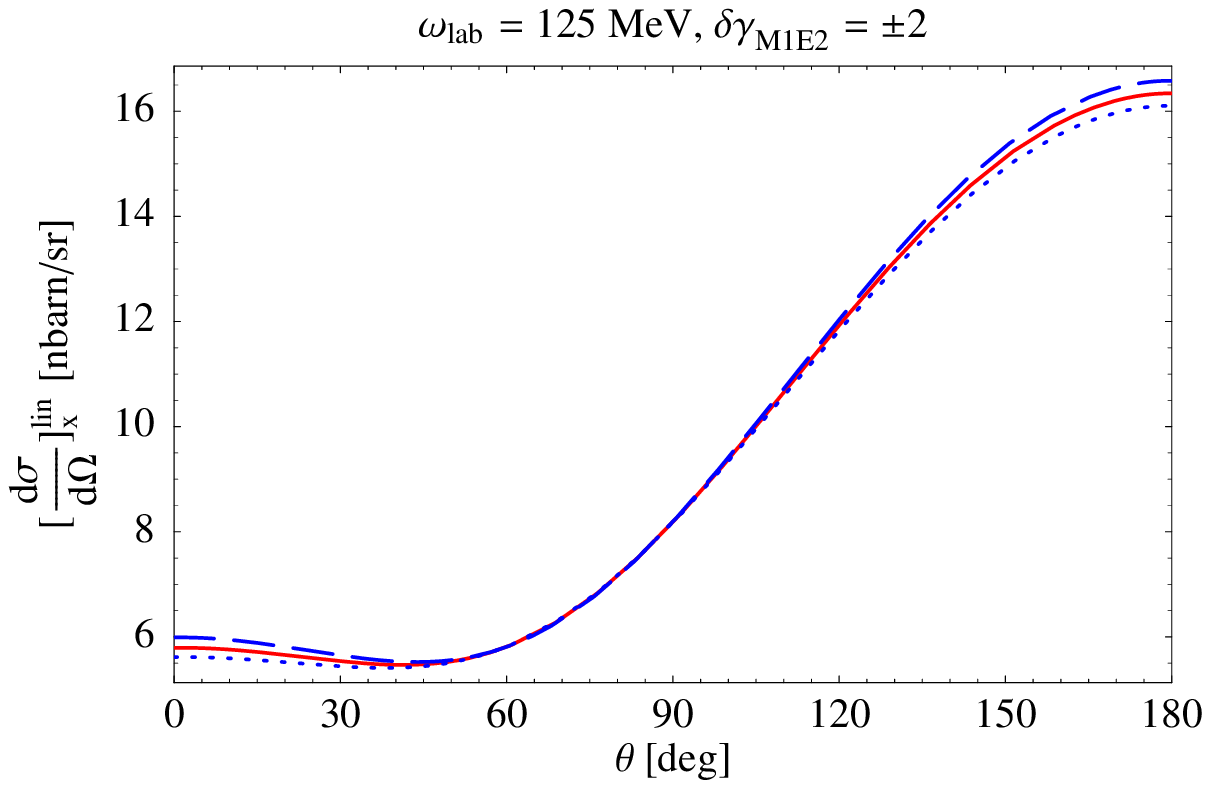}
    \caption {(Colour online) Dependence of $\left[
        \frac{\dd\sigma}{\dd\Omega}\right]^\text{lin}_{x}$ on the dipole
      spin-polarisabilities at $\omega_\text{lab}=125$~MeV around the complete
      $\mathcal{O}(\epsilon^3)$ calculation. Variation of
      $\delta\gamma_{E1E1}=\pm2$ (top left); $\delta\gamma_{M1M1}=\pm2$ (top
      right); $\delta\gamma_{E1M2}=\pm2$ (bottom left);
      $\delta\gamma_{M1E2}=\pm2$ (bottom right). Notation as in
      Fig.~\ref{fig:dcsx_ab}. Notice the scale offset.}
  \label{fig:dcsx_gs}
\end{center}
\end{figure}

\begin{figure}[!htb]
  \begin{center}
    \includegraphics*[width=0.48\linewidth]{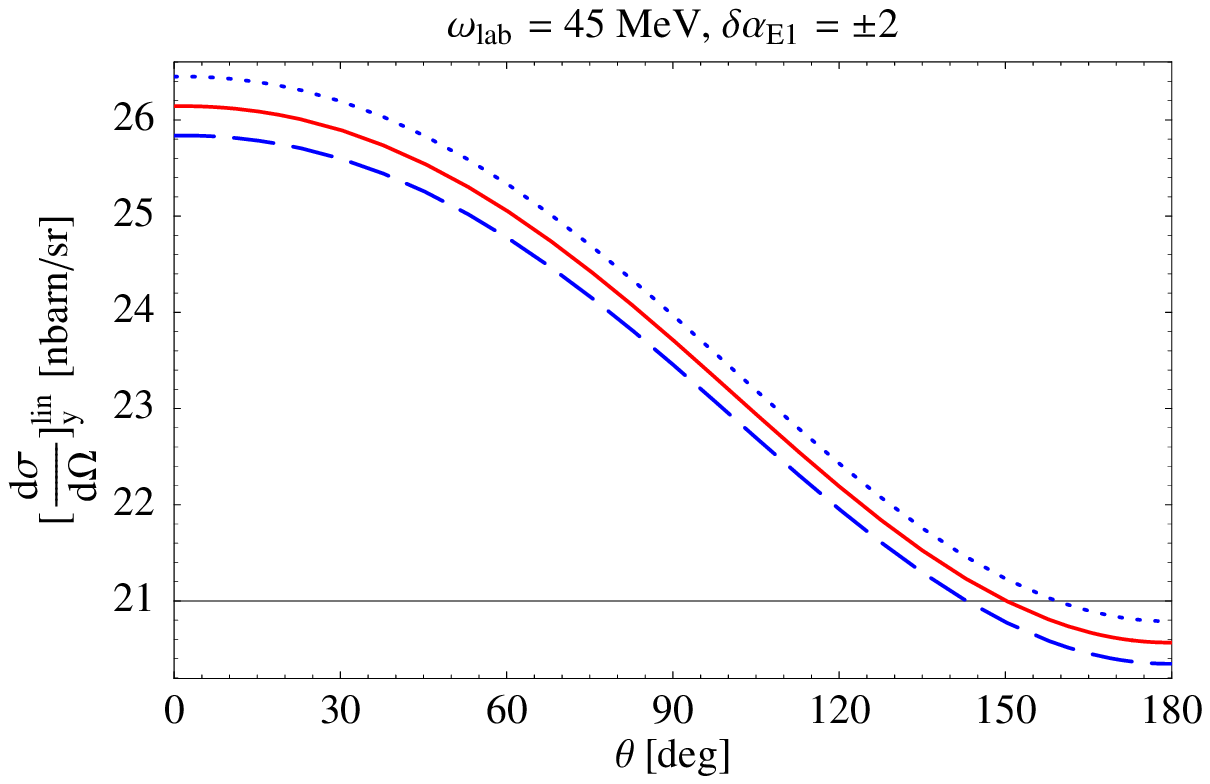}
    \hq\hq
    \includegraphics*[width=0.48\linewidth]{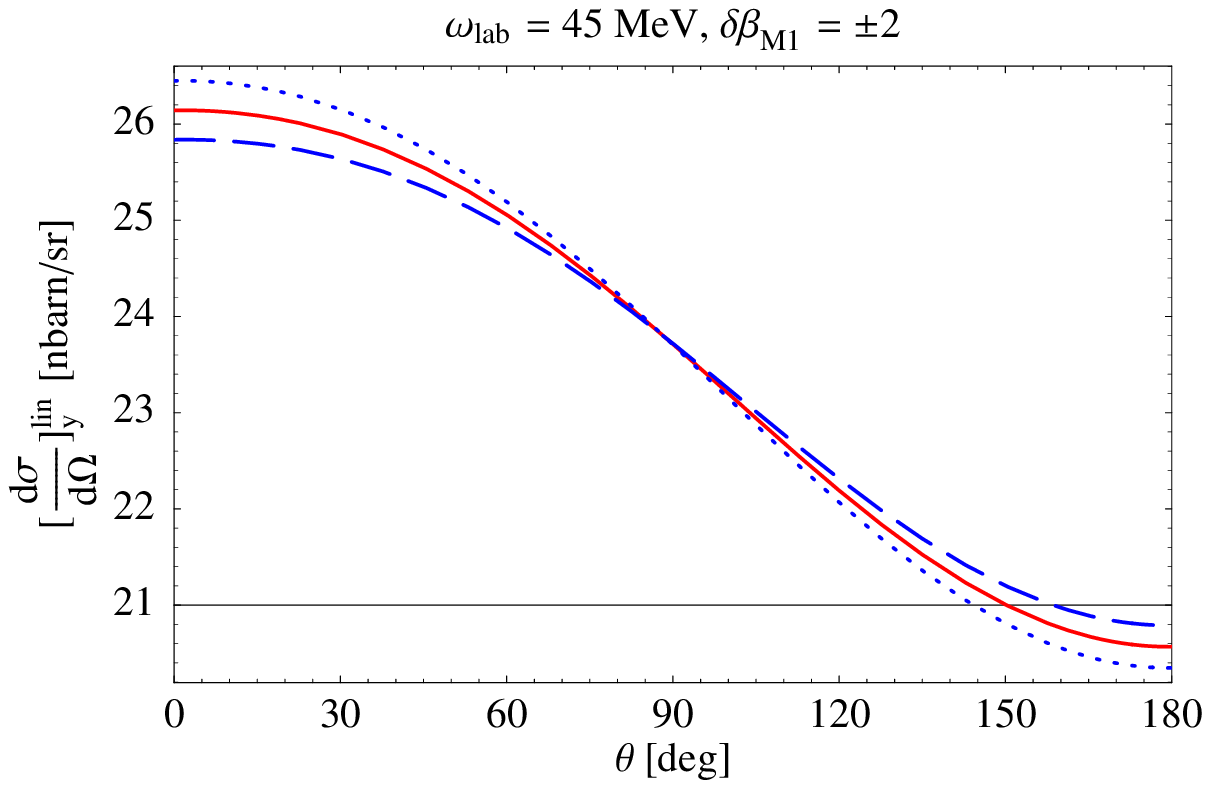}\\[1.5ex]
    \includegraphics*[width=0.48\linewidth]{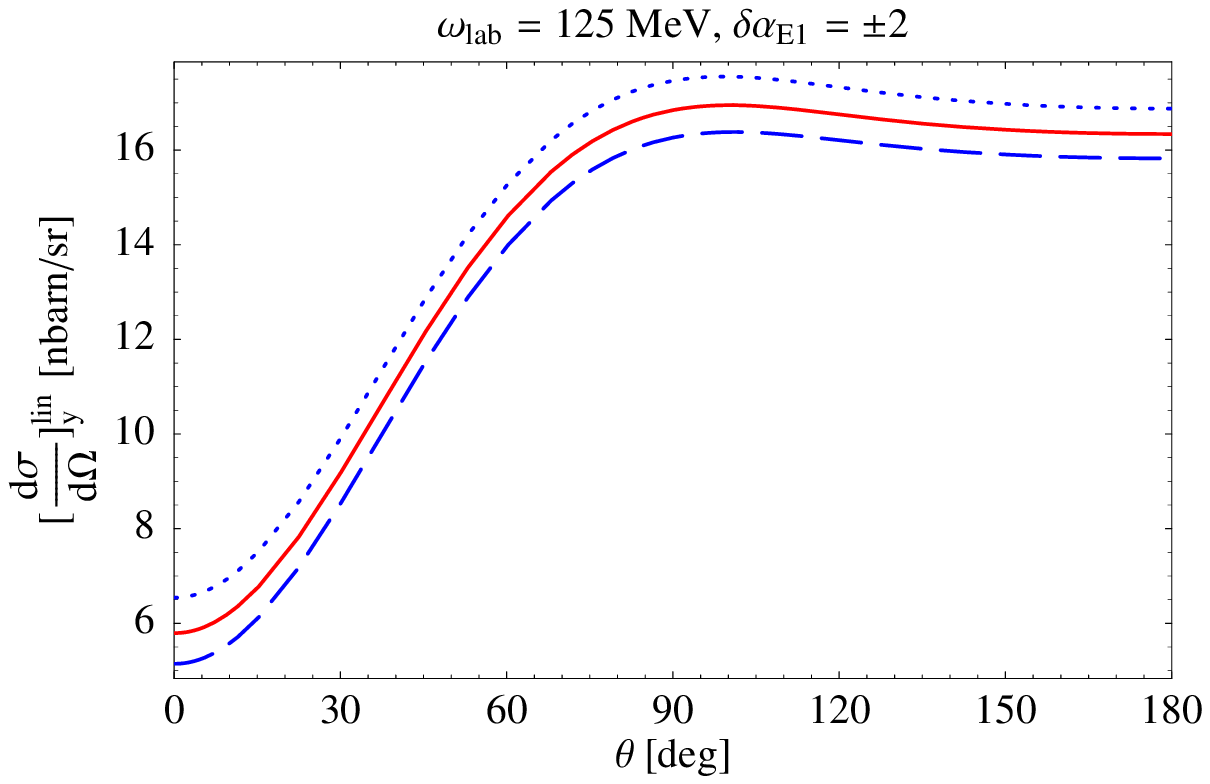}
    \hq\hq
    \includegraphics*[width=0.48\linewidth]{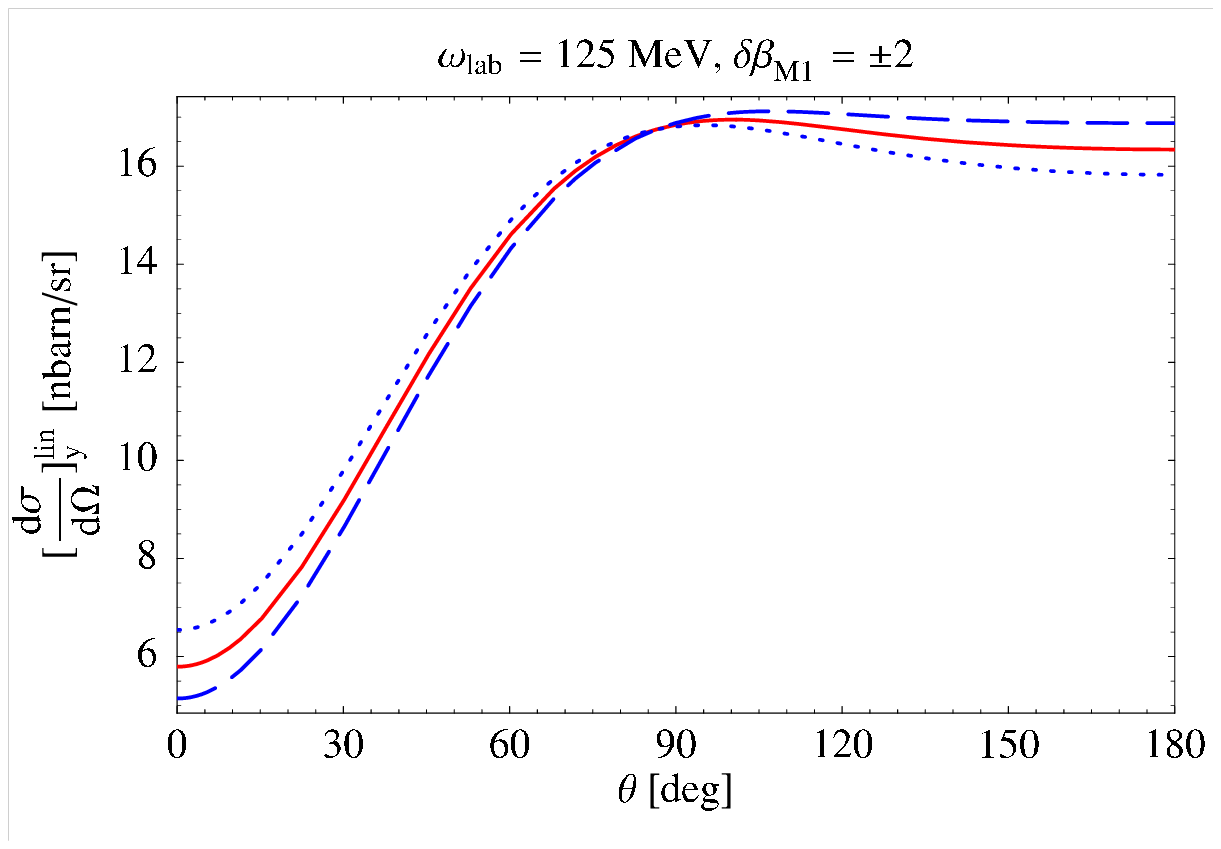}
    \caption {(Colour online) Dependence of $\left[
        \frac{\dd\sigma}{\dd\Omega}\right]^\text{lin}_{y}$ on the spin-independent
      dipole polarisabilities at $\omega_\text{lab}=45$~MeV (top) and
      $\omega_\text{lab}=125$~MeV (bottom). Notation as in
      Fig.~\ref{fig:dcsx_ab}. Notice the scale offset.}
\label{fig:dcsy_ab}
\end{center}
\end{figure}

\begin{figure}[!htb]
  \begin{center}
    \includegraphics*[width=0.48\linewidth]{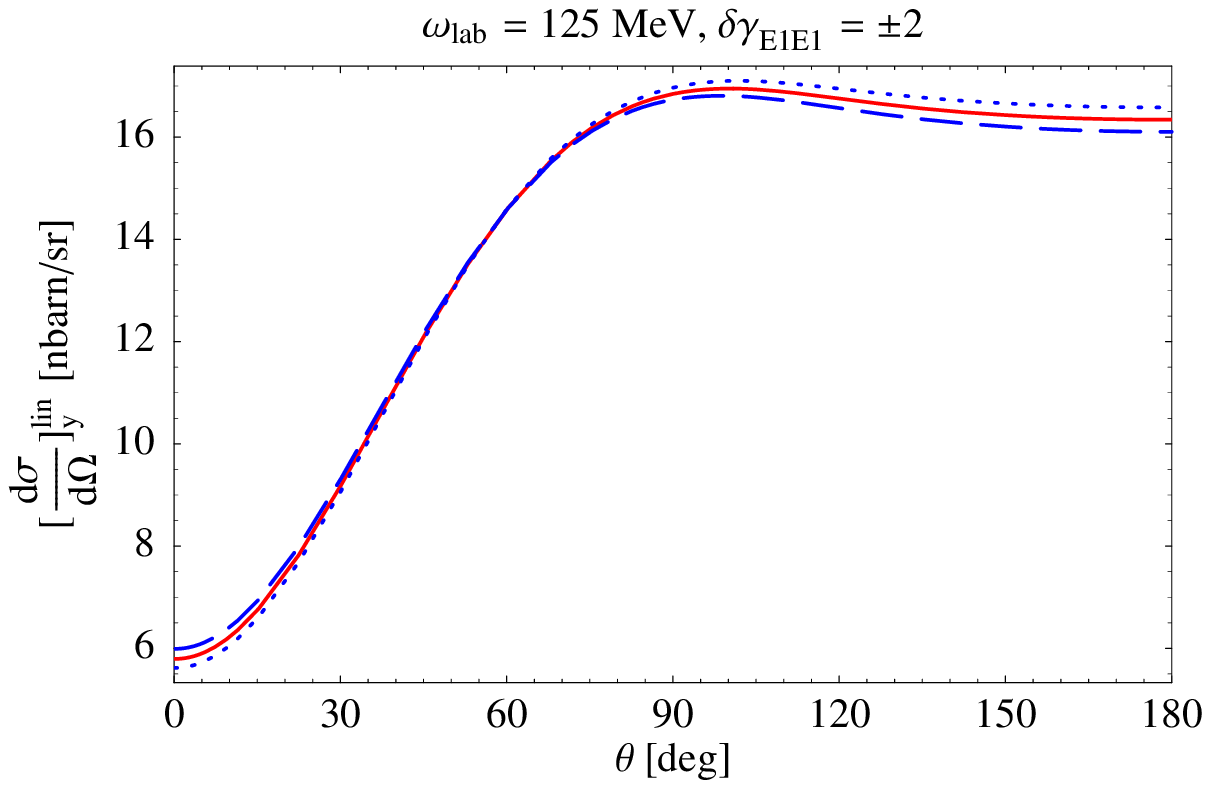}
    \hq\hq
    \includegraphics*[width=0.48\linewidth]{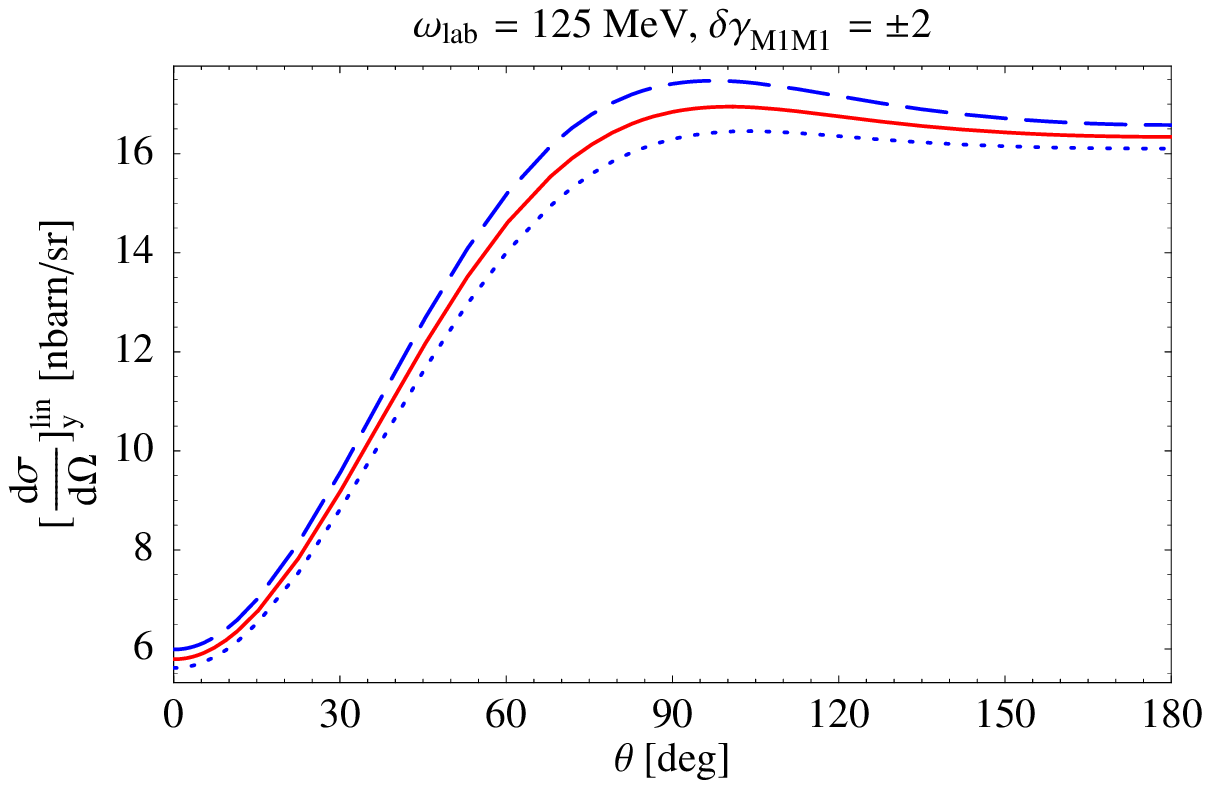}\\[1.5ex]
    \includegraphics*[width=0.48\linewidth]{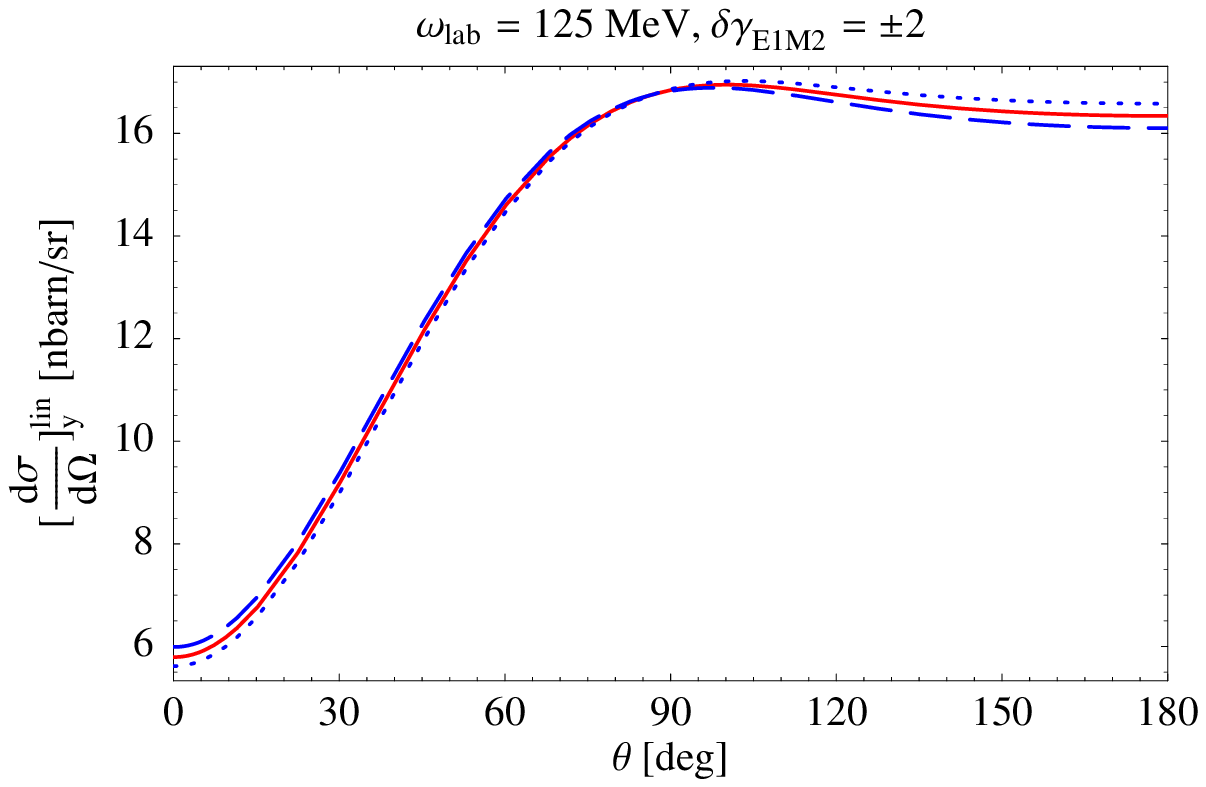}
    \hq\hq
    \includegraphics*[width=0.48\linewidth]{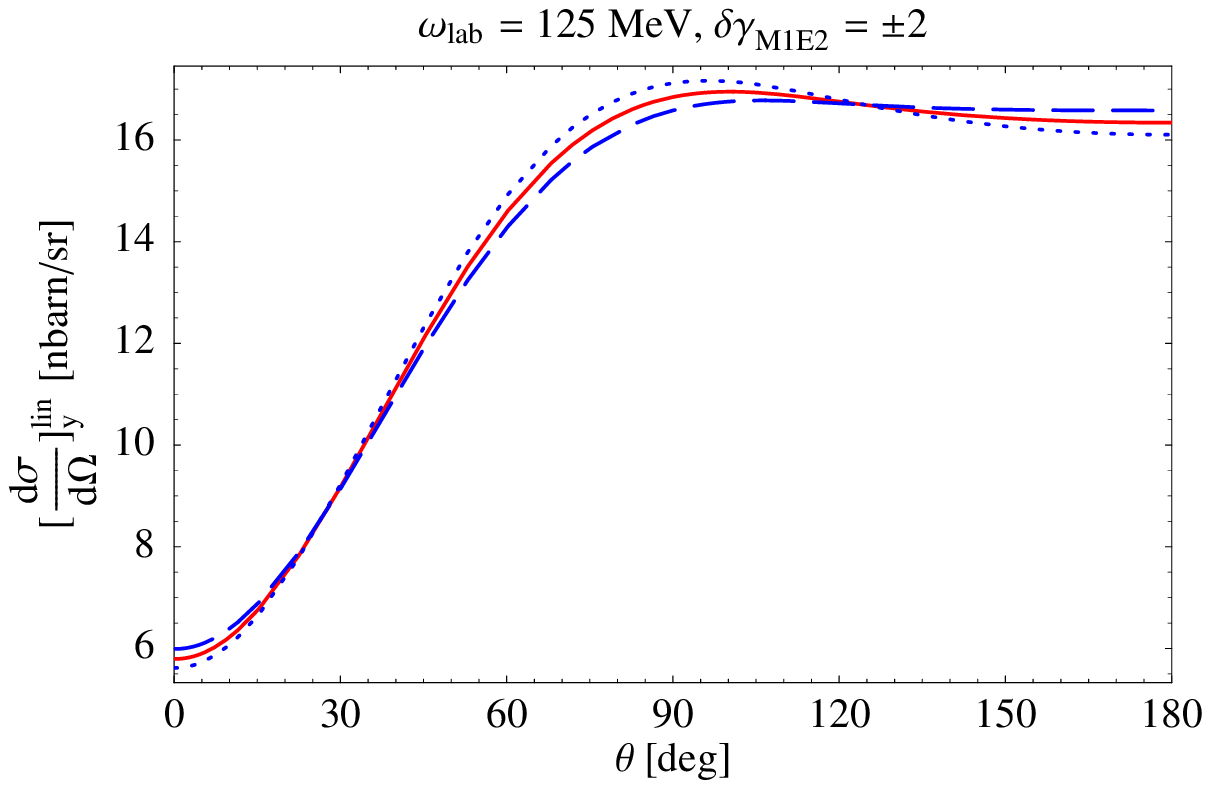}
    \caption {(Colour online) Dependence of $\left[
        \frac{\dd\sigma}{\dd\Omega}\right]^\text{lin}_{y}$ at
      $\omega_\text{lab}=125$~MeV on the dipole spin-polarisabilities.
      Notation as in Figs.~\ref{fig:dcsx_ab} and~\ref{fig:dcsx_gs}. Notice the
      scale offset.}
\label{fig:dcsy_gs}
\end{center}
\end{figure}

For Compton scattering on the deuteron, the relative motion of the $\gamma N$
cm system must be taken into account. As the two nucleons are in a relative
$S$ or $D$ wave in the deuteron, the conclusions remain unchanged, see
Figs.~\ref{fig:dcsx_ab} to \ref{fig:dcsy_gs}. 

Such arguments can of course be turned around: At $\theta_\text{cm}=90^\circ$,
$\left[\frac{\dd\sigma}{\dd\Omega}\right]^\text{lin}_{x}$ should be most
sensitive to $\beta_{M1}$, and
$\left[\frac{\dd\sigma}{\dd\Omega}\right]^\text{lin}_{y}(\theta_\text{cm}=90^\circ)$
to $\alpha_{E1}$, etc. However, embedding the nucleon into the deuteron
changes these predictions. For example, the greatest sensitivity to
$\beta_{M1}$ is for $\left[\frac{\dd\sigma}{\dd\Omega}\right]^\text{lin}_{x}$
actually found at $\theta\to180^\circ$, see Fig.~\ref{fig:dcsx_ab}.

Turning first to the initial photon polarised in the scattering plane, i.e.~to
$\left[\frac{\dd\sigma}{\dd\Omega} \right]_x^\text{lin}$ \eqref{eq:dcsx}, we
consider the two energies $\omega_\text{lab}=$45~MeV and 125~MeV. In
Fig.~\ref{fig:dcsx_ab}, the spin-independent polarisabilities are varied
independently by $\pm2$ units around the SSE values~\eqref{eq:cval}.
The effect at 45~MeV does not exceed $\pm0.2\;\mathrm{nbarn/sr}$, even at both
forward and back angles where the sensitivity is most noticeable. At 125~MeV,
there is with $\pm0.4$~nbarn/sr limited but anti-correlated sensitivity to
both $\alpha_{E1}$ and $\beta_{M1}$. As predicted above,
$\left[\frac{\dd\sigma}{\dd\Omega} \right]_x^\text{lin}$ is indeed insensitive
to $\alpha_{E1}$ at $\theta_\text{cm}=90^\circ$,
i.e.~$\theta_\text{lab}=88.65^\circ$ at $\omega_\text{lab}=45$~MeV and
$86.41^\circ$ at $125$~MeV. At 45~MeV, a variation of the spin
polarisabilities by $\pm$2 induces negligible effects
($\lesssim\pm0.02$~nbarn/sr) and hence is not shown. This is expected because
the spin-polarisabilities appear at higher order in energy than $\alpha_{E1}$
and $\beta_{M1}$. Figure~\ref{fig:dcsx_gs} reports the sensitivity on varying
the spin-polarisabilities one by one by $\pm$2 at $\omega_\text{lab}=125$~MeV.
Although visible, this causes a maximum change of $\pm 0.25$~nbarn/sr. Note
that $\gamma_{M1E2}$ does not contribute at $\theta_\text{cm}=90^\circ$. That
there is neither dependence on $\gamma_{E1E1}$ at
$\theta_\text{lab}\approx45^\circ$ nor on $\gamma_{E1M2}$ at
$\theta_\text{lab}\approx90^\circ$, seems to be a fortunate accident,
un-related to intuitive arguments like those above.  A simultaneous experiment
at various angles can thus dis-entangle individual spin-polarisabilities, with
some of the experimental systematics cancelling.  However, the small
variations may make clear signals difficult to obtain.

Figure~\ref{fig:dcsy_ab} shows plots analogous to Fig.~\ref{fig:dcsx_ab} when
the initial photons are polarised perpendicular to the scattering plane,
i.e.~for $\left[\frac{\dd\sigma}{\dd\Omega} \right]_y^\text{lin}$
\eqref{eq:dcsy}).
At 45~MeV, varying $\alpha_{E1}$ or $\beta_{M1}$ by $\pm$2 induces an effect
of $\approx\pm$0.4~nbarn/sr effect, while the maximum sensitivity at 125~MeV is
increased to $\approx\pm$0.7~nbarn/sr. Sensitivity to $\alpha_{E1}$ is quite
uniform at all angles, while that to $\beta_{M1}$ vanishes at
$\theta_\text{cm}=90^\circ$, as predicted.

A simultaneous extraction of both the electric and magnetic polarisabilities
may thus be possible from the same observable. Particularly promising is
measuring
$\left[\frac{\dd\sigma}{\dd\Omega}\right]^\text{lin}_{y}(\theta_\text{cm}=90^\circ)$,
which gives $\alpha_{E1}$ and can be used as an input for the extraction of
$\beta_{M1}$ from the same observable at other angles. Alternatively, one may
use the same detector configuration $\theta_\text{cm}=90^\circ$ but change the
linear beam polarisation from in-scattering-plane (insensitive to
$\alpha_{E1}$) to perpendicular-to-scattering-plane (insensitive to
$\beta_{M1}$), taking advantage of the cancellation of some systematic
experimental uncertainties~\cite{Weller}.

Analogously to Fig.~\ref{fig:dcsx_gs}, we also present the same observable at
$\omega_\text{lab}=$125 MeV in Fig.~\ref{fig:dcsy_gs} when the spin
polarisabilities are varied.
The maximum sensitivity of $\approx\pm$0.7~nbarn/sr is to $\gamma_{M1M1}$
around $90^\circ$, comparable to that on $\alpha_{E1}$. Since dependence on
$\gamma_{E1E1}$ and $\gamma_{E1M2}$ is negligible there and that on
$\gamma_{M1E2}$ is smaller ($\approx\pm$0.3~nbarn/sr), one can hope to extract
a linear combination of only two spin-polarisabilities, dominated by
$\gamma_{M1M1}$. Now, $\gamma_{E1M2}$ does not contribute at
$\theta_\text{cm}=90^\circ$, $\gamma_{E1E1}$ not at
$\theta_\text{lab}\approx60^\circ$ and $\gamma_{M1E2}$ not at
$\theta_\text{lab}\approx30^\circ$ and $\approx120^\circ$. 

\begin{figure}[!htb]
  \begin{center}
    \includegraphics*[width=0.48\linewidth]{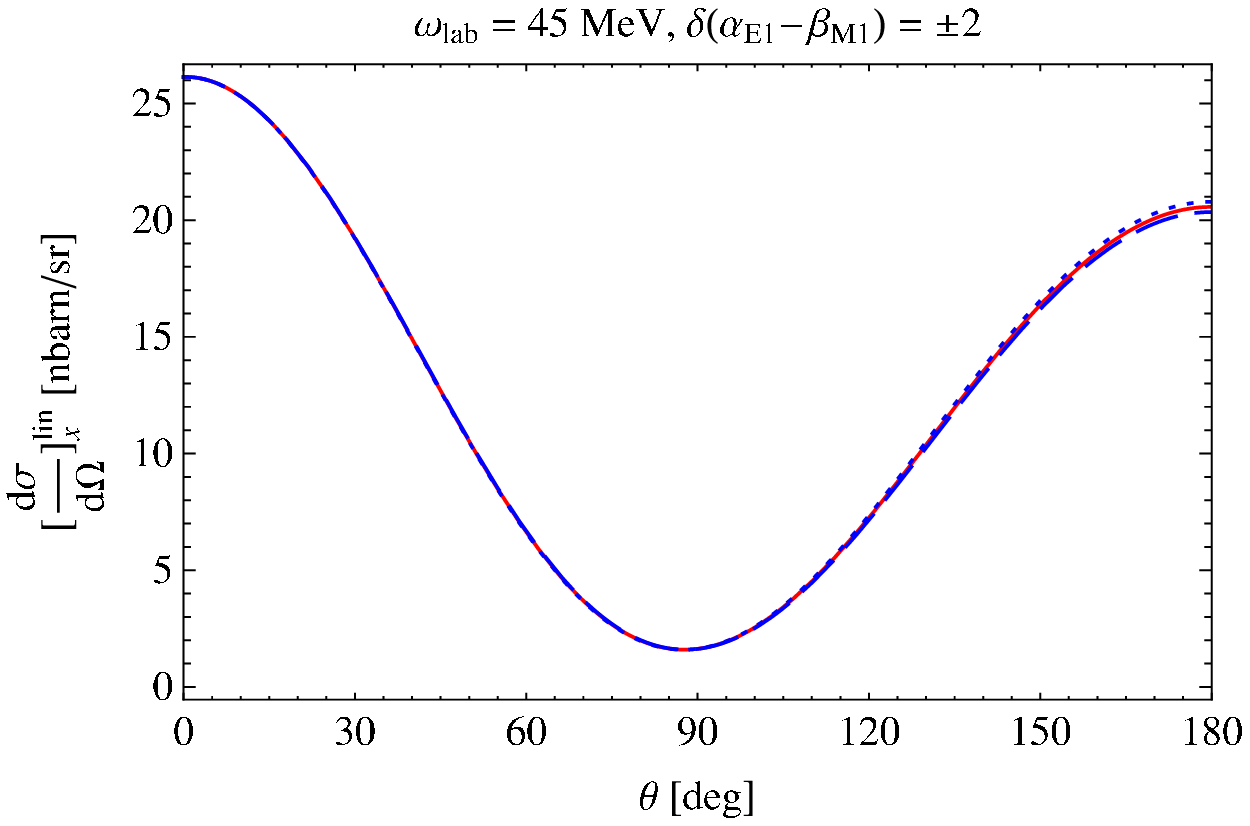}
    \hq\hq
    \includegraphics*[width=0.48\linewidth]{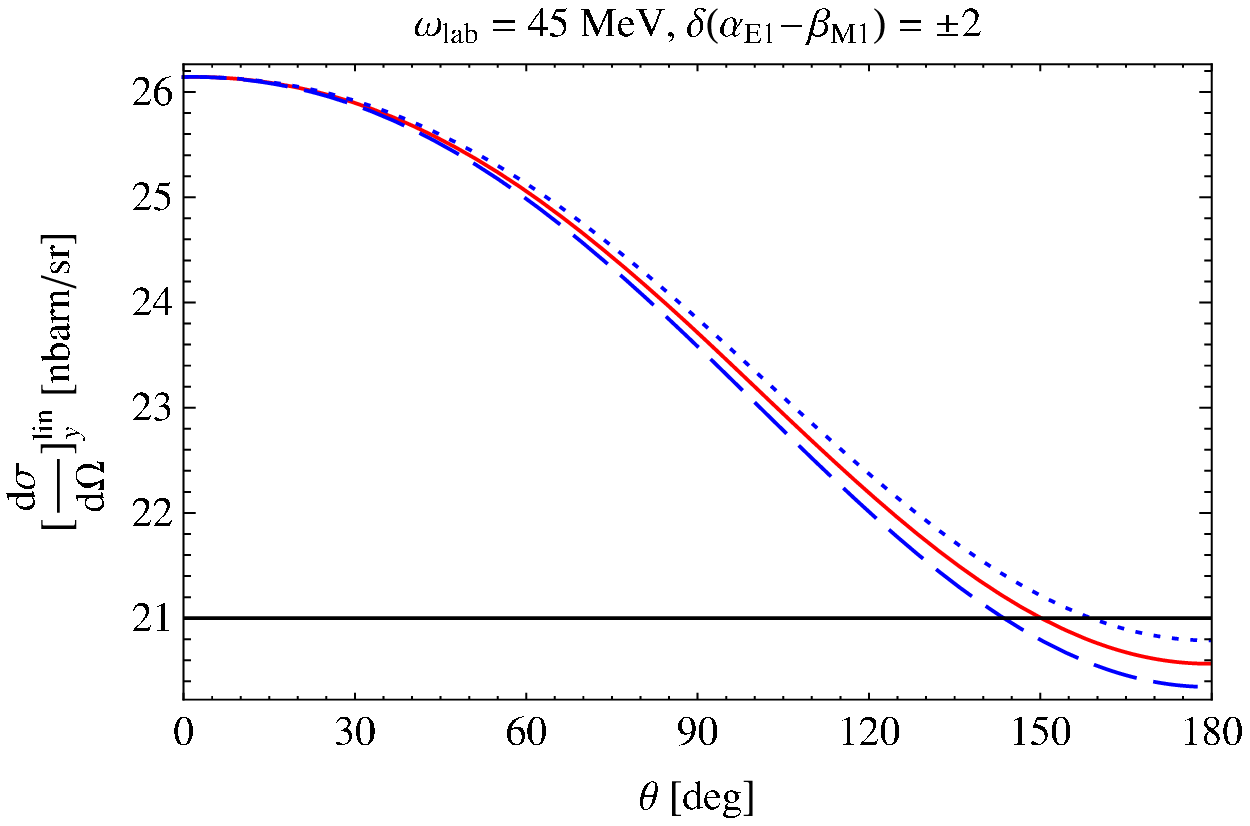}\\[1.5ex]
    \includegraphics*[width=0.48\linewidth]{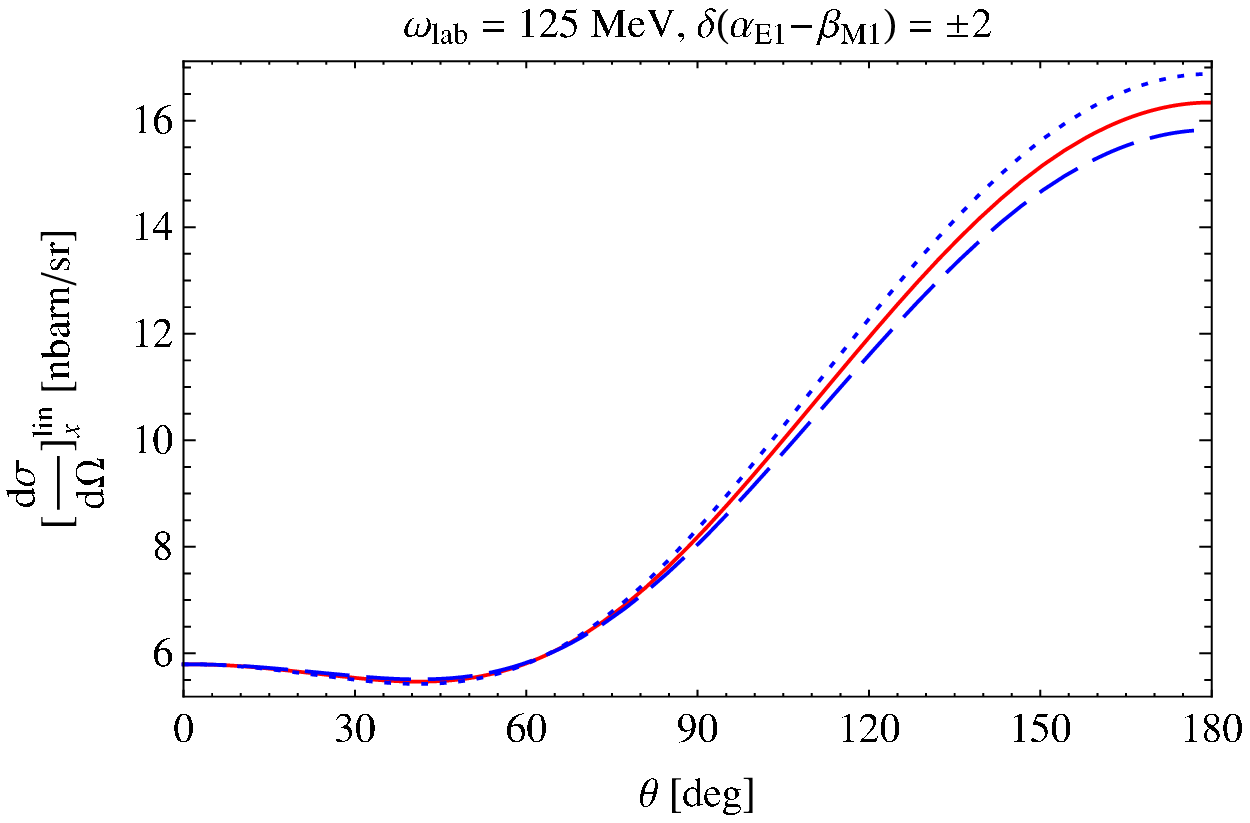}
    \hq\hq
    \includegraphics*[width=0.48\linewidth]{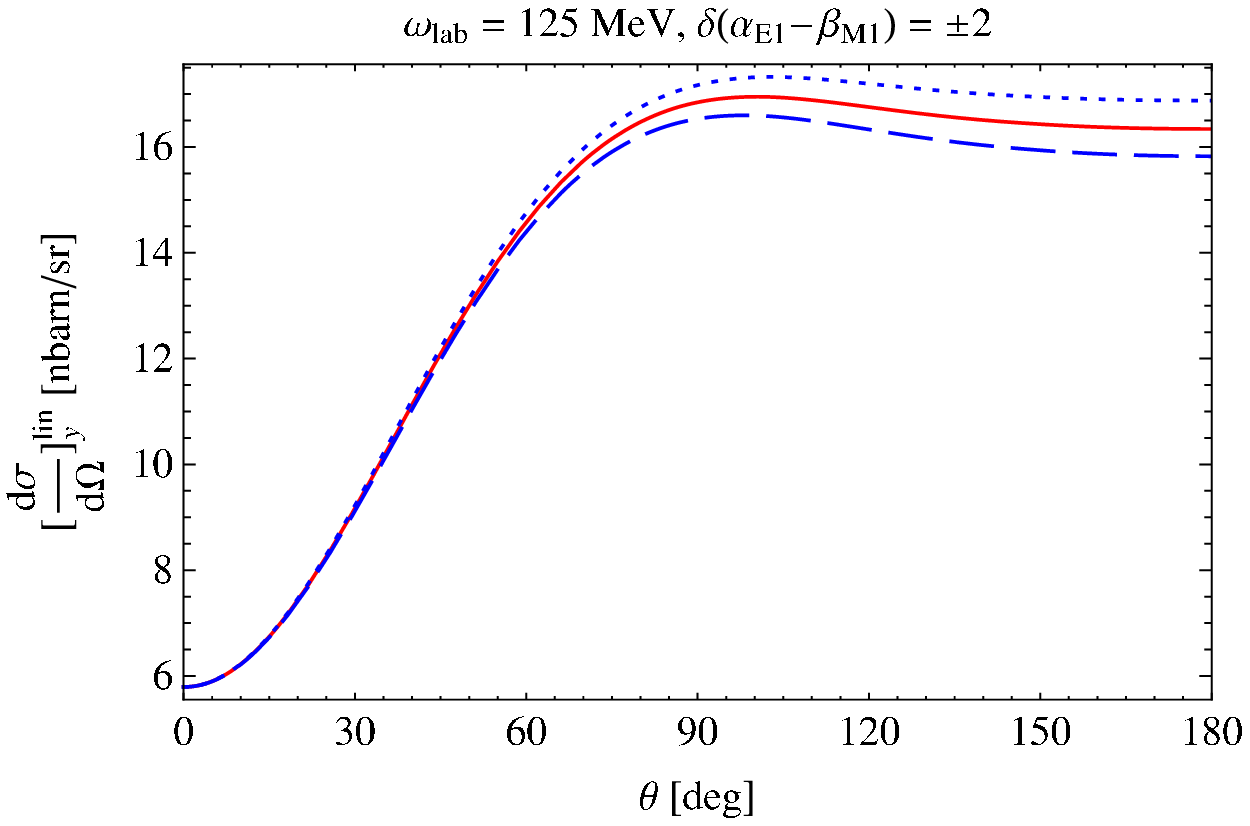}
    \caption {(Colour online) Dependence of $\left[
        \frac{\dd\sigma}{\dd\Omega}\right]^\text{lin}_{x}$ (left) and $\left[
        \frac{\dd\sigma}{\dd\Omega}\right]^\text{lin}_{y}$ (right) at
      $\omega_\text{lab}=45$~MeV (top) and $125$~MeV (bottom) on
      $\delta(\alpha_{E1}-\beta_{M1})=\pm2$, with their sum constrained by the
      Baldin sum-rule.  Notation as in Fig.~\ref{fig:dcsx_ab}. Notice the
      scale offset.}
\label{fig:dcsxy_ab}
\end{center}
\end{figure}

Finally, we present an example in which an additional theoretical constraint
is taken into account. Figure \ref{fig:dcsxy_ab} shows that both observables
change by $\pm$0.25~nbarn/sr at backward angles when $\alpha_{E1}-\beta_{M1}$
while the Baldin sum-rule constraint \eqref{eq:baldin} is imposed for
$\alpha_{E1}+\beta_{M1}$. As sensitivity to $\alpha_{E1}-\beta_{M1}$ remains
more-or-less constant with energy, one could extract the difference at
relatively low energies, where spin-polarisabilities have negligible effects.

In summary, $\left[\frac{\dd\sigma}{\dd\Omega} \right]_x^\text{lin}$ does not
appear to be a good observable to extract even the largest of the
polarisabilities, while $\left[\frac{\dd\sigma}{\dd\Omega}
\right]_y^\text{lin}$ proves to be an effective observable to extract both
$\alpha_{E1}$ and $\beta_{M1}$, or the combination $\alpha_{E1}-\beta_{M1}$.
It also shows sensitivity to mostly $\gamma_{M1M1}$ at higher energies. With
the increased flux at \HIGS, such effects should be measurable.

\section{Double-polarisation Observables}
\label{sec:doublepol}

New experimental techniques make it feasible to measure observables involving
linearly or circularly polarised photons and a deuteron target that can be
vector-polarised either along the beam axis or perpendicular to it, see
eqs.~\eqref{eq:deltaxlin} to \eqref{eq:deltaz}. A first qualitative study of
some effects was given in ``pion-less'' EFT in Ref.~\cite{Chen:2004wwa}.  We
extend the analysis of polarisation observables for circularly polarised
photons by \ChiEFT in~\cite{Ch05,mythesis} to include the sizable
contributions from $NN$-rescattering and dynamical $\Delta$ effects. In
addition, we demonstrate, to our knowledge for the first time, that linearly
polarised beams provide an avenue to extract polarisabilities in deuteron
Compton scattering.

Consider first again the effective Lagrangean \eqref{eq:pols-lag} with a
configuration in which the target is polarised parallel to the linear
polarisation of the beam, cf.~Fig.~\ref{fig:switchingoffpols}. Now,
$\gamma_{E1E1}$ does not contribute to scattering under any scattering angle.
Similarly, scattering off a target polarised perpendicular to a linearly
polarised beam is under any angle insensitive to $\gamma_{M1M1}$.
No such relations are found for circularly polarised
photons on vector polarised targets. Double polarisation observables are
however formulated using differences of cross-sections with different
polarisations, and no straight-forward argument exists that one or more of the
dipole polarisabilities as insensitive to $\Delta^\text{lin/circ}_{x/z}$ or
$\Sigma^\text{lin/circ}_{x/z}$. Recall that differences and asymmetries are
favoured in small count-rates since some systematic effects cancel.

As done for single-polarisation observables, we consider first sensitivity of
each observable to the spin-independent polarisabilities at
$\omega_\text{lab}=45$ and $125$~MeV, followed by sensitivity to the
spin-polarisabilities at $125$~MeV.  We also re-iterate that while only
cross-section differences $\Delta^\text{circ/lin}_{x/z}$ are presented in the
following, we also investigated asymmetries $\Sigma^\text{circ/lin}_{x/z}$,
with results available in a \emph{Mathematica} notebook. In general,
sensitivity to polarisabilities are decreased in asymmetries, relative to
signals in differences.

\subsection{Polarised Target and Linearly-polarised Photons}
\label{sec:ldpol}

The observable $\Delta^\text{lin}_{z}$, \eqref{eq:deltazlin}, describes
scattering linearly polarised photons off a target polarised parallel to the
beam. The best signal is obtained when the beam polarisation lies in the
scattering plane.
\begin{figure}[!htb]
  \begin{center}
    \includegraphics*[width=0.48\linewidth]{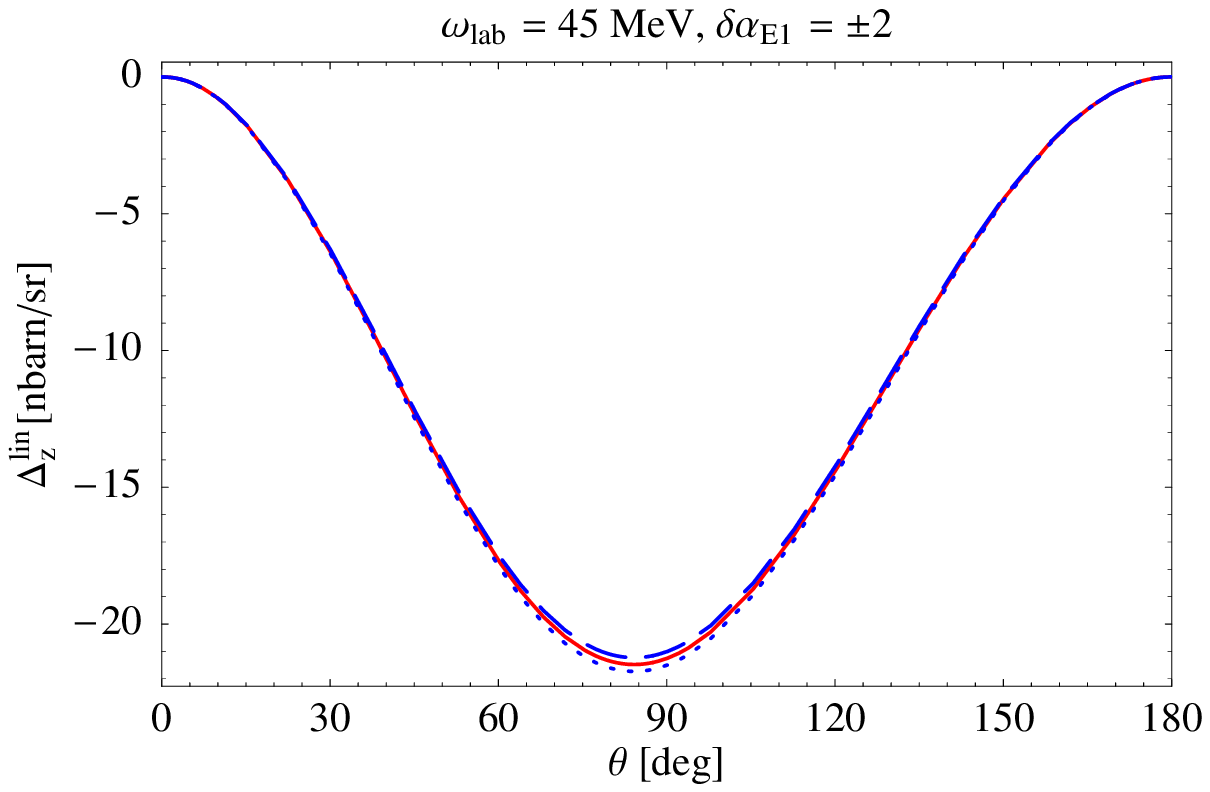}
    \hq\hq
    \includegraphics*[width=0.48\linewidth]{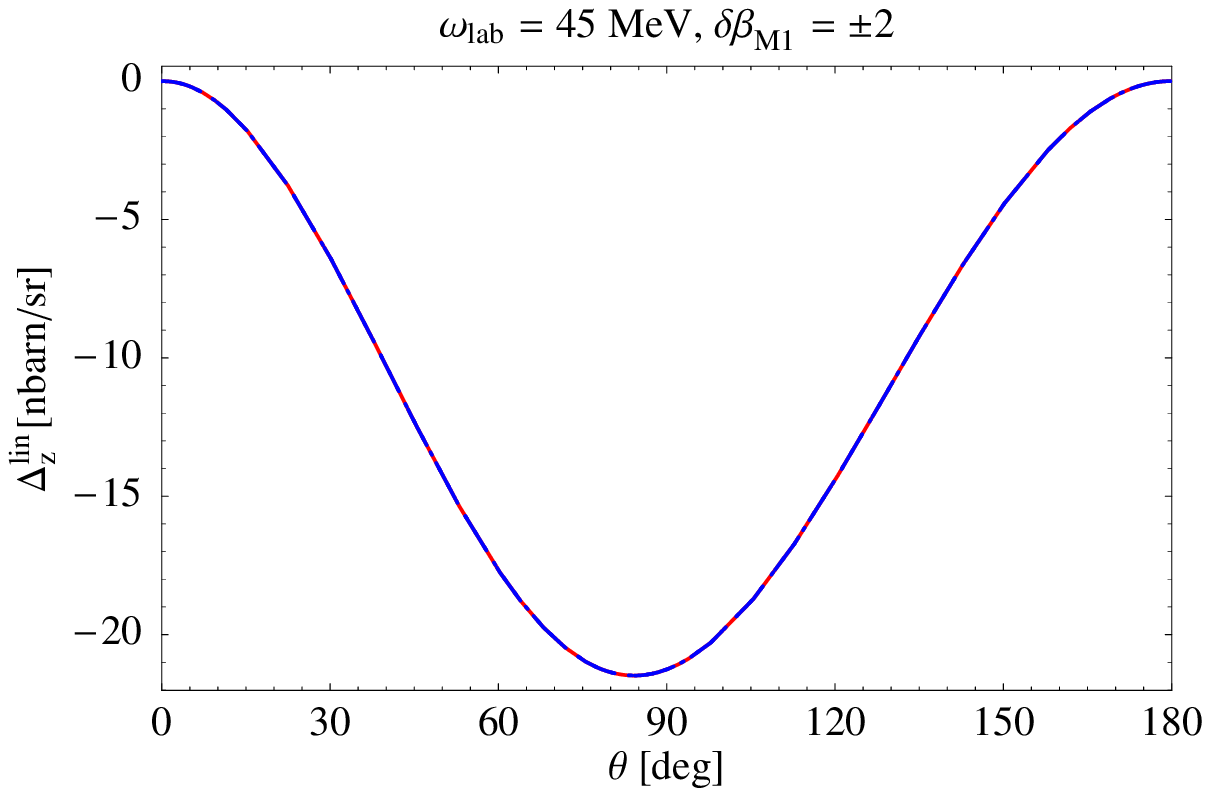}\\[1.5ex]
    \includegraphics*[width=0.48\linewidth]{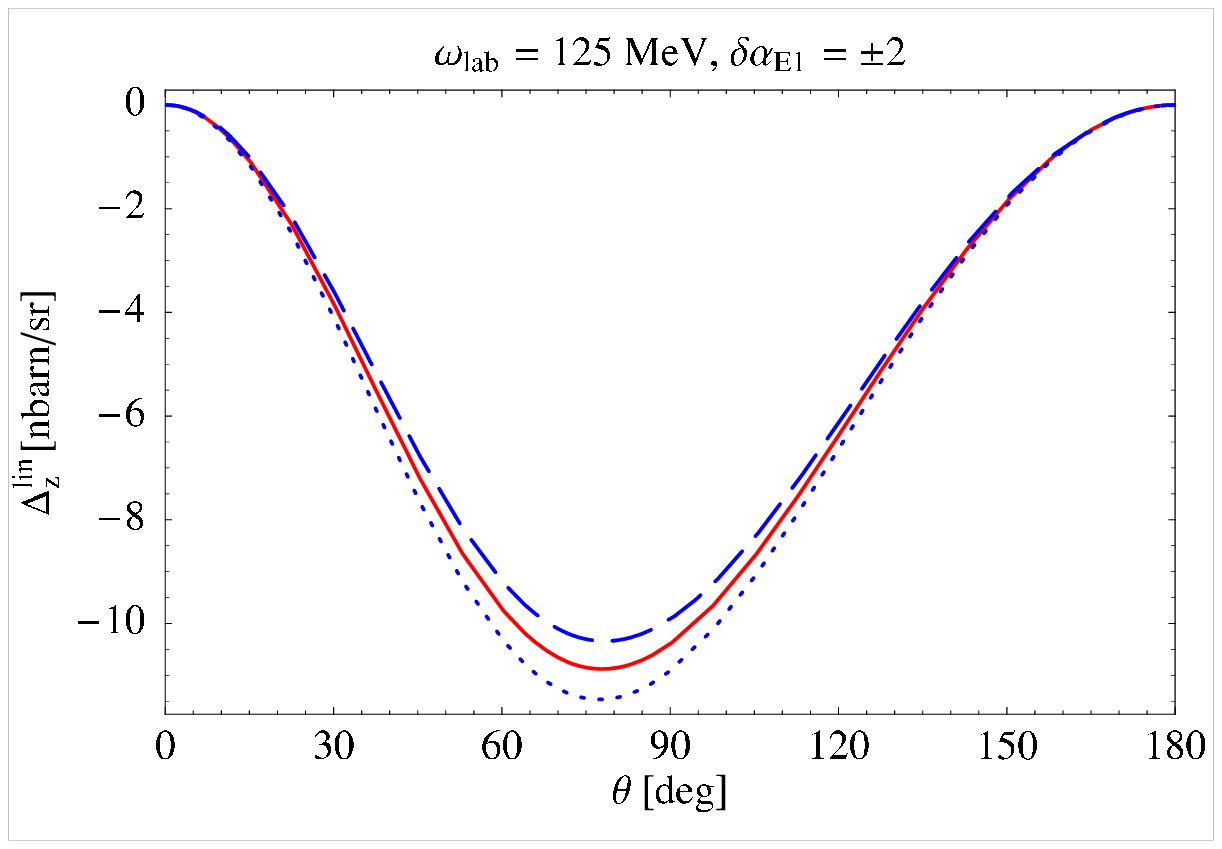}
    \hq\hq
    \includegraphics*[width=0.48\linewidth]{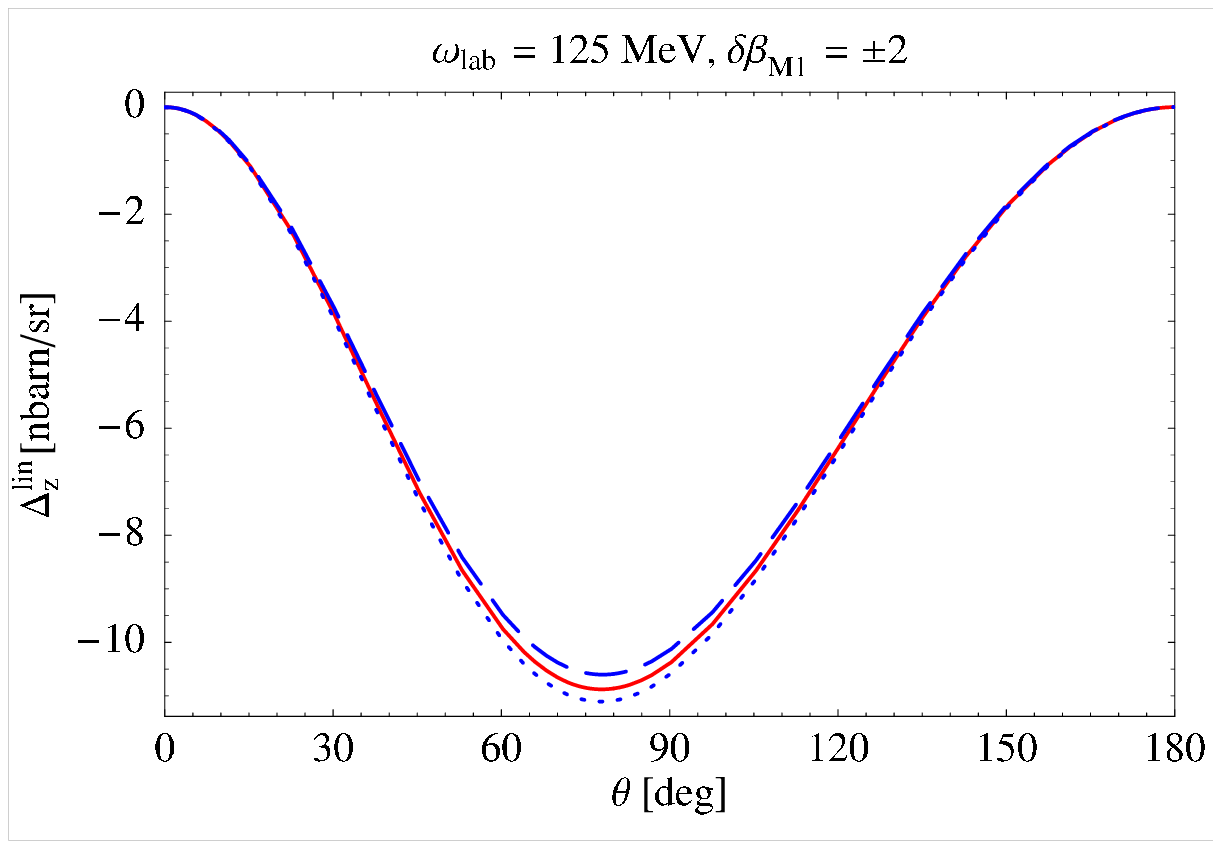}
    \caption {(Colour online) Dependence of $\Delta^\text{lin}_{z}$ on the
      spin-independent dipole polarisabilities at $\omega_\text{lab}=45$~MeV
      (top) and $\omega_\text{lab}=125$~MeV (bottom). Notation as in
      Fig.~\ref{fig:dcsx_ab}.}
  \label{fig:deltazlin_ab}
\end{center}
\end{figure}
At 45~MeV, varying $\delta \alpha_{E1}=\pm2$ causes an effect of
$\lesssim\pm0.25$~nbarn/sr at $\theta_\text{lab}\approx90^\circ$, while this
doubles at 125~MeV to a maximum of $\pm0.5$~nbarn/sr, see
Fig.~\ref{fig:deltazlin_ab}.  The effect of $\beta_{M1}$ is insignificant at 45~MeV
and $\lesssim\pm0.25$~nbarn/sr at 125~MeV, anti-correlated to that of
$\alpha_{E1}$.
\begin{figure}[!htb]
  \begin{center}
    \includegraphics*[width=0.48\linewidth]{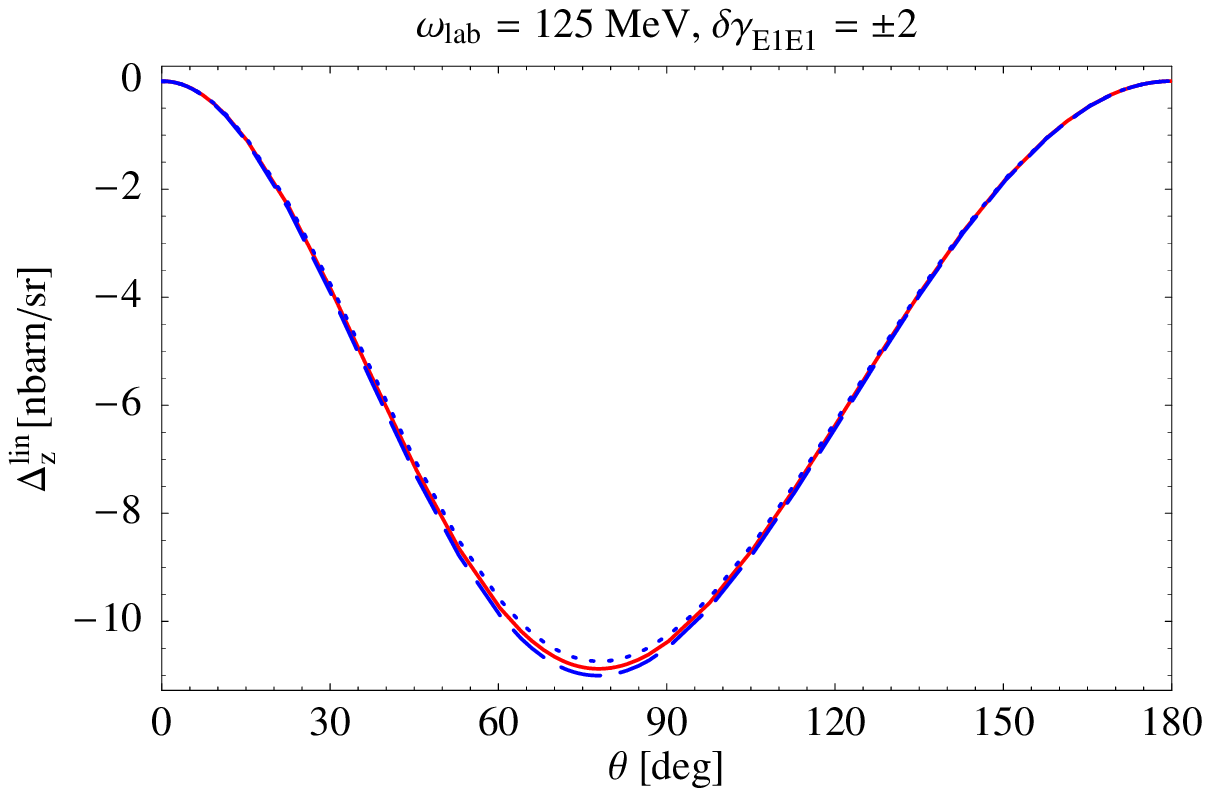}
    \hq\hq
    \includegraphics*[width=0.48\linewidth]{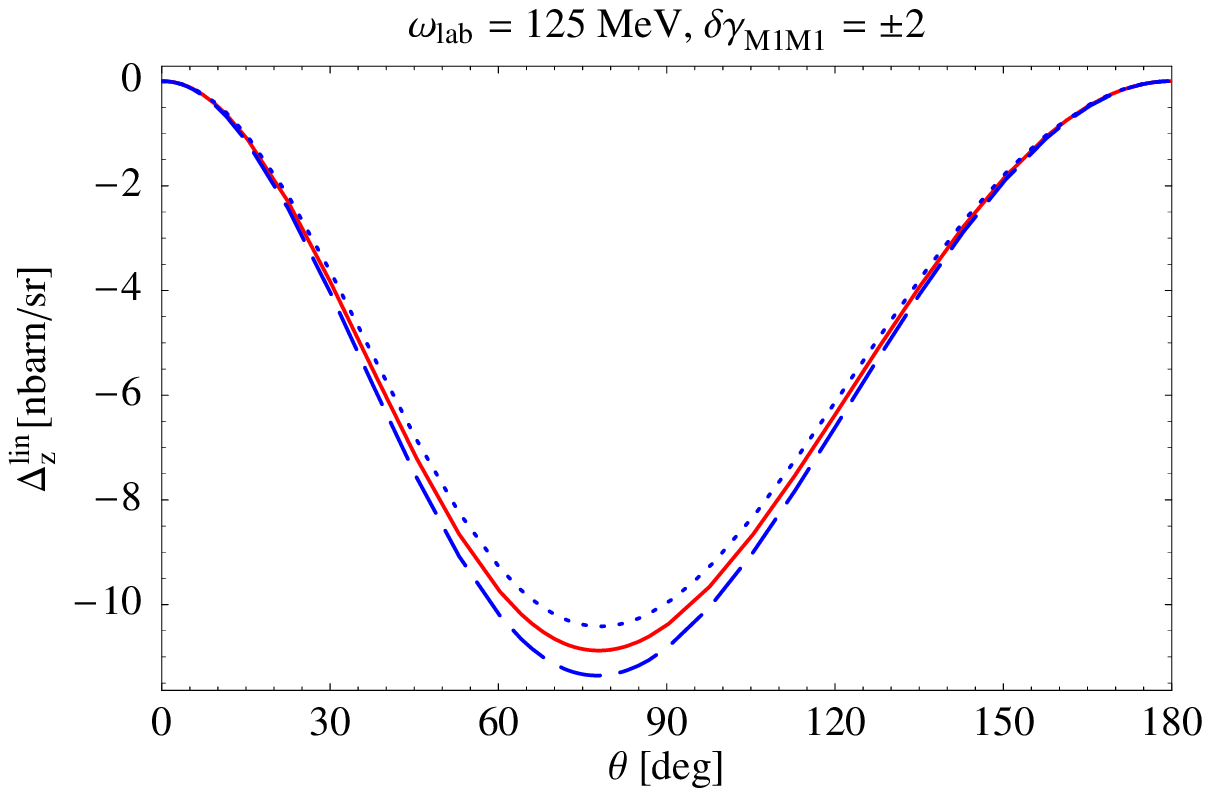}\\[1.5ex]
    \includegraphics*[width=0.48\linewidth]{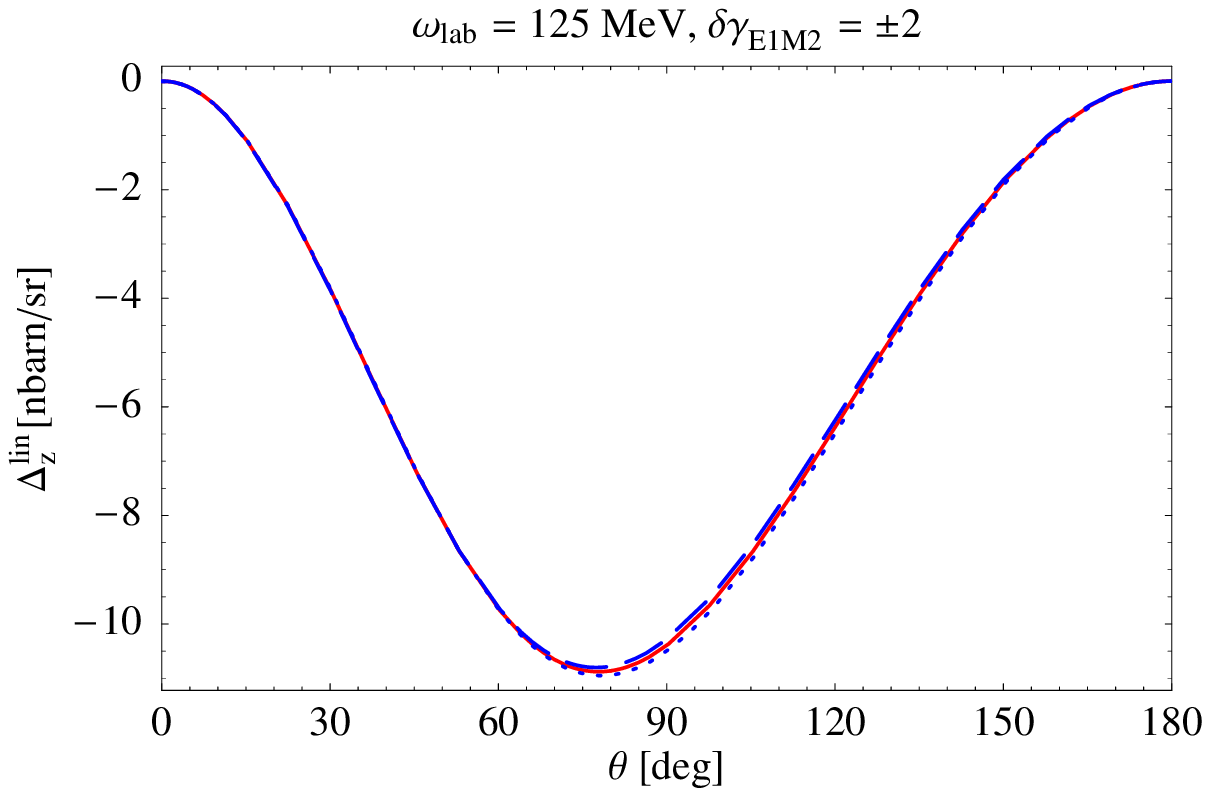}
    \hq\hq
    \includegraphics*[width=0.48\linewidth]{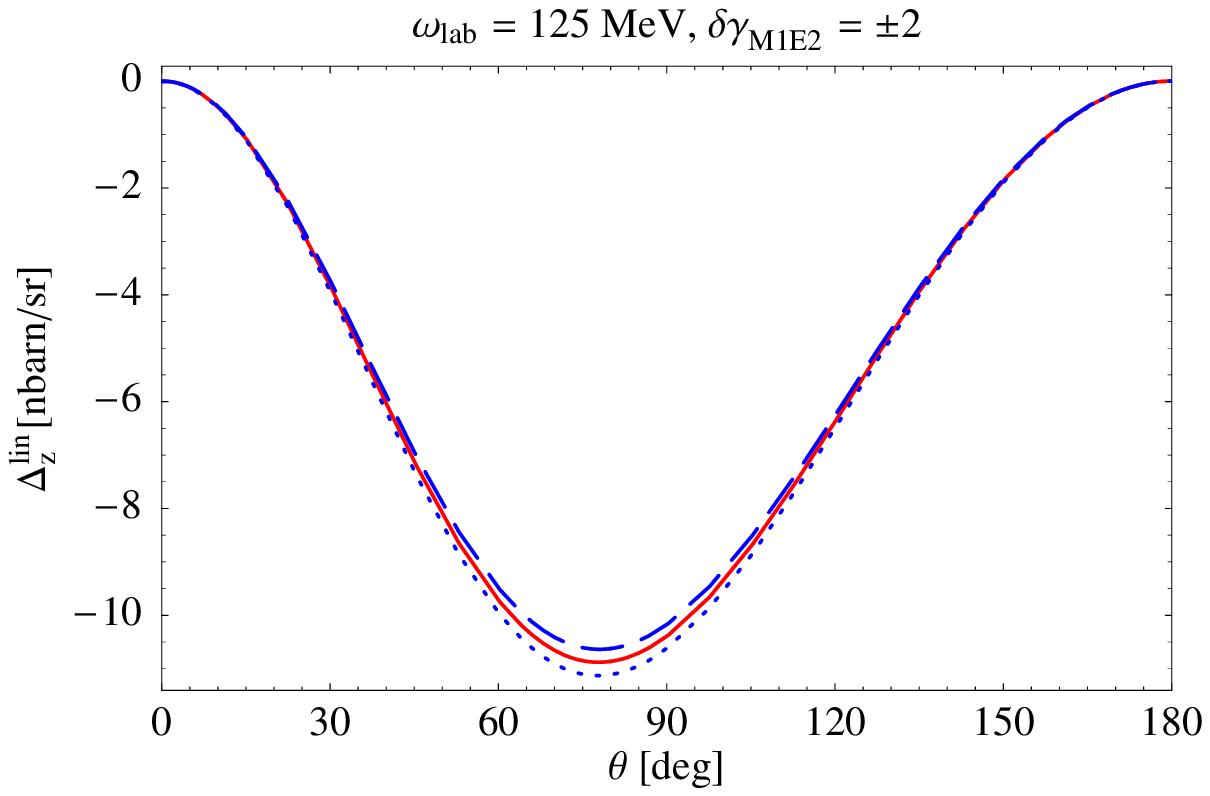}
    \caption {(Colour online) Dependence of $\Delta^\text{lin}_{z}$ at
      $\omega_\text{lab}=125$~MeV on the dipole spin-polarisabilities.
      Notation as in Figs.~\ref{fig:dcsx_ab} and~\ref{fig:dcsx_gs}.}
\label{fig:deltazlin_gs}
\end{center}
\end{figure}
When the spin-polarisabilities are varied one by one by $\pm2$ units around
their central values at $\omega_\text{lab}=125$~MeV in
Fig.~\ref{fig:deltazlin_gs}, sensitivity to $\gamma_{M1M1}$ around
$\theta_\text{lab}\approx 75^\circ$ provides with $\pm$0.5~nbarn/sr a
measurably large effect, comparable but anti-correlated to that of
$\alpha_{E1}$. However, a weak dependence on $\gamma_{M1E2}$
($\pm$0.2~nbarn/sr) cannot be discounted. Since sensitivity on the other
spin-polarisabilities is negligible, one can extract a linear combination of
the anti-correlated polarisabilities $\gamma_{M1M1}$ and $\gamma_{M1E2}$, with
the former dominating. These polarisabilities are expected to be dominated by
the strongly para-magnetic $\gamma\Delta N$ coupling.

\begin{figure}[!htb]
  \begin{center}
    \includegraphics*[width=0.48\linewidth]{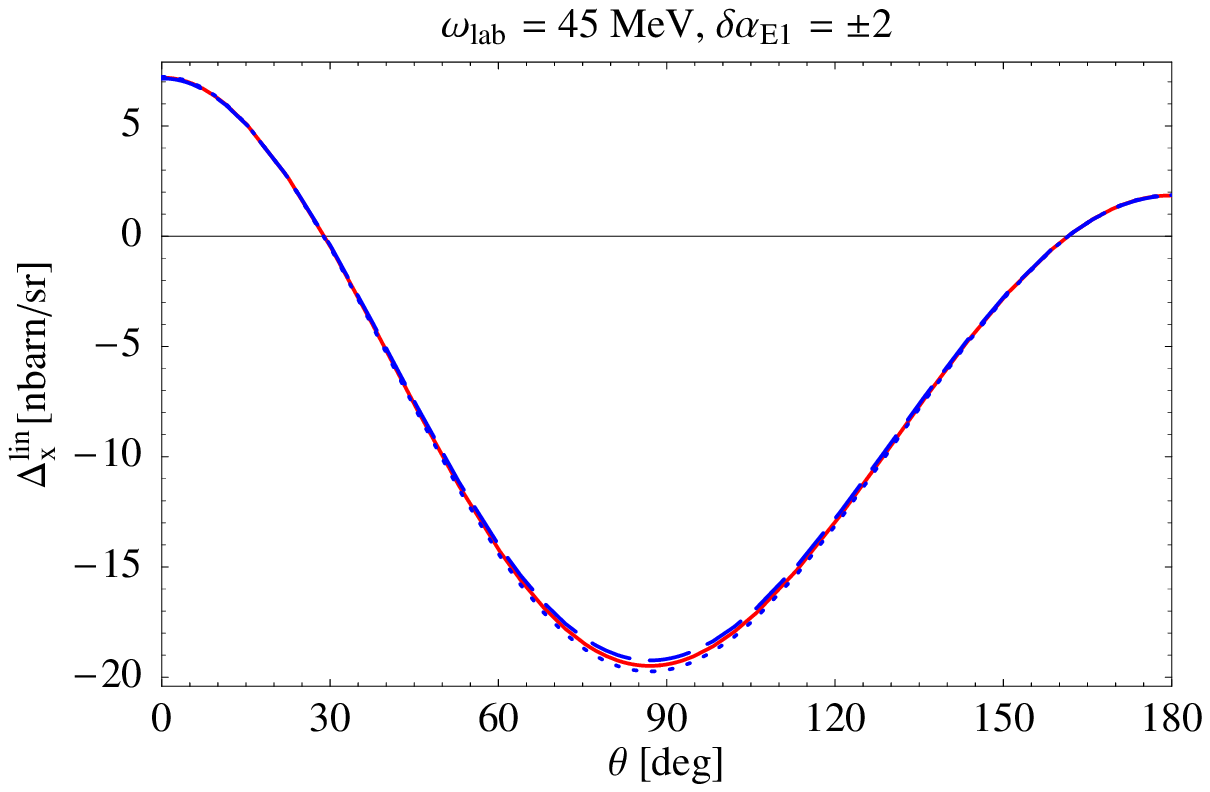}
    \hq\hq
    \includegraphics*[width=0.48\linewidth]{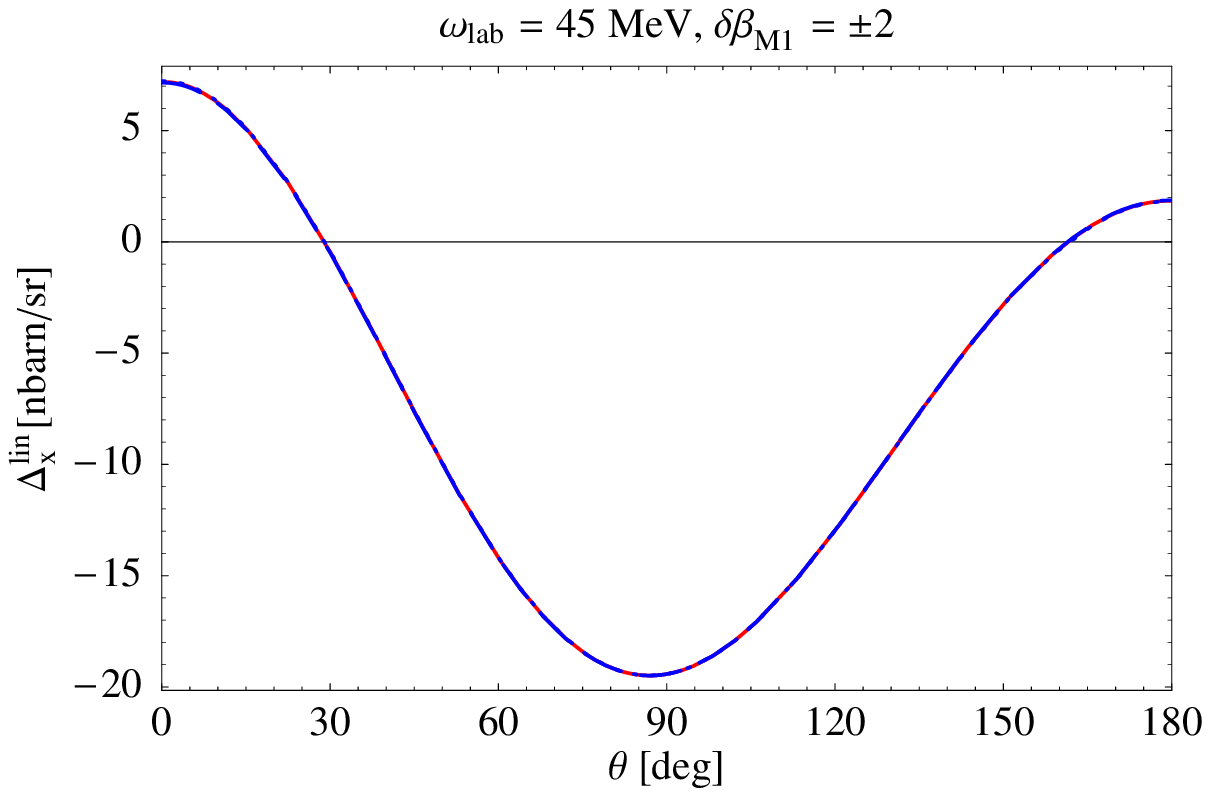}\\[1.5ex]
    \includegraphics*[width=0.48\linewidth]{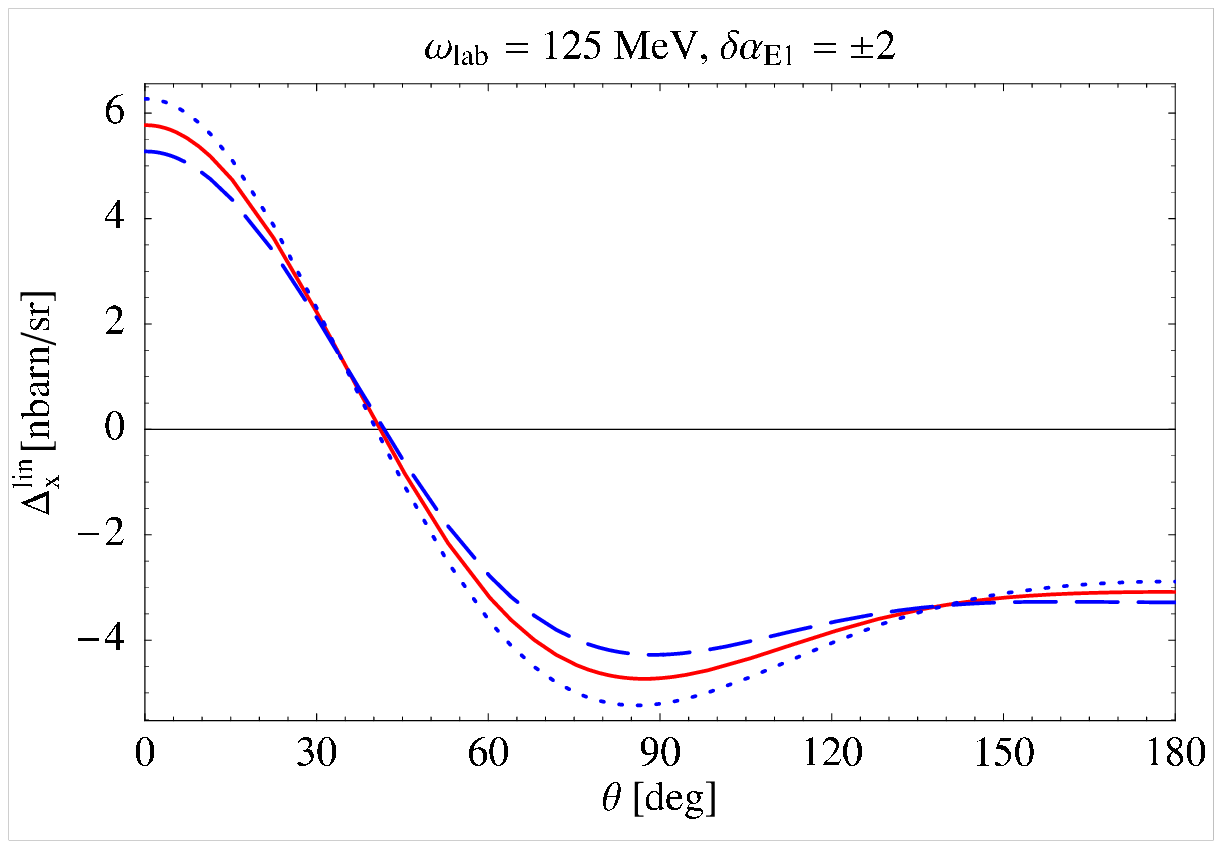}
    \hq\hq
    \includegraphics*[width=0.48\linewidth]{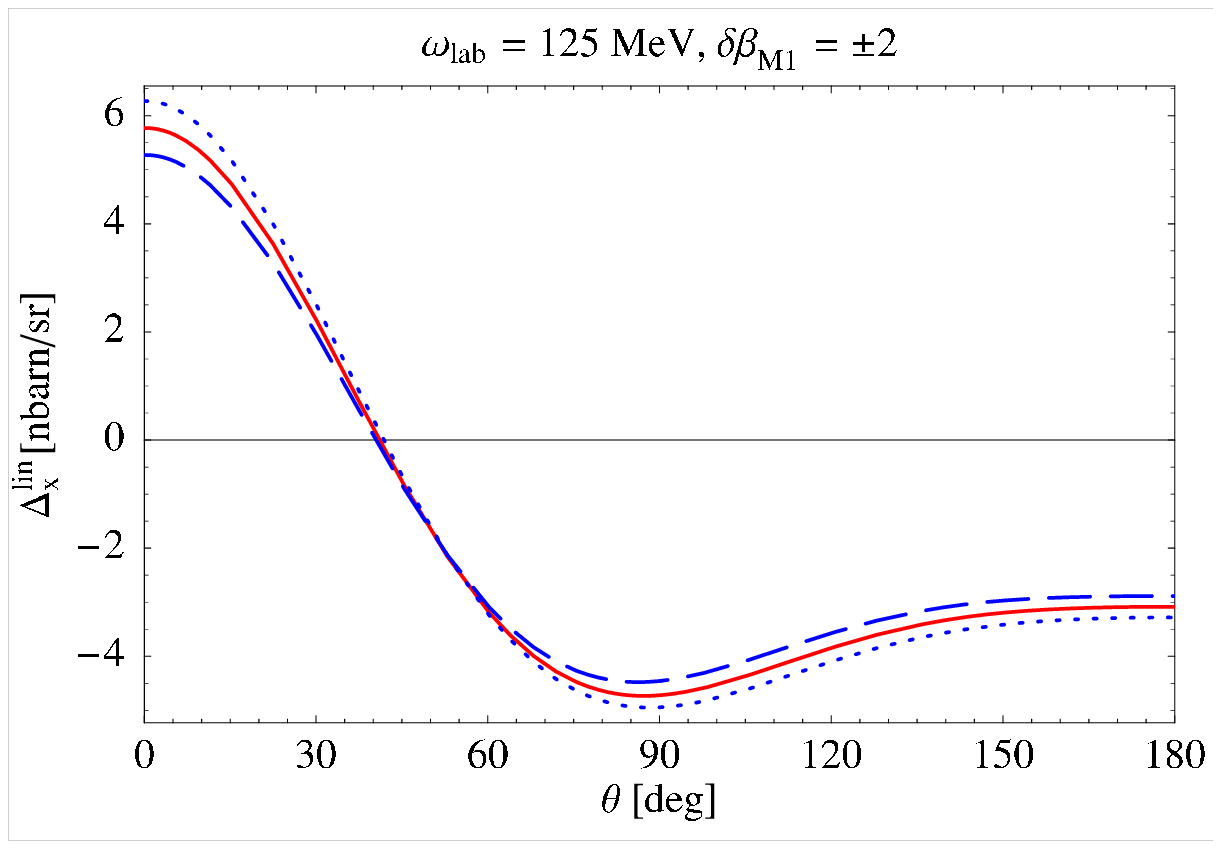}
  \caption {(Colour online) Dependence of $\Delta^\text{lin}_{x}$ on the
      spin-independent dipole polarisabilities at $\omega_\text{lab}=45$~MeV (top) and
      $\omega_\text{lab}=125$~MeV (bottom). Notation as in
      Fig.~\ref{fig:dcsx_ab}.}
\label{fig:deltaxlin_ab}
\end{center}
\end{figure}

\begin{figure}[!htb]
  \begin{center}
    \includegraphics*[width=0.48\linewidth]{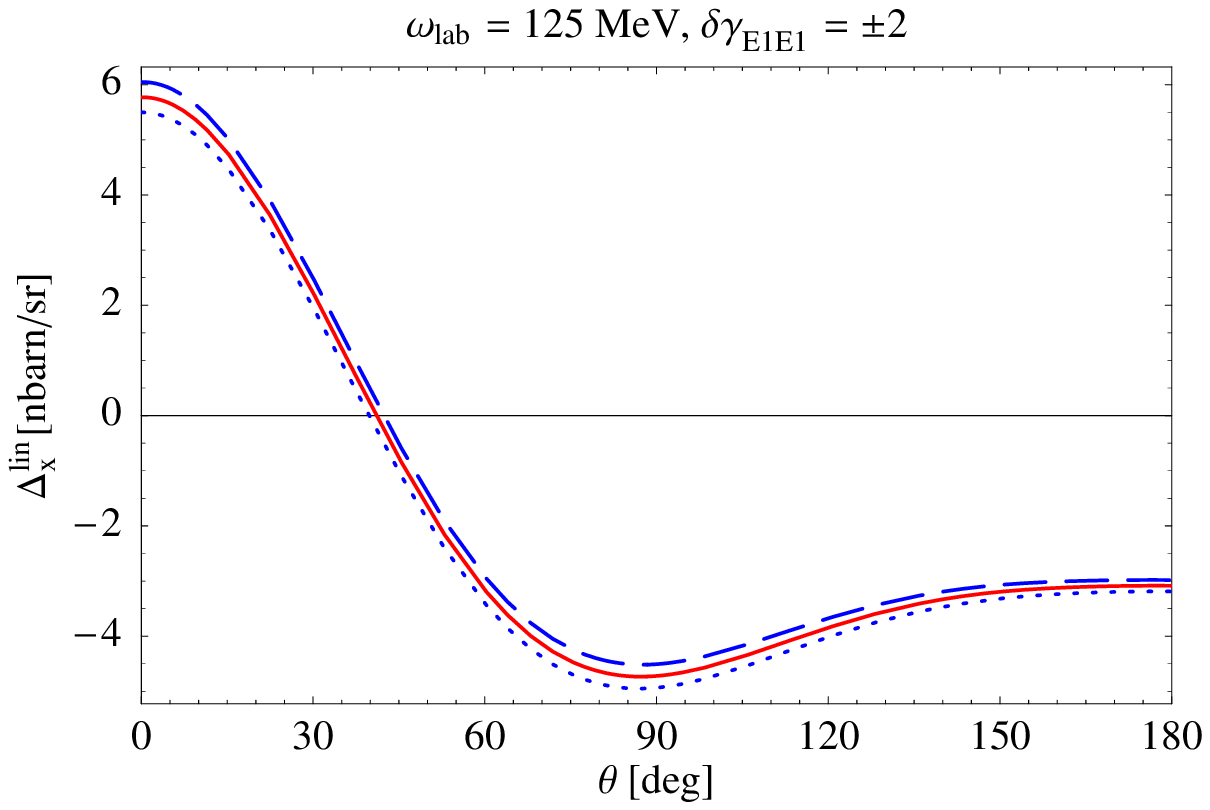}
    \hq\hq
    \includegraphics*[width=0.48\linewidth]{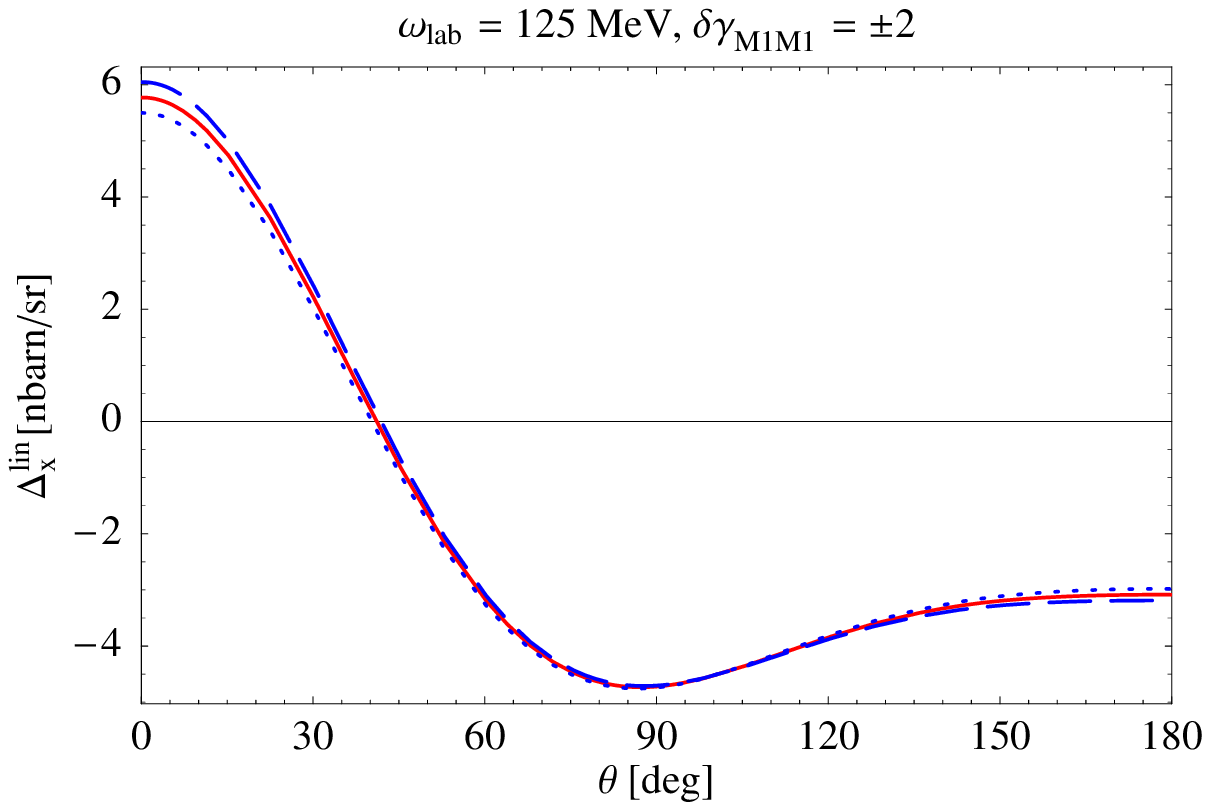}\\[1.5ex]
    \includegraphics*[width=0.48\linewidth]{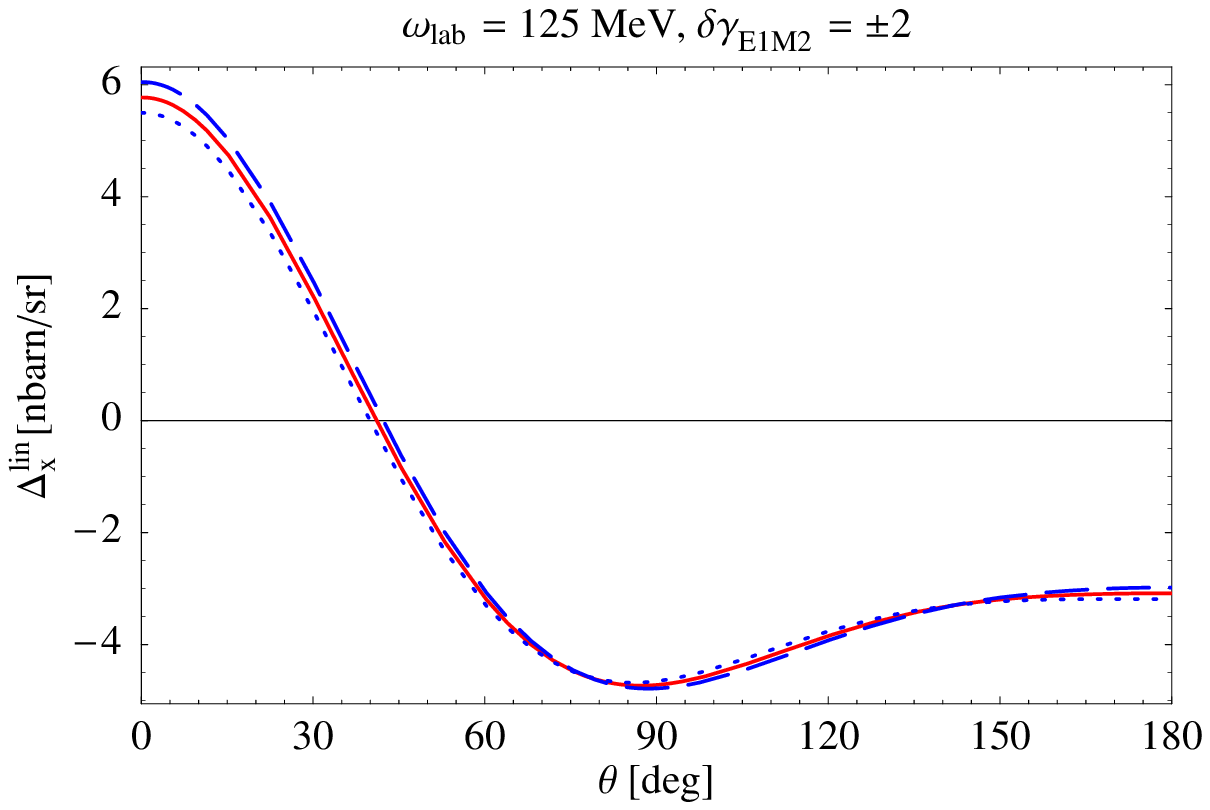}
    \hq\hq
    \includegraphics*[width=0.48\linewidth]{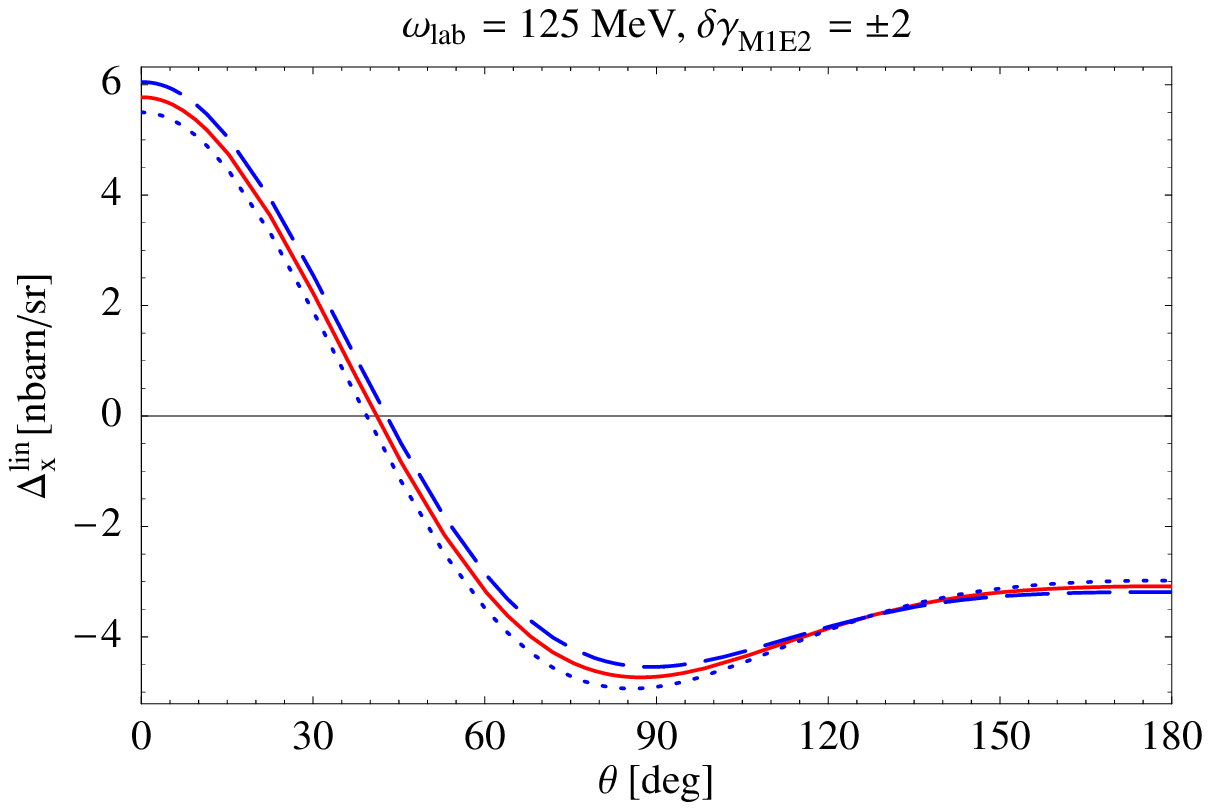}
  \caption {(Colour online)  Dependence of $\Delta^\text{lin}_{x}$ at
      $\omega_\text{lab}=125$~MeV on the dipole spin-polarisabilities.
      Notation as in Figs.~\ref{fig:dcsx_ab} and~\ref{fig:dcsx_gs}.} 
\label{fig:deltaxlin_gs}
\end{center}
\end{figure}

A complementary picture emerges in Figs.~\ref{fig:deltaxlin_ab} and
\ref{fig:deltaxlin_gs} for $\Delta^\text{lin}_{x}$, in which the target spin
is perpendicular to the beam direction, \eqref{eq:deltaxlin}.
At 45~MeV, varying $\alpha_{E1}$ leads to an effect of
$\lesssim\pm0.2$~nbarn/sr, whereas the effect due to $\delta \beta_{M1}$ is
indiscernible. At 125~MeV the maximum sensitivity to $\alpha_{E1}$ increases
to $\approx\pm0.5$~nbarn/sr, twice of the dependence on $\beta_{M1}$. At
forward angles, $\Delta^\text{lin}_{x}$ at $\omega_\text{lab}=125$~MeV is
equally sensitive to all spin-polarisabilities. For
$\theta_\text{lab}\in[90^\circ;120^\circ]$, contributions from $\gamma_{E1E1}$
dominate (sensitivity $\lesssim\pm0.25$~nbarn/sr), with a modest admixture of
$\gamma_{M1E2}$ which decreases to zero around $120^\circ$. One may thus
extract a linear combination of only two spin-polarisabilities, complementary
to that from $\Delta^\text{lin}_{z}$.

To summarise, $\alpha_{E1}$ has a sizable effect on both
$\Delta^\text{lin}_{z}$ and $\Delta^\text{lin}_{x}$, bigger than the effect of
$\beta_{M1}$ or any of the spin-polarisabilities.  $\Delta^\text{lin}_{z}$ is
sensitive to a combination of $\gamma_{M1M1}$ and $\gamma_{M1E2}$, while
$\Delta^\text{lin}_{x}$ can be used to extract a combination of
$\gamma_{E1E1}$ and $\gamma_{M1E2}$. Both observables are virtually
in-sensitive to $\gamma_{E1M2}$.

\subsection{Polarised Target and Circularly-polarised Photons}
\label{sec:cpol}

Calculations for the circularly-polarised photon observables
$\Delta_z^\text{circ}$ and $\Delta_x^\text{circ}$ at $\mathcal{O} (Q^3)$ for
$\omega\sim\mpi$ and without dynamical $\Delta$ were first reported in
Refs.~\cite{Ch05,mythesis}. The spin-polarisabilities were given not in the
basis of photon multipolarities used here but by $\gamma_{1,2,3,4}$ of the
Ragusa basis~\cite{Ragusa:1993rm}; see e.g.~Ref.~\cite[App.~A]{Babusci:1998ww}
for a translation between the two.  It was shown that $\Delta_x^\text{circ}$
had considerable sensitivity to some spin-polarisabilities. As discussed in
Sec.~\ref{sec:comparison}, Figs.~\ref{fig:comp45} and \ref{fig:comp125}, the
dynamical $\Delta$ increases these effects. 

\begin{figure}[!htb]
  \begin{center}
    \includegraphics*[width=0.48\linewidth]{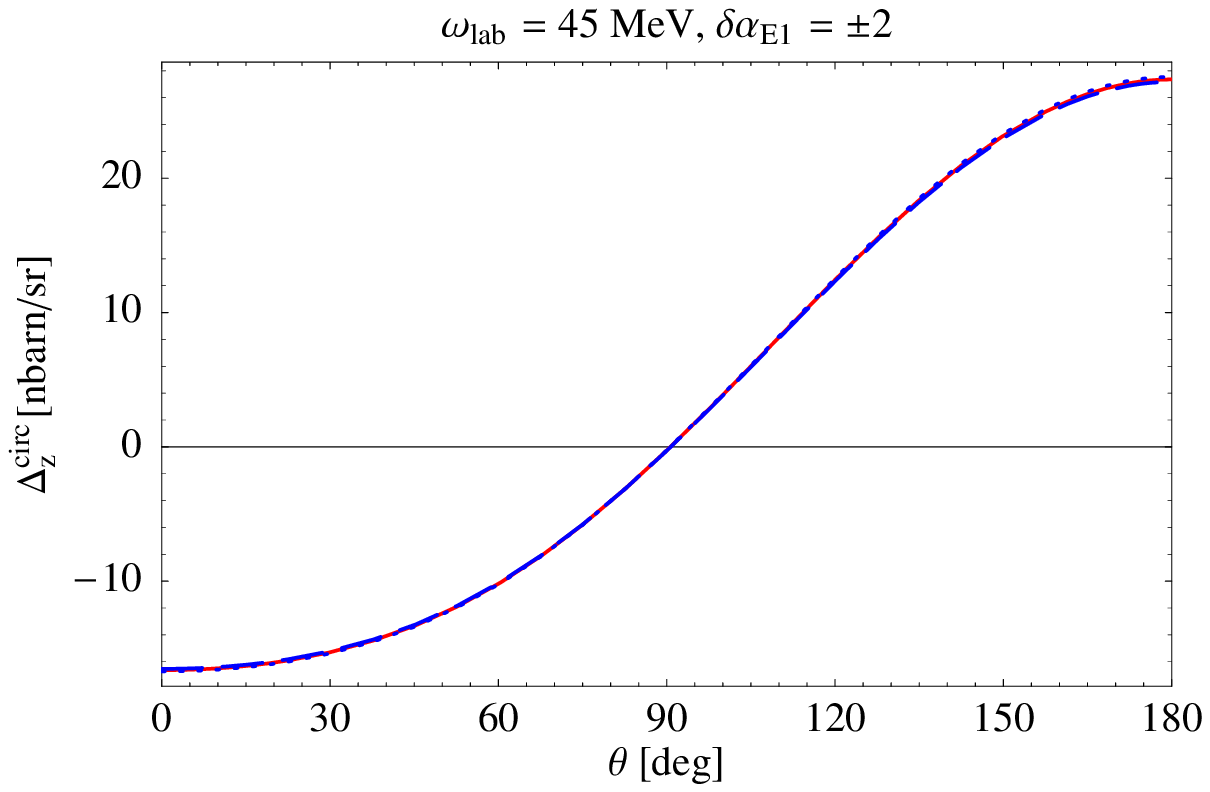}
    \hq\hq
    \includegraphics*[width=0.48\linewidth]{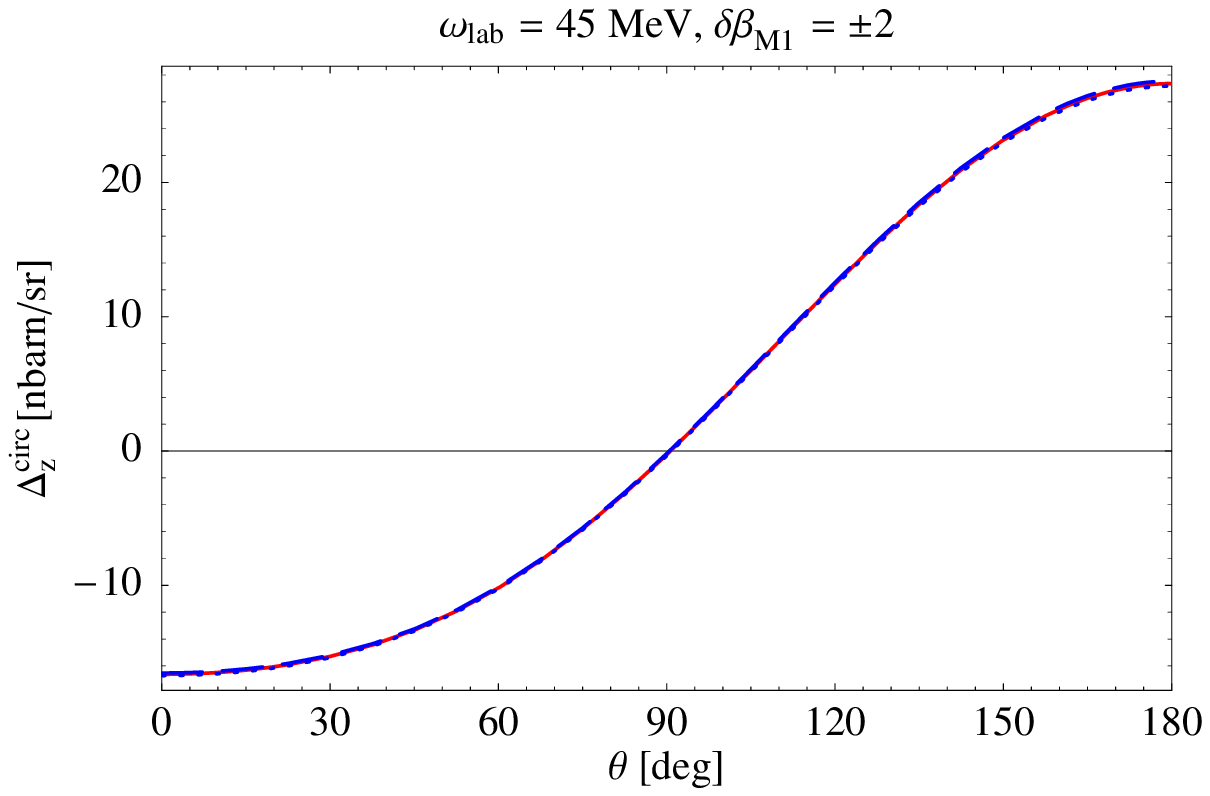}\\[1.5ex]
    \includegraphics*[width=0.48\linewidth]{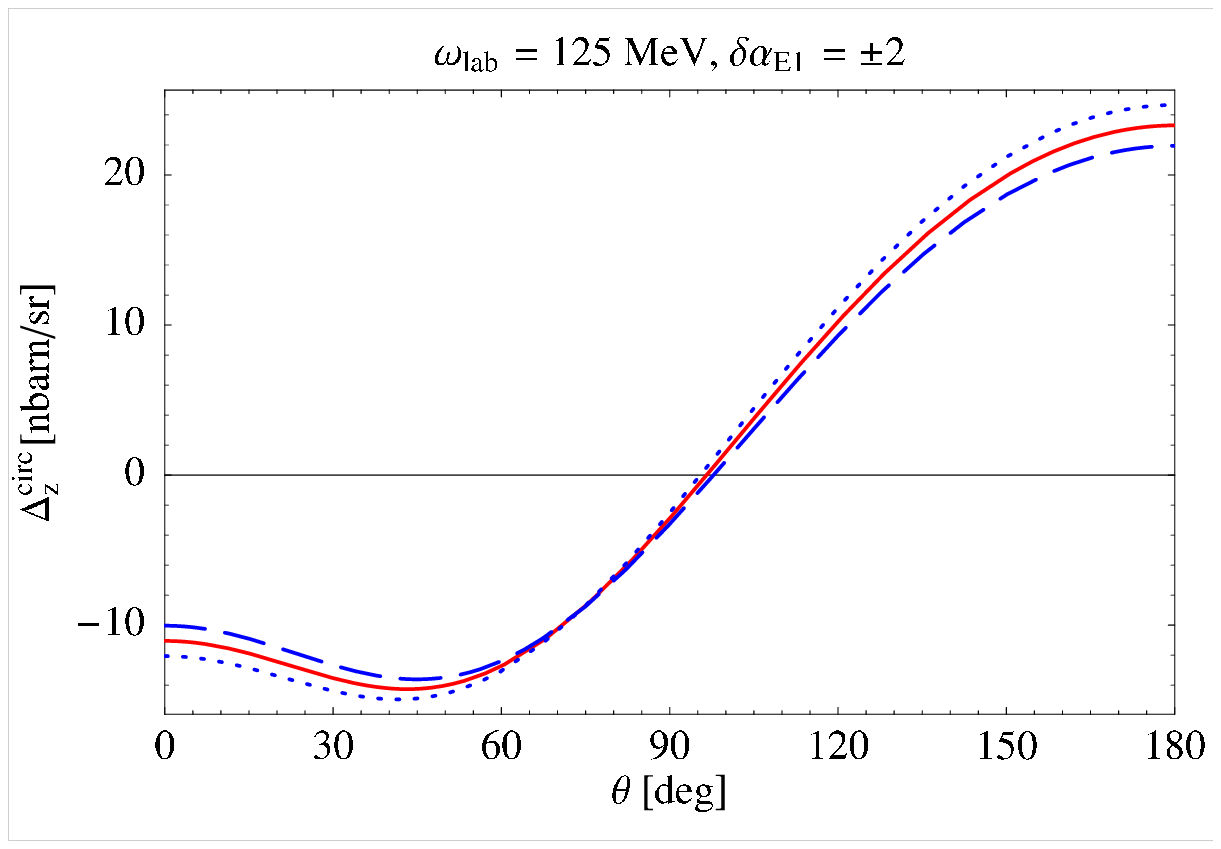}
    \hq\hq
    \includegraphics*[width=0.48\linewidth]{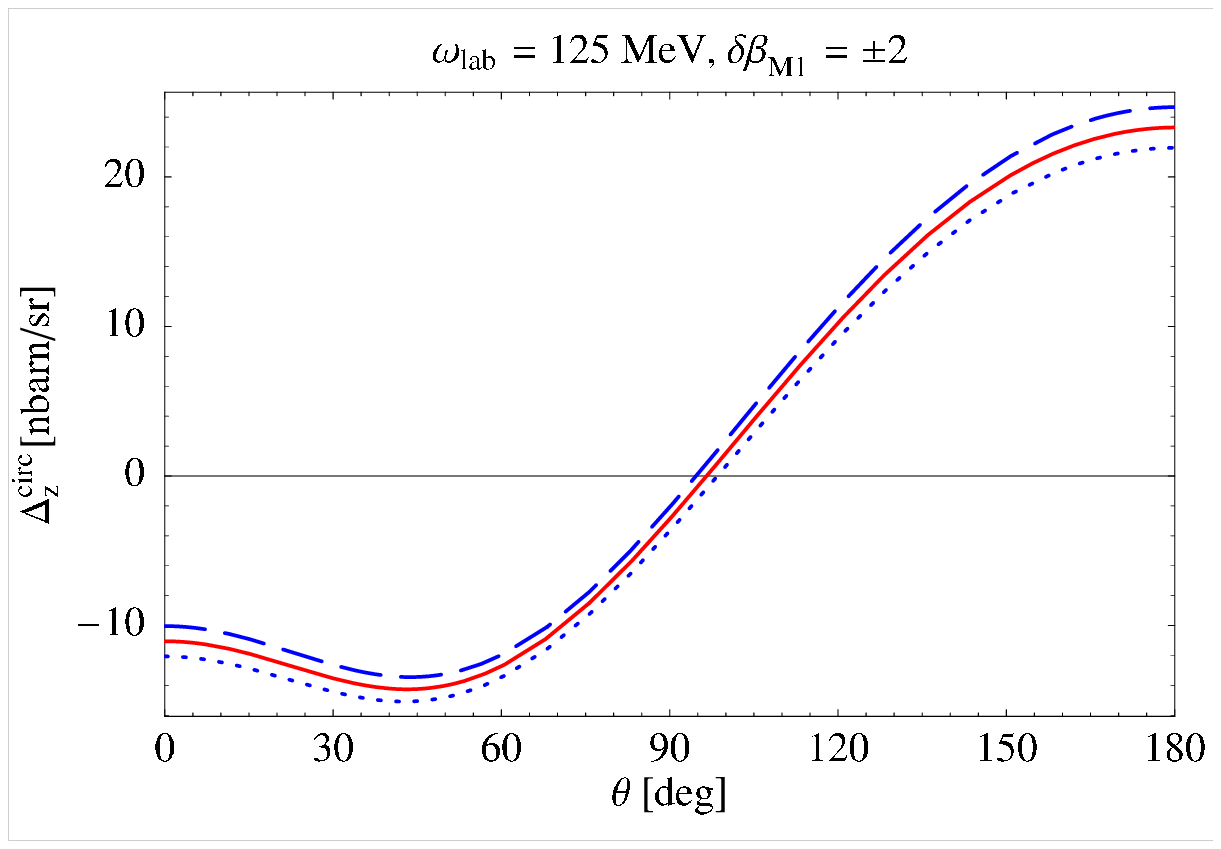}
  \caption {Colour online) Dependence of $\Delta^\text{circ}_{z}$  on the
      spin-independent dipole polarisabilities at $\omega_\text{lab}=45$~MeV (top) and
      $\omega_\text{lab}=125$~MeV (bottom). Notation as in
      Fig.~\ref{fig:dcsx_ab}.}
  \label{fig:deltaz_ab}
\end{center}
\end{figure}

\begin{figure}[!htb]
\begin{center}
\includegraphics*[width=0.48\linewidth]{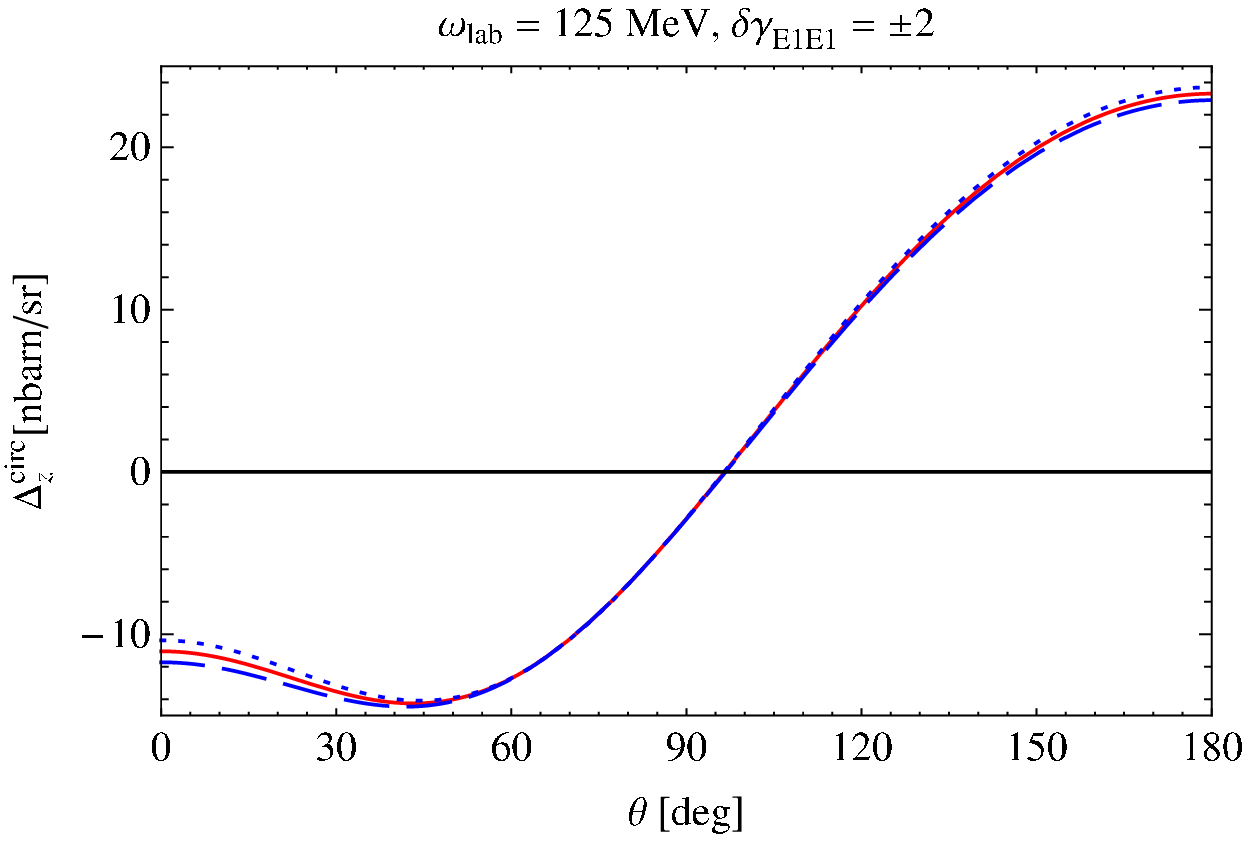}
\hq\hq
\includegraphics*[width=0.48\linewidth]{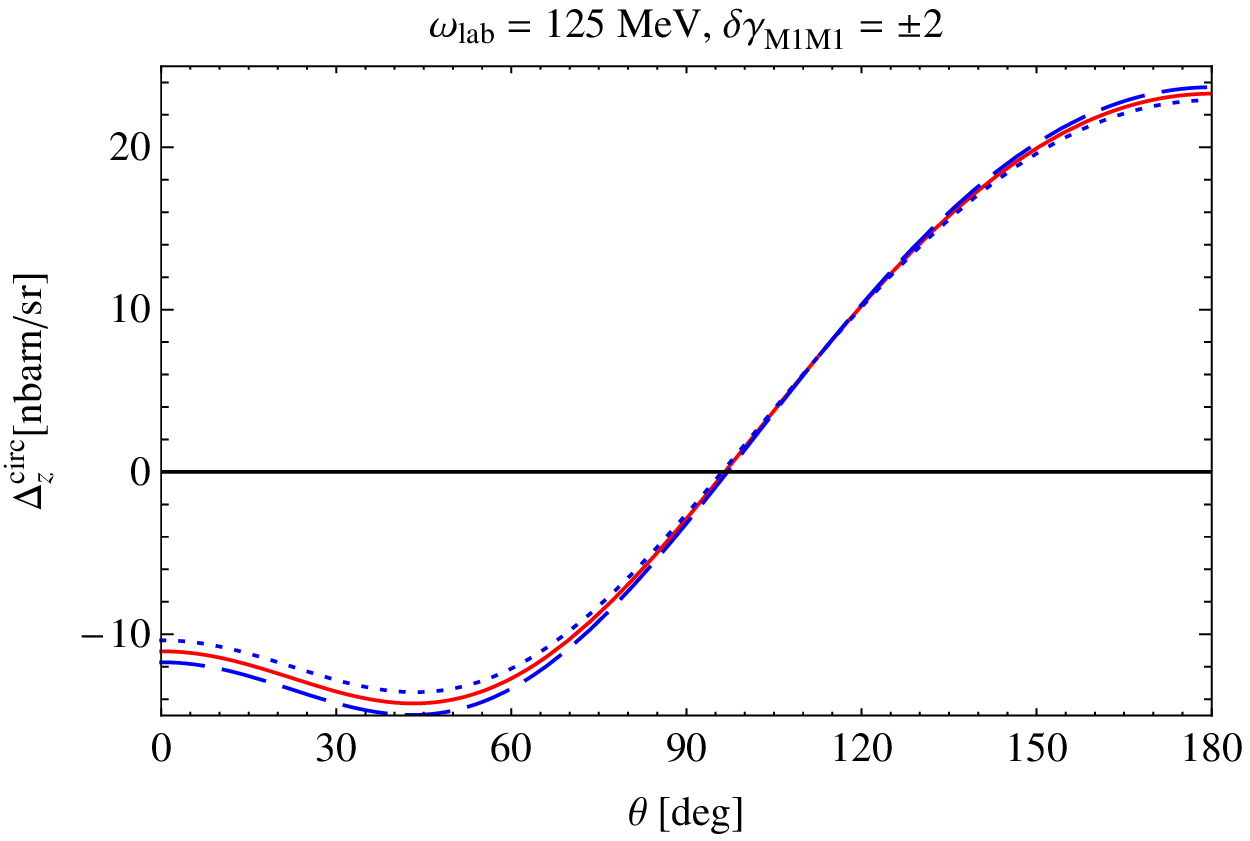}\\[1.5ex]
\includegraphics*[width=0.48\linewidth]{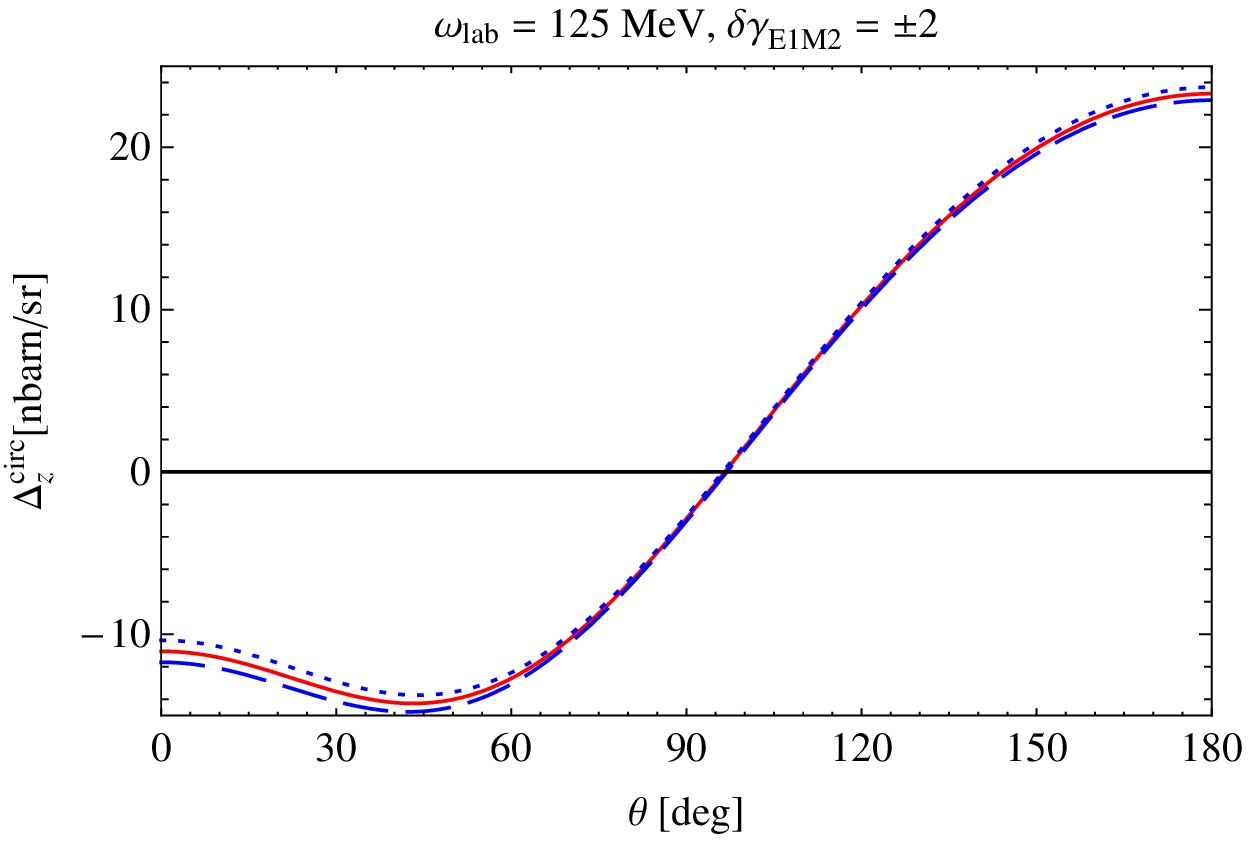}
\hq\hq
\includegraphics*[width=0.48\linewidth]{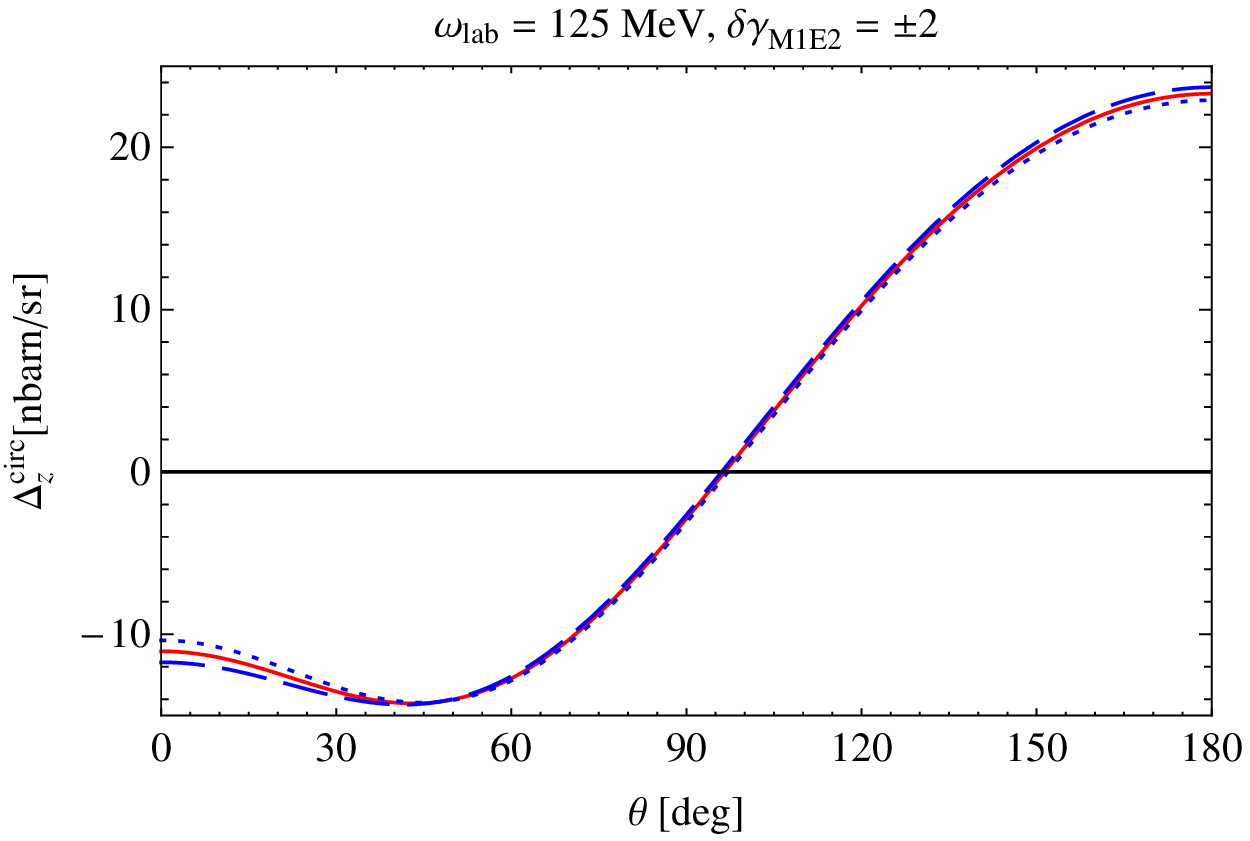}
\caption {(Colour online) Dependence of $\Delta^\text{circ}_{z}$ at
  $\omega_\text{lab}=125$~MeV on the dipole spin-polarisabilities.  Notation
  as in Figs.~\ref{fig:dcsx_ab} and~\ref{fig:dcsx_gs}.}
  \label{fig:deltaz_gs}
\end{center}
\end{figure}

Figures~\ref{fig:deltaz_ab} and \ref{fig:deltaz_gs} show the dependence of the
parallel polarisation asymmetry $\Delta_z^\text{circ}$ on the spin-independent
and spin-polarisabilities in SSE \ChiEFT.
At 45~MeV, the sensitivity to any polarisability is less than the line
thickness. At 125~MeV, sensitivity to both $\alpha_{E1}$ and $\beta_{M1}$
grows to $\pm2.5$~nbarn/sr, with different angular dependence. 

Looking back at the contribution of $\approx7$~nbarn/sr which a dynamical
$\Delta$ provides overall according to Fig.~\ref{fig:comp125}, it seems
surprising that the sensitivity on varying $\delta\beta_{M1}=\pm2$ in
Fig.~\ref{fig:dcsx_ab} is only $\pm2.5$~nbarn/sr. This serves however to
illustrate an important point on the energy-dependence of the polarisabilities
and the interpretation of the variation parameters
$\delta(\alpha_{E1},\beta_{M1},\gamma_i)$. Bear in mind that the value of the
\emph{dynamical polarisability} is at this energy
$\beta_{M1}(\omega=120\;\MeV)\approx 7$ as seen~in Ref.~\cite[Fig.~8]{Hi04}.
This is about $3$ to $4$ times the static value $\beta_{M1}\approx[2\dots3]$
or its near-identical dynamical value at that energy in the version without
explicit $\Delta$. Recall that $\beta_{M1}(\omega)$ is only very weakly
$\omega$-dependent without $\Delta$s. Switching off effects from the magnetic
polarisability completely does therefore at this energy \emph{not} correspond
to choosing $\delta\beta_{M1}=-2$ or $-3$, but to $\delta\beta_{M1}=-7$.
Switching off the $\Delta$-effects corresponds to $\delta\beta_{M1}=-5$.
Re-scaling the variation in Fig.~\ref{fig:deltaz_ab} accordingly and keeping
in mind that the amplitudes are to a good approximation linear in the
polarisabilities, one indeed recovers the result of Fig.~\ref{fig:comp125}.
We therefore re-iterate what we already noted in Sec.~\ref{sec:stg}: Varying
the polarisabilities by values $\delta(\alpha_{E1},\beta_{M1},\gamma_i)$ is
equivalent to varying the static polarisabilities by the same amount
\emph{only if} the energy-dependence of the polarisabilities follows the
\ChiEFT prediction. If not, the variations test deviations to experiments at
fixed energy.

Sensitivity of $\Delta_z^\text{circ}$ to the spin-polarisabilities is
$\lesssim\pm0.7$~nbarn/sr at 125~MeV, most notably at forward and backward
angles. Angular dependence does not provide a clean tool to dis-entangle
different spin-polarisabilities. Sensitivity to $\gamma_{E1E1}$ and
$\gamma_{E1M2}$ is again equal but opposite, while that to $\gamma_{M1M1}$,
$\gamma_{E1M2}$ and $\gamma_{M1E2}$ is largely additive.

The dependence of $\Delta_z^\text{circ}$ was also studied in ``pion-less''
EFT~\cite{Chen:2004wwa} by comparing to the case when the
spin-polarisabilities are absent.  The range of applicability of this theory
in which the pion is integrated out as heavy is however limited to typical
momenta well below the pion mass and thus to typical photon energies
$\omega\lesssim\mpi^2/M\approx20\;\MeV$. As noted by the authors, their
predictions at cm energies of $70$ and $90$~MeV are therefore only of
qualitative interest.  At $70$~MeV, the discrepancy to the results presented
here rises to $\gtrsim20\%$ and is most pronounced at back-angles. This may be
attributed to the fact that the dynamical $\Delta$ increases
$\Delta_z^\text{circ}$, see Fig.~\ref{fig:comp125}.  It is interesting that
their results at $30$ and $50$~MeV are still very similar to those with
dynamical pions and $\Delta$s, even thought these energies are strictly
speaking beyond the breakdown scale of the EFT without pions. This holds also
for their results on $\Sigma_{z}^\text{circ}$. Their $\Sigma_{x}^\text{circ}$
is however larger by about $20$\% in the peak and has a less pronounced
concave flank at forward angles, cf.~Fig.~\ref{fig:comp45}.

\begin{figure}[!htb]
  \begin{center}
    \includegraphics*[width=0.48\linewidth]{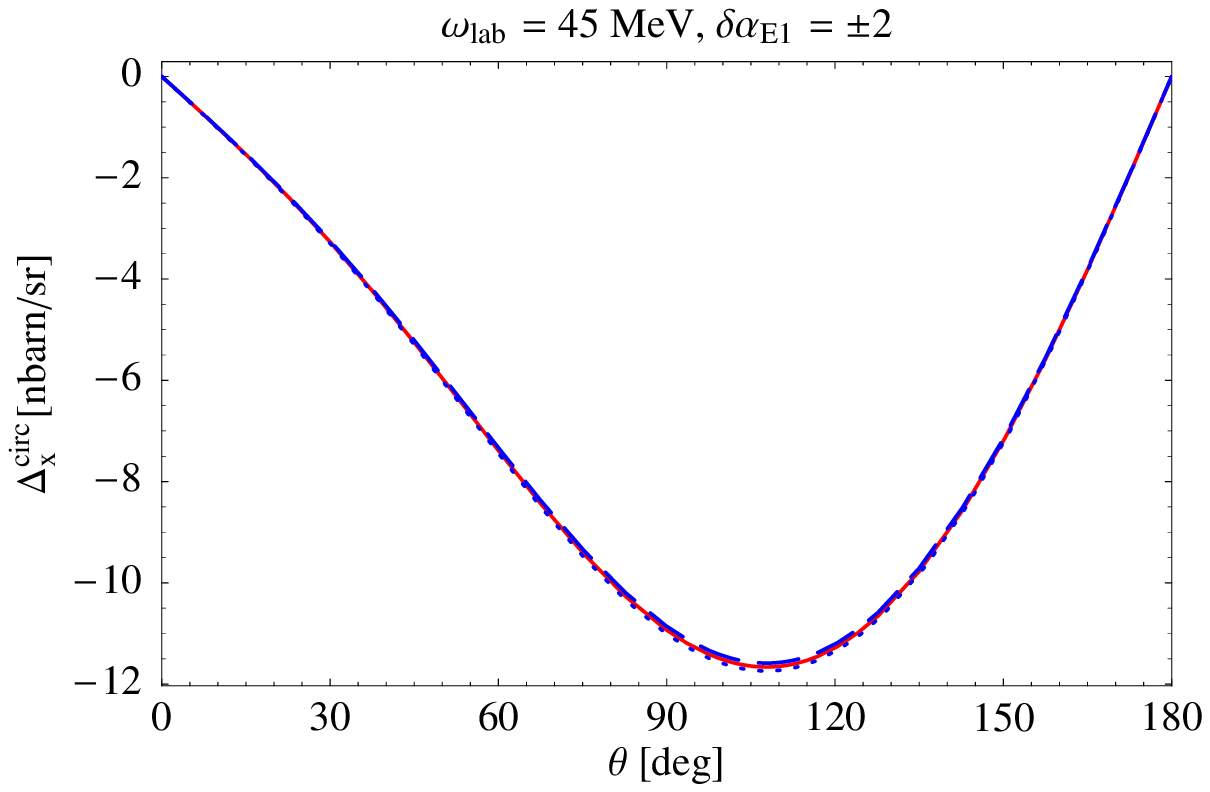}
    \hq\hq
    \includegraphics*[width=0.48\linewidth]{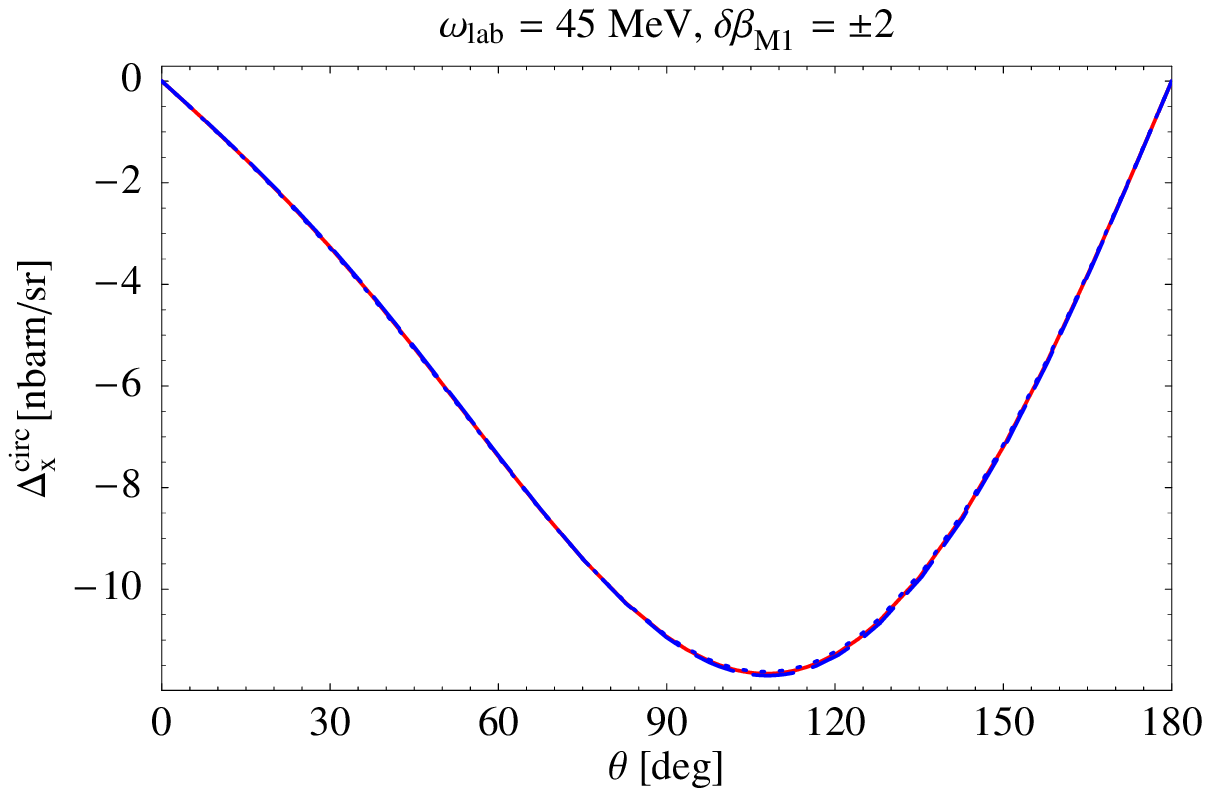}\\[1.5ex]
    \includegraphics*[width=0.48\linewidth]{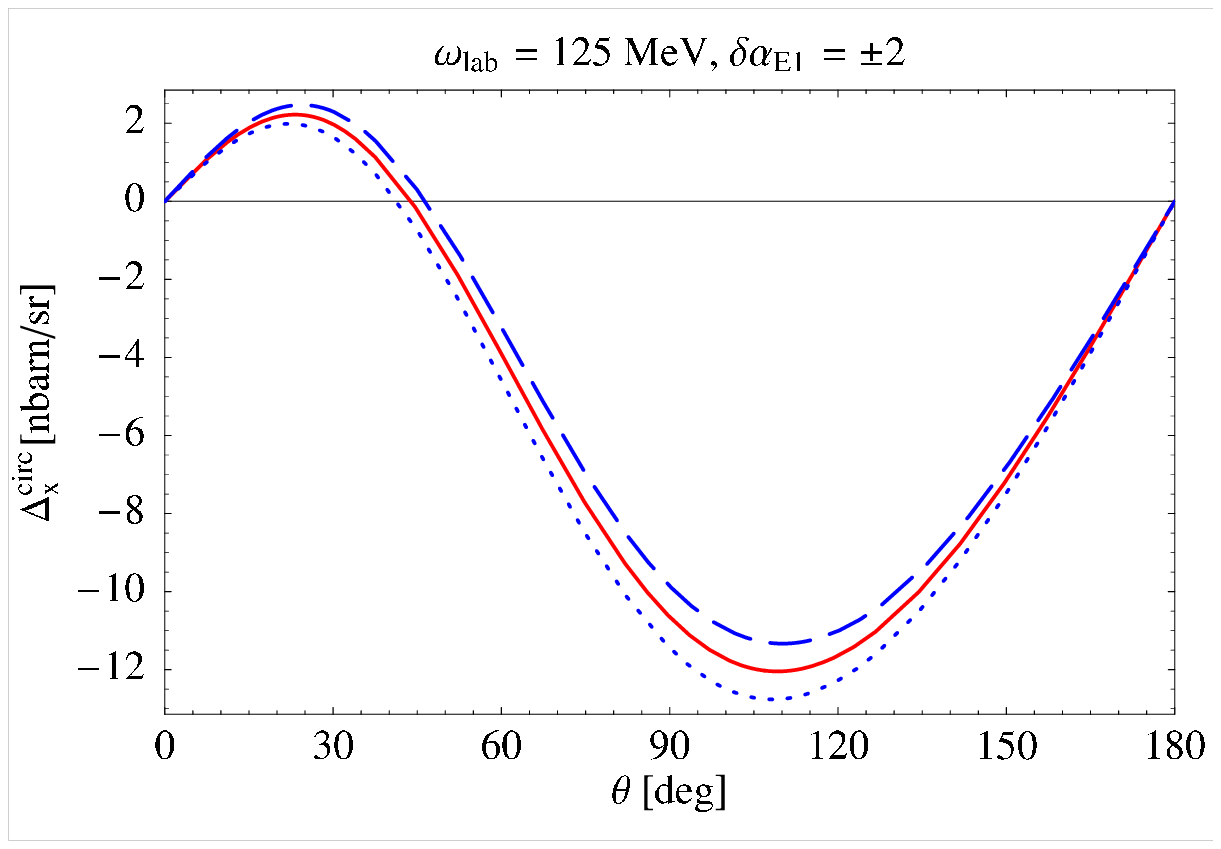}
    \hq\hq
    \includegraphics*[width=0.48\linewidth]{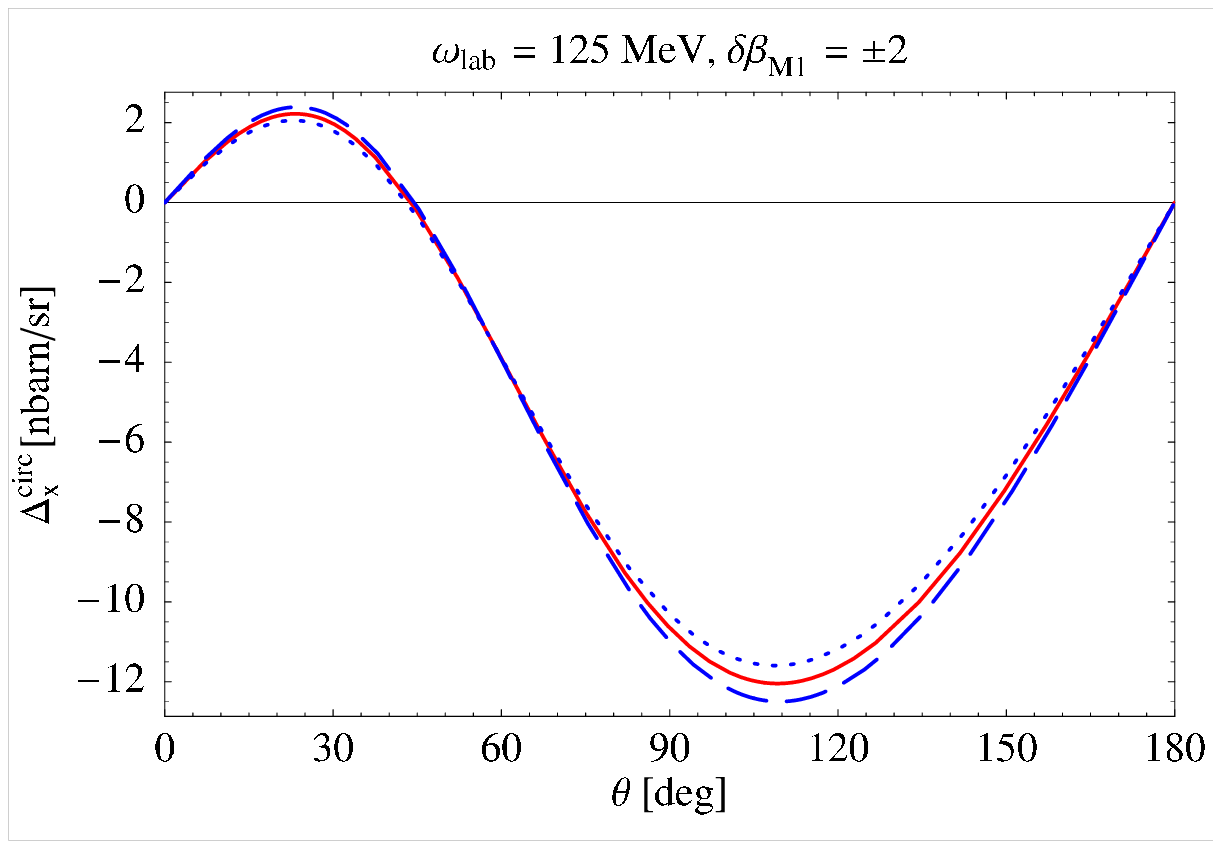}
  \caption {(Colour online) Dependence of $\Delta^\text{circ}_{x}$  on the
      spin-independent dipole polarisabilities at $\omega_\text{lab}=45$~MeV (top) and
      $\omega_\text{lab}=125$~MeV (bottom). Notation as in
      Fig.~\ref{fig:dcsx_ab}.} 
\label{fig:deltax_ab}
\end{center}
\end{figure}

\begin{figure}[!htb]
  \begin{center}
    \includegraphics*[width=0.48\linewidth]{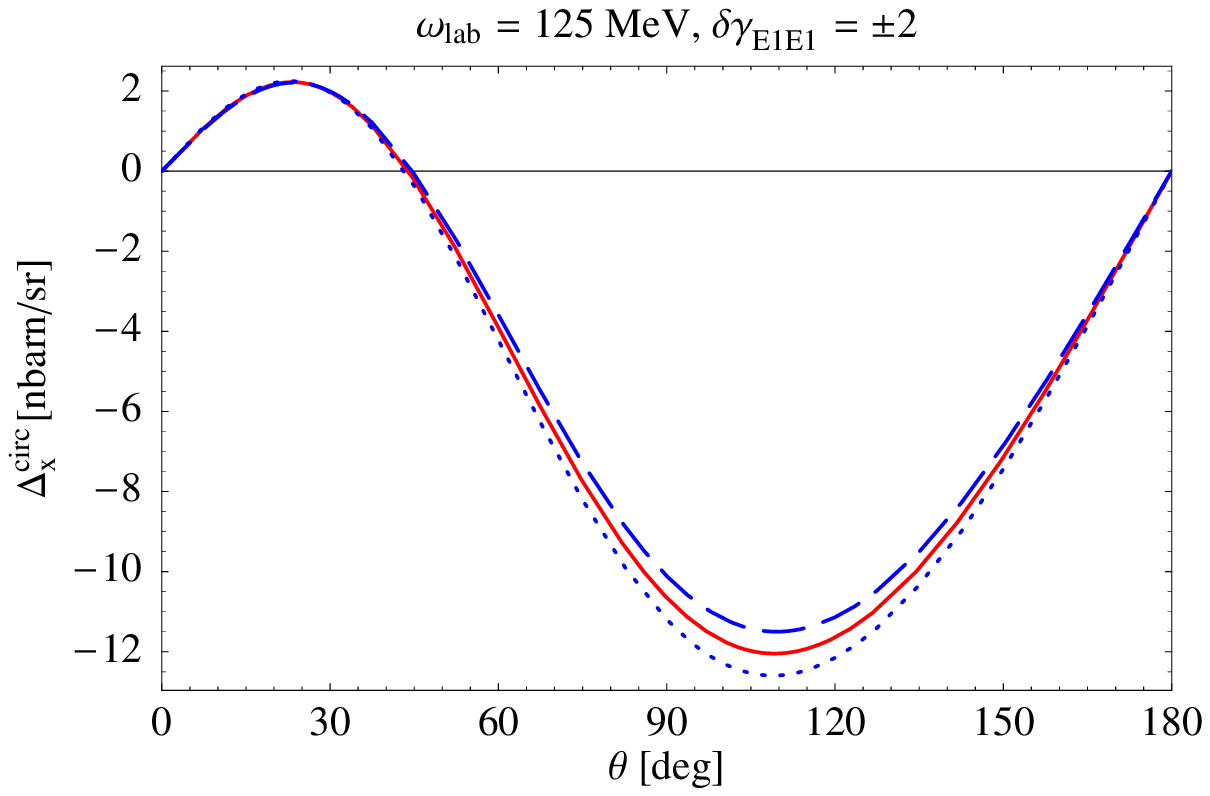}
    \hq\hq
    \includegraphics*[width=0.48\linewidth]{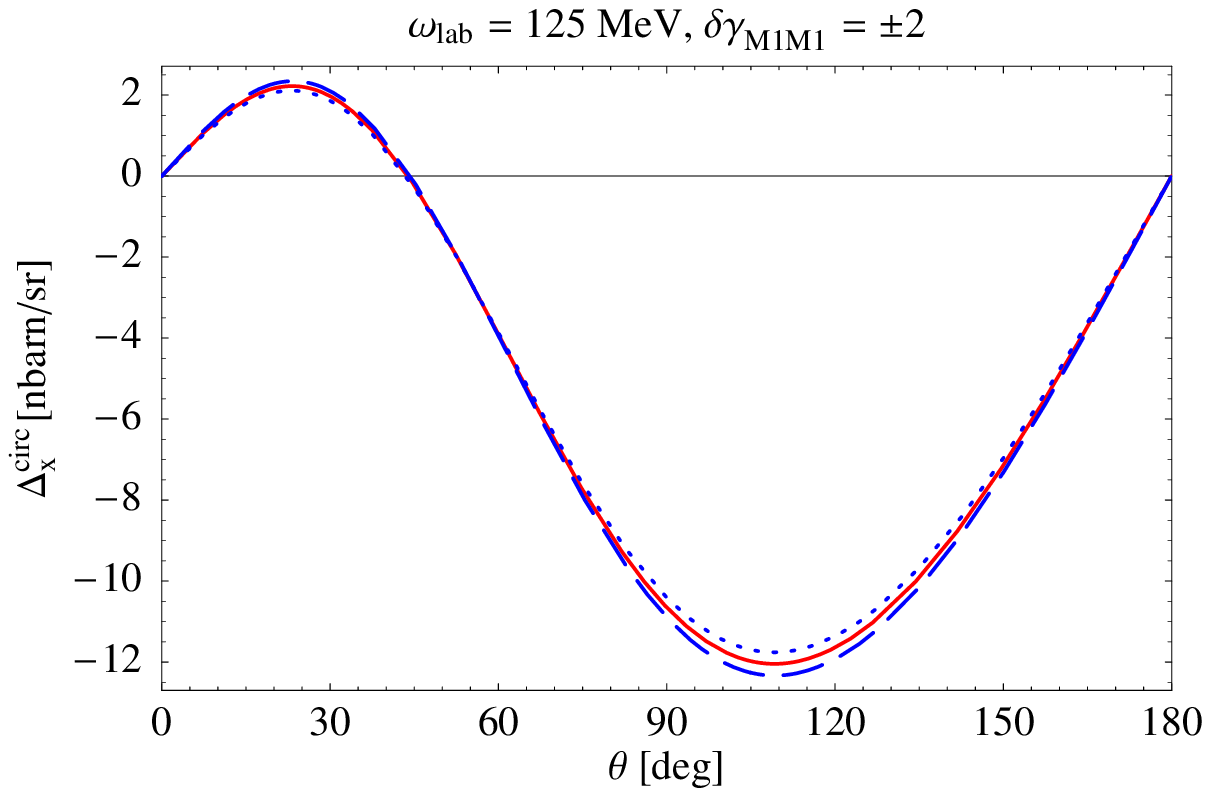}\\[1.5ex]
    \includegraphics*[width=0.48\linewidth]{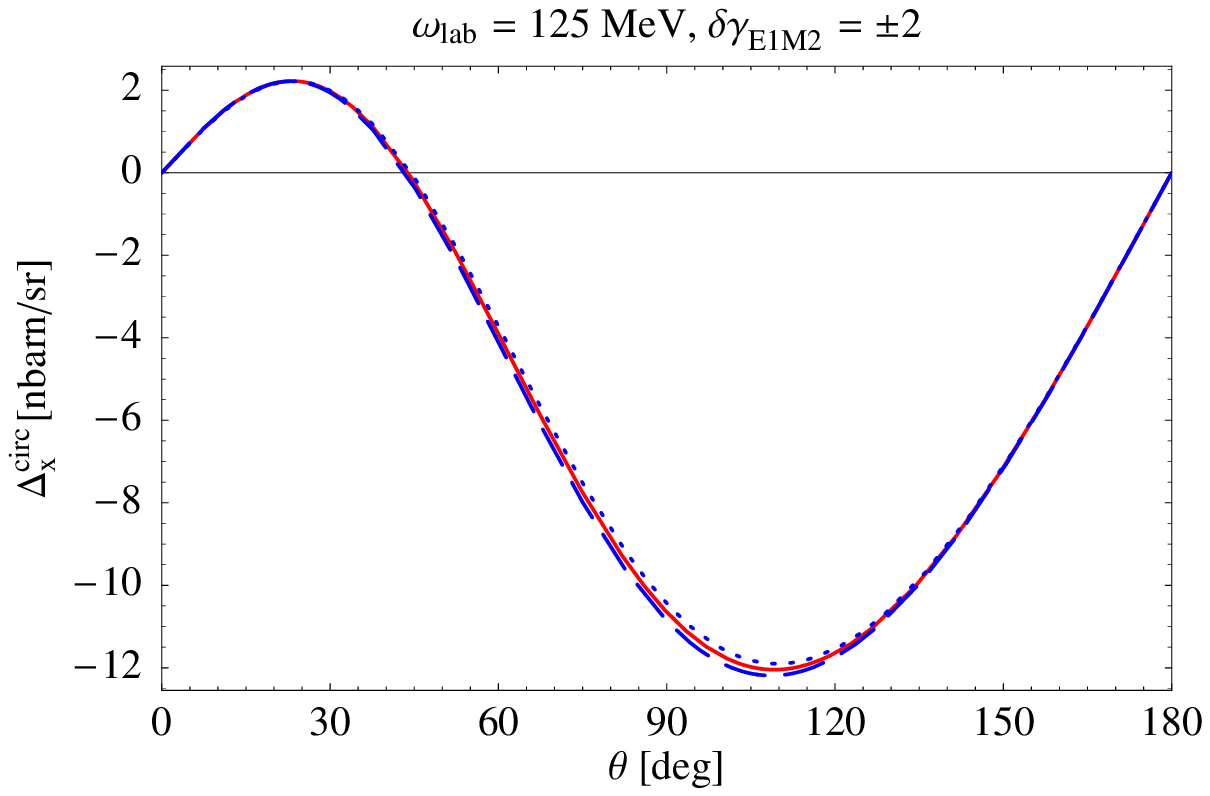}
    \hq\hq
    \includegraphics*[width=0.48\linewidth]{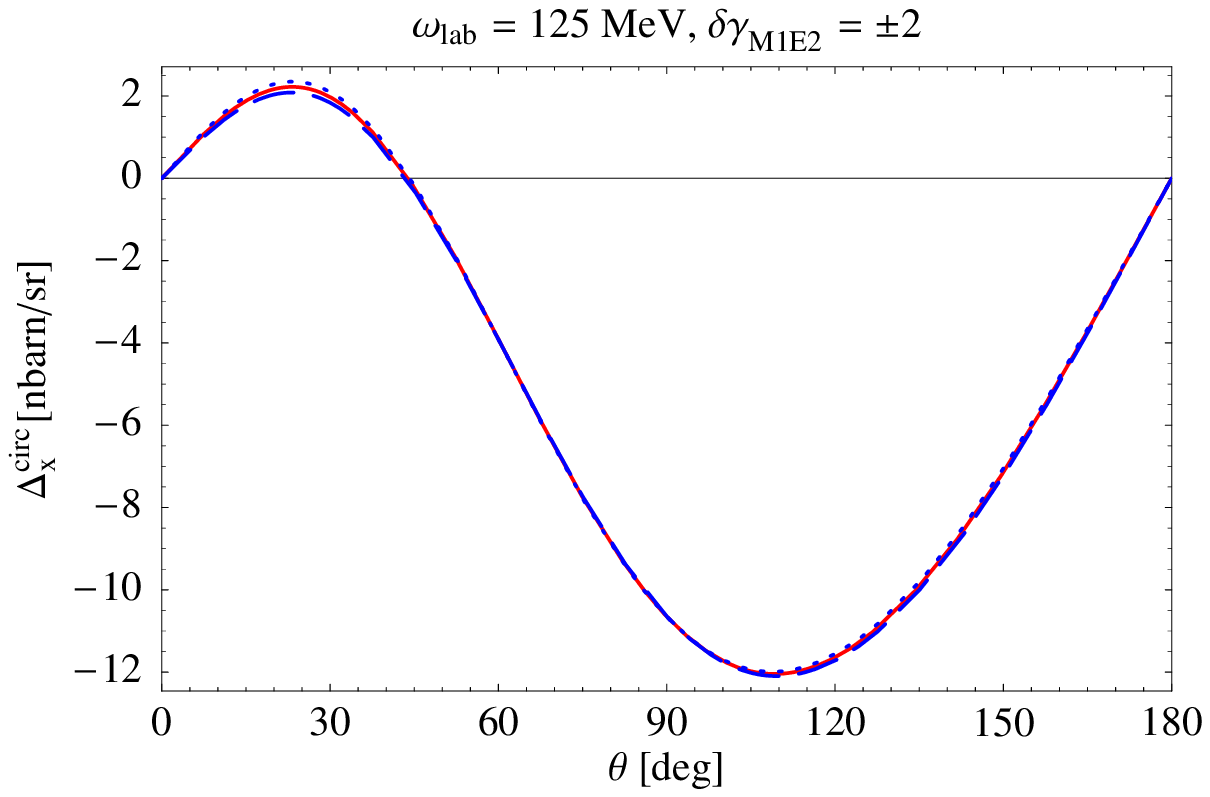}
    \caption {(Colour online) Dependence of $\Delta^\text{circ}_{x}$ at
      $\omega_\text{lab}=125$~MeV on the dipole spin-polarisabilities.
      Notation as in Figs.~\ref{fig:dcsx_ab} and~\ref{fig:dcsx_gs}.}
\label{fig:deltax_gs}
\end{center}
\end{figure}

The parallel polarisation asymmetry $\Delta_x^\text{circ}$ is only minimally
sensitive to polarisabilities at $45$~MeV, see Figs.~\ref{fig:deltax_ab} and
\ref{fig:deltax_gs}.
At 125~MeV, dependence on $\alpha_{E1}$ is $\lesssim\pm0.8$~nbarn/sr, and on
$\beta_{M1}$ $\lesssim\pm0.6$~nbarn/sr. The two can be dis-entangled, as
contributions from $\beta_{M1}$ are zero at
$\theta_\text{lab}\approx60^\circ$. Of the spin-polarisabilities, $\delta
\gamma_{E1E1}$ causes a maximum change of $\lesssim\pm0.6$~nbarn/sr,
i.e.~about $\frac{3}{4}$ of changing $\alpha_{E1}$, but with opposite sign.
Varying $\gamma_{M1M1}$ induces a $\lesssim\pm0.3$~nbarn/sr effect, equal in
magnitude and sign to that of $\beta_{M1}$. A weak but noticeable angular
dependence may allow dis-entangling $\gamma_{M1E2}$ at forward angles.

Asymmetries with circularly polarised beams offer thus a sensitivity to the
spin-polarisabilities which is weaker or at most comparable to those of
linearly polarised ones. There is no realistic opportunity to measure an
observable which is sensitive to the linear combination of less than three
spin-polarisabilities. In contradistinction, linear beam polarisation offers
the chance to measure linear combinations of only two spin-polarisabilities,
with the others being practically absent. For all double-polarisation
observables, the spin-independent dipole polarisabilities $\alpha_{E1}$ and
$\beta_{M1}$ must be known reliably, as their effects over-shadow those of the
spin-polarisabilities at all energies.

\section{Summary and Outlook}
\label{sec:summary}

We investigated elastic deuteron Compton scattering with a polarised beam
and/or target at next-to-leading order, $\mathcal{O}(\epsilon^3)$ within
Chiral Effective Field Theory in the photon energy range between zero and
$125$ MeV (lab), using the power-counting scheme of Hildebrandt et
al.~\cite{Hi05, Hi05b}. This systematic, model-independent approach contains
dynamical $\Delta(1232)$ degrees of freedom in the Small Scale Expansion and
contributions from intermediate-state $NN$-rescattering, leading to strong
para-magnetic effects and the correct Thomson limit, respectively. Both have
sizeable effects on polarisation observables: The $\Delta$-isobar is
significant at higher energies, while the intermediate $NN$-rescattering
states are significant throughout the energy range. These are thus important
ingredients of any theoretical description of deuteron Compton scattering at
these energies. They are needed both for consistency of the theory and for
matching to well-established data.

Single and double polarisation observables are quite sensitive to the electric
and magnetic polarisabilities, as shown in Sec.~\ref{sec:doublepol}.  Since
these can also be directly extracted from unpolarised scattering, as shown
e.g.~in Refs.~\cite{Be02, Be04, Hi05, Hi05b}, it is imperative that they be
determined to better accuracy so as not to taint an extraction of the so-far
experimentally nearly un-determined spin-polarisabilities from polarised
experiments. Our results also show that some of the double-polarisation
observables are sensitive to different linear combinations of the
spin-polarisabilities at energies $\gtrsim80$~MeV. The linear-beam
polarisation observables $\Delta^\text{lin}_{x,z}$ with target polarisations
parallel or perpendicular to the beam proved most promising from the
theorist's point of view, since they are each at high energies and for
certain, experimentally feasible angles dominated by linear combinations of
only two spin-polarisabilities.

The cross-section of a linearly polarised photon on an un-polarised target can
be used to ``switch off'' or maximise dependence on one of the
spin-independent polarisabilities, as demonstrated in Sec.~\ref{sec:lpol}.
When the beam polarisation is perpendicular to the scattering plane, the
signal of a combination of spin-polarisabilities is particularly strong.  Even
then, the spin-independent polarisabilities must be known with sufficient
accuracy.

In view of these findings, we advocate the following strategy:
\begin{enumerate}
\item Since an accurate extraction of the electric and magnetic
  polarisabilities is central to extracting the spin-polarisabilities, it is
  crucial to perform a number of relatively low-energy experiments ($\lesssim
  70$~MeV).  In these, spin-polarisabilities are negligible but
  spin-independent polarisabilities can be determined to high accuracy.
  Besides an unpolarised deuteron Compton scattering experiment at MAXlab
  whose analysis is in progress~\cite{Feldman:2008zz,Feldman2}, experiments at \HIGS
  have been approved~\cite{Weller,Weller:2009zza} using a polarised beam. Some
  of the single and double-polarisation observables can also be used to that
  purpose. Such data will not only serve as stepping-stone for determinations
  of spin-polarisabilities; comparing the iso-scalar spin-independent
  polarisabilities thus extracted with the spin-independent polarisabilities
  of the proton will also reveal differences between the proton and neutron
  response to external electro-magnetic fields.

\item With $\alpha_{E1}$ and $\beta_{M1}$ better known, a series of concurrent
  polarised and unpolarised measurements at higher energies will allow the
  extraction of the spin-polarisabilities. The best chances are presumably
  offered by double-polarisation observables at $\gtrsim$ 100 MeV but below
  the pion-production threshold. For example, $\Delta_x^\text{lin}$ and
  $\Delta_x^\text{circ}$ are sensitive to different linear combinations of the
  four spin-polarisabilities; and at some angles, $\Delta_x^\text{lin}$ and
  $\Delta_z^\text{lin}$ can be used to determine linear combinations of only
  two spin-polarisabilities. Measuring several linear combinations makes it
  possible to disentangle the four independent spin-polarisabilities.
\end{enumerate}

Ideally, a multipole-analysis of $4+1$ experiments at different angles
suffices to over-determine the $4$ spin-polarisabilities -- if the data are of
an unprecedented high accuracy which is presumably not feasible today.
$\gamma_{E1E1}$ and $\gamma_{M1M1}$ can to a good approximation be extracted
uniquely from $\Delta_x^{circ}$ and $\Delta_z^{lin}$, respectively. However, a
larger number of independent data will be necessary to account for the high
complexity of these experiments and for the fact that no clear-cut observables
exist for the ``mixed'' spin-polarisabilities $\gamma_{E1M2}$ and
$\gamma_{M1E2}$. In each experiment, the accuracy achievable and the
observables and kinematics most suited strongly depend on geometry and
acceptance of the setup.

A concerted effort of planned and approved experiments at
$\omega_\text{lab}\lesssim200\;\mathrm{MeV}$ is indeed under way,
e.g.~polarised photons on polarised protons, deuterons and ${}^3$He at
TUNL/HI$\gamma$S~\cite{Weller:2009zza,Weller,Miskimen,Miskimentalk,Gao,Ahmed};
polarised photons on polarised protons at MAMI~\cite{AhrensBeckINT08}. The
unpolarised experiment on the deuteron at MAXlab over a wide range of energies
and angles is being analysed~\cite{Feldman:2008zz,Feldman2}. At present, only 28
(un-polarised) data exist for the deuteron in an energy range
$\omega_\text{lab}\in[49;94]$ MeV and with error-bars on the order of $15\%$.
It was shown in Refs.~\cite{mythesis, Ch07, Sh08} that one can in addition
extract a subset of the neutron polarisabilities from elastic Compton
scattering off ${}^3$He, with an experiment at \HIGS approved~\cite{Gao}.
Since the total charge is doubled in this target, sensitivity to
polarisabilities is increased by larger interference of non-structure and
polarisability amplitudes. On the other hand, restoration of the Thomson limit
needs to be addressed for ${}^3$He. Other avenues may include quasi-free
measurements on light nuclei.

To facilitate planning and analysis of such experiments, we have made our
detailed results for unpolarised and polarised observables in deuteron Compton
scattering available as interactive \emph{Mathematica 7.0} notebook (email to
hgrie@gwu.edu). It lists and plots both energy- and angle-dependences of
cross-sections and all the asymmetries as well as cross-section differences
from $10$ to $\approx120$~MeV in both the cm and lab systems, including the
sensitivities to varying the spin-independent and spin-dependent dipole
polarisabilities independently and with the Baldin sum rule constraint. It
thus provides a more thorough and efficient tool to investigate sensitivities
and impacts both of experimental constraints and of theoretical information
like the Baldin sum rule~\eqref{eq:baldin}.

In the long run, a global multipole-analysis at fixed energies of a database
that includes both polarised and un-polarised elastic Compton scattering
high-accuracy measurements on the proton, deuteron and ${}^3$He, from low to
high energies at various angles, gives an unambiguous extraction of the two
spin-independent and four spin-dependent polarisabilities, for both the proton
and neutron. That such a multipole-analysis is feasible has been advocated
in~\cite{hgrieproc,Miskimentalk}. First results were reported based on the
wealth of unpolarised proton data~\cite{hgrieproc}. High-quality data will
allow one to zoom in on the proton-neutron differences, which according to
\ChiEFT stem from chiral-symmetry breaking pion-nucleon interactions; and
provide information on the spin-polarisabilities which parameterise the
detailed response of the nucleon spin degrees of freedom in external
electro-magnetic fields.

Future improvement in our deuteron Compton scattering calculations include:
transition to a chirally fully consistent deuteron wave-function and
$NN$-potential; implementing the kinematically correct pion-production
threshold to extend our reach into the $\Delta$-resonance region; and a
detailed assessment of residual theoretical uncertainties. The work presented
here is part of a wider effort to describe elastic Compton scattering on the
proton, deuteron and ${}^3$He from the Thomson limit to well into the
$\Delta$-resonance region in one model-independent, unified framework. We focus
at present on improving the accuracy of current calculations by a full
next-to-next-to-leading order calculation with nucleons, pions and the
$\Delta(1232)$ as dynamical, effective degrees of
freedom~\cite{Griesshammer:2009pq,McGovern:2009sw,allofus}.


\section*{Acknowledgements}
We thank R.~Hildebrandt, J.~McGovern and D.~R.~Phillips for handy discussions
and helpful suggestions. The input and encouragement of our experimental
colleagues M.~W.~Ahmed, G.~Feldman, K.~Fissum, R.~Miskimen, L.~Myers and
H.~Weller was particularly important. HWG is grateful for the kind hospitality
of the Institut f\"ur Theoretische Physik III at Universit\"at
Erlangen-N\"urnberg, of the Institut f\"ur Theoretische Physik (T39) at TU
M\"unchen, of the Nuclear Experiment group of the Institut Laue-Langevin
(Grenoble, France), and to the organisers and participants of the INT workshop
08-38: ``Soft Photons and Light Nuclei'' and of the INT programme 10-01:
``Simulations and Symmetries'', which also provided financial support. This
work was carried out under National Science Foundation \textsc{Career} award
PHY-0645498 and US-Department of Energy grants DE-FG02-95ER-40907 (HWG and DS)
and DE-FG02-97ER41019 (DS).


\end{document}